\documentclass[useAMS, usenatbib]{mnras}
\usepackage{aas_macros}
\usepackage{natbib}

\usepackage{amsmath}
\usepackage{flexisym}
\usepackage{graphicx}

\usepackage{booktabs}
\usepackage{tabularx}
\usepackage[table]{xcolor}
\usepackage{multirow}

\def \arcsec {{\rm ~arcsec}}
\def \MHz {{\rm ~MHz}}
\def \GHz {{\rm ~GHz}}
\def \mJy {{\rm ~mJy}}

\title[LOFAR window on SF galaxies and AGN]
{The LOFAR window on star-forming galaxies and AGN -- curved radio SEDs and IR-radio correlation at $0<z<2.5$}

\author[]{G. Calistro Rivera,$^{1}$\thanks{Contact e-mail: \href{mailto:calistro@strw.leidenuniv.nl}{calistro@strw.leidenuniv.nl}}
W. L. Williams,$^{2}$  
M. J. Hardcastle,$^{2}$ 
K. Duncan,$^{1}$\newauthor
H. J. A. R\"ottgering,$^{1}$
P. N. Best,$^{3}$
M. Br\"uggen,$^{5}$ 
K. T. Chy\.zy,$^{6}$
C. J. Conselice,$^{7}$\newauthor
F. de Gasperin,$^{1}$
D. Engels,$^{5}$
G. G\"urkan$^{2,4},$
H. T. Intema,$^{1}$
M. J. Jarvis$^{8,9},$\newauthor
E. K. Mahony,$^{10,11}$
G. K. Miley,$^{1}$
L.K. Morabito,$^{8}$
I. Prandoni,$^{12}$
J. Sabater,$^{3}$\newauthor
D. J. B. Smith,$^{2}$
C. Tasse $^{13,14},$
P.P. van der Werf,$^{1}$
G.J. White$^{15,16}$ \\
$^{1}$Leiden Observatory, Leiden University, P.O. Box 9513, 2300 RA Leiden, The Netherlands\\
$^{2}$School of Physics, Astronomy and Mathematics, University of Hertfordshire, College Lane, Hatfield AL10 9AB, UK\\
$^{3}$  SUPA, Institute for Astronomy, Royal Observatory, Blackford Hill, Edinburgh, EH9 3HJ, UK\\
$^{4}$ CSIRO Astronomy and Space Science, 26 Dick Perry Avenue, Kensington, WA 6151, Australia \\
$^{5}$  Universit\"at Hamburg, Hamburger Sternwarte, Gojenbergsweg 112, 21029 Hamburg, Germany\\
$^{6}$ Astronomical Observatory, Jagiellonian University, ul. Orla 171, 30-244 Krak\'ow, Poland\\
$^{7}$ School of Physics and Astronomy, University of Nottingham, NG7 2RD, UK\\
$^{8}$ Astrophysics, University of Oxford, Denys Wilkinson Building, Keble Road, Oxford, OX1 3RH\\
$^{9}$ Physics and Astronomy Department, University of the Western Cape, Bellville 7535, South Africa \\
$^{10}$Sydney Institute for Astronomy, School of Physics A28, The University of Sydney, NSW 2006, Australia\\
$^{11}$ARC Centre of Excellence for All-Sky Astrophysics (CAASTRO), Australia\\
$^{12}$INAF-Istituto di Radioastronomia, Via P. Gobetti 101, 40129 Bologna, Italy \\
$^{13}$GEPI, Observatoire de Paris, CNRS, Universite Paris Diderot, 5 place Jules Janssen, 92190 Meudon, France \\
$^{14}$Department of Physics and Electronics, Rhodes University, PO Box 94, 6140 Grahamstown, South Africa \\
$^{15}$Department of Physics and Astronomy, The Open University, Walton Hall, Milton Keynes, MK7 6AA, England\\
$^{16}$RAL Space, STFC Rutherford Appleton Laboratory, Chilton, Didcot, Oxfordshire, OX11 0QX, England\\
}

\begin{document}

\date{Accepted 1988 December 15. Received 1988 December 14; in original form 1988 October 11}

\pagerange{\pageref{firstpage}--\pageref{lastpage}} \pubyear{2002}

\maketitle

\label{firstpage}

\begin{abstract}

We present a study of the low-frequency radio properties of star forming (SF) galaxies and active galactic nuclei (AGN) up to redshift $z=2.5$. The new spectral window probed by the Low Frequency Array (LOFAR) allows us to reconstruct the radio continuum emission from 150 MHz to 1.4 GHz to an unprecedented depth for a radio-selected sample of $1542$ galaxies in $\sim 7~ \rm{deg}^2$ of the LOFAR Bo\"otes field.
Using the extensive multi-wavelength dataset available in Bo\"otes and detailed modelling of the FIR to UV spectral energy distribution (SED), we are able to separate the star-formation (N=758) and the AGN (N=784) dominated populations.
We study the shape of the radio SEDs and their evolution across cosmic time and find significant differences in the spectral curvature between the SF galaxy and AGN populations. 
While the radio spectra of SF galaxies exhibit a weak but statistically significant flattening, AGN SEDs show a clear trend to become steeper towards lower frequencies.  
No evolution of the spectral curvature as a function of redshift is found for SF galaxies or AGN. 
We investigate the redshift evolution of the infrared-radio correlation (IRC) for SF galaxies  and find that the ratio of total infrared to 1.4 GHz radio luminosities decreases with increasing redshift: $ q_{ 1.4\GHz} = (2.45 \pm 0.04)\times(1+z)^{-0.15\pm0.03}$.
 Similarly, $ q_{150\MHz}$ shows a redshift evolution following $ q_{150\GHz} = (1.72 \pm 0.04)\times(1+z)^{-0.22\pm0.05}$. 
Calibration of the 150 $\MHz$ radio luminosity as a star formation rate tracer suggests that a single power-law extrapolation from $q_{1.4\GHz}$ is not an accurate approximation at all redshifts.

\end{abstract}


%%%%%%%%%%%%%%%%%%%%%%%%%%%%%%%%%%%%%%%%%%%%%%%%%%%%%%%%
%%%%%%%%%%%%%%%%%%%%%%%%%%%%%%%%%%%%%%%%%%%%%%%%%%%%%%%%

\section{Introduction }

%%%%%%%%%%%%%%%%%%%%%%%%%%%%%%%%%%%%%%%%%%%%%%%%%%%%%%%%
%%%%%%%%%%%%%%%%%%%%%%%%%%%%%%%%%%%%%%%%%%%%%%%%%%%%%%%%

%% INTRO
Radio selected samples of galaxies primarily consist of two populations: star-forming galaxies (hereafter SF galaxies) and active galactic nuclei (AGN).
As both star formation activity and black-hole growth in AGN are processes closely related to the overall mass growth of galaxies \citep{shapley11, bestANDheckman12}, radio-emitting galaxies therefore represent a unique laboratory for investigating the epoch of peak galaxy assembly, $1 < z < 3$ .
While studies of these galaxies from the far-infrared (FIR) to the ultraviolet (UV)  have contributed greatly to our understanding of galaxy evolution, statistically significant samples of radio-selected galaxies across cosmic time have only recently become available for exploration thanks to a new generation of radio facilities.
At low radio frequencies the unprecedented sensitivity and resolution of the Low Frequency Array \citep[LOFAR,][]{LOFAR} opens a new window for galaxy evolution studies.

%% DEFINITION: SYNCHROTON EMISSION
Low frequency radio emission in SF galaxies and AGN have different physical origins, although both are thought to be dominated by synchrotron emission. 
In SF galaxies the synchrotron emission is powered by high-energy electrons and positrons (cosmic rays, CRs), accelerated in supernova remnants (SNRs), that emit when interacting with the diffuse magnetic field of the galaxy \citep{condon92}.
Due to the short lifetime of the massive stars producing Type II and Type Ib supernovae, the synchrotron emission in SF galaxies is closely related to recent star formation, so that its emission (e.g. at 1.4 GHz) is widely used as a star formation tracer \citep{condon92, bell03, schmitt06, murphy11}.

In radio-selected AGN, the synchrotron radiation is ultimately powered by the central accreting black hole. However, the observed emission is believed to be emitted from regions which differ depending on the nature of the AGN \citep[see][for a review of AGN radio classifications]{smolcic16}.
Fundamental physical differences have been seen between radio AGN classified as high-excitation and low-excitation radio galaxies \citep[HERGs and LERGs, ][]{ hardcastle07, bestANDheckman12}, where HERGs appear to be the dominant population at high radio luminosities ($L_{1.4 \rm GHz} > 10^{26} \, \rm W\,Hz^{-1}$), while LERGs appear to dominate below this limit.
Most HERGs are observed to consist of three large-scale structures: jets, hotspots and lobes (consistent with \cite{fanaroff74} class II, FRII galaxies).
High-energy CRs first emit while being accelerated in the relativistic \emph{jet}, which transports them from the central AGN to shock regions called \emph{hotspots} \citep[e.g.][]{meisenheimer89}. Finally, the CRs expand away from the hotspot center, forming the large \emph{lobes} of synchrotron emission \citep[e.g.][]{krause12}.

%% PROBLEM

The synchrotron emission in all of these processes is commonly described by a power-law function $S(\nu)\propto \nu^{\alpha}$, where $S(\nu)$ is the radio flux density at a given frequency, $\nu$, and $\alpha$ the corresponding power-law slope.
Nevertheless, there are processes which can alter the shape of the spectra.
While in SF galaxies the synchrotron emission is typically observed to follow a slope of $\alpha\sim-0.7$ \citep[e.g.][]{gioia82, condon92, yun01}, intrinsic changes in the CR energy distributions or environmental and ISM processes (free-free absorption, ionization losses, and synchrotron self-absorption) can significantly alter the spectral shape of this emission \citep[][and references therein]{lacki13}, intrinsic changes in the CR energy distributions or environmental and ISM process . 
In AGN, both spectral ageing \citep[e.g.][]{harwood15} and the relative brightness of the different components (e.g. core, jets, hotspots and lobes) play an important role in shaping the integrated radio SEDs of radio AGN \cite[e.g.][]{hardcastle09}.

%% RECENT RESULTS
Multi-frequency spectral studies of the integrated radio SED of AGN are not widespread in the literature \citep[though there are a few exceptions,][]{laing80, ker12, singh13, kharb16, mahoney16}.
Recent literature has focused on morphological studies through spectral index maps \citep[e.g.][]{harwood15, vardoulaki15}.
A luminosity dependence of the spectral shape of the integrated radio spectrum was studied e.g. by \cite{laing80}, where they suggest radio AGN with $L_{1.4\, \rm GHz}<10^{25}\rm\, W\,Hz^{-1}$ have spectra that steepen at low frequencies, while brighter AGN show the opposite trend.  
Similarly, \citet{whittam16} observed spectral flattening for radio-selected galaxies at the high-frequency end (15.7 GHz) and suggest this may be due to the cores of \citet{fanaroff74} class I sources (FRI) becoming dominant at these high frequencies.
Deep multi-frequency radio observations of representative samples of galaxies are needed to study the radio SED and understand the physical processes shaping it.

Observational studies of the spectral properties of radio SF galaxies have so far focused on the local universe as sensitivity usually prevents the detection of statistically significant samples of galaxies at higher redshifts. 
An exception to this is the radio spectral slope study presented by  \citet[][]{ibar09, ibar10} for a sample of submillimetre galaxies (SMGs). 
They found no redshift evolution of the spectral indices for SMGs, although due to the sparse coverage of the radio SED they could not rule out the presence of curvature.
For local SF galaxies, spectral flattening towards low radio frequencies has been observed ($\nu < 1\rm \,GHz$), with thermal absorption or intrinsic synchrotron curvature as plausible explanations for this \citep{ israelmahoney90, clemens10,marvil14}.
Theoretical models explaining an alternative picture to the simple power-law shape which includes spectral curvature have also been developed \citep[e.g.][]{lacki13}.  

%% DEFINITION
In SF galaxies the radio emission is typically calibrated to trace star formation based on a tight empirical correlation with the IR radiation \citep[IR-radio correlation (IRC),][]{dejong85, helou85}.
In the local universe the IRC has been observed to be roughly linear across more than three orders of magnitude in FIR-luminosity ($10^9 < L_{\mathrm{IR, (8-1000} \mu m)} [\rm L_{\odot}]<10^{12.5}$) \citep{yun01, magnelli15}, for different galaxy classes (from dwarf to ultra-luminous IR galaxies, ULIRGs), and in star-forming regions within galaxies \citep[e.g.][]{dumas11, tabatabaei13}.
The basic understanding of the IRC relies on both the cold dust IR emission and the radio emission being tracers of recent star formation \citep[e.g. calorimeter model and conspiracy model,][respectively]{voelk89, lacki10}.
The possibility of a redshift evolution of the IRC has motivated an extensive debate from the observational point of view: while several studies claim the existence of significant redshift evolution\citep{seymour08, ivison10a, magnelli15, delhaize17}, a few studies suggest that such an evolution is a product of selection biases \citep[e.g.][]{appleton04, ibar08, jarvis10, sargent10b, bourne11} or a dust temperature dependence \citep[e.g.][]{smith14}.
In theoretical studies different trends as a function of redshift have also been discussed \citep[e.g.][]{schober16, schleicher13, lacki10}. 

%% QUESTIONS FOR THIS PAPER
In this work we take advantage of the unique sensitivity and resolution of LOFAR \citep{LOFAR, rottgering11}, combining deep 150 MHz radio observations \citep{williams16} and the wealth of ancillary data available in the Bo\"{o}tes field to address some of the key outstanding questions regarding the radio emission of galaxies, namely, what are the star formation and AGN contributions to the radio continuum at low frequencies?; does the low-frequency radio emission of SF galaxies exhibit a correlation with the IR luminosity as tight as that observed at 1.4 GHz?; and if so, how does the IRC evolve with redshift at both 1.4 GHz and 150 MHz?
Answering these questions is crucial to test the reliability of photometric-redshift estimations based on the radio--IR SEDs \citep[e.g.][]{yun02, dacunha15}. Finally, this study will allow us to investigate the low-frequency radio emission as a SFR tracer of galaxy populations at higher redshifts.

%% STRUCTURE
This paper is structured as follows. 
In section \ref{sec:data} we present the multi-frequency radio and FIR to UV data used for this study, while section \ref{sec:selection} describes the selection strategy for our sample. 
In section \ref{sec:sedfitting} we discuss the SED analysis and classification into SF galaxies and AGN populations. 
The analysis of the spectral slope and curvature for the two populations is described in section
\ref{sec:radiocontinuum}.
Next, the IRC in our data is investigated in section \ref{sec:IRC} and section \ref{sec:SFRtracer} discusses the low frequency radio emission as a SFR diagnostic. 
Finally, section \ref{sec:summary} summarizes our findings. 
Throughout the paper we adopt a concordance flat $\Lambda$-CDM cosmology with $H_0=70\, \rm km \,s^{-1}\, Mpc^{-1}$, $\Omega_{\mathrm{m}}=0.3$, and $\Omega_{\Lambda}=0.7$ \citep{komatsu09,planck14} and all quoted magnitudes assume the AB system \citep{okegunn83} unless otherwise specified.

%%%%%%%%%%%%%%%%%%%%%%%%%%%%%%%%%%%%%%%%%%%%%%%%%%%%%%%%
%%%%%%%%%%%%%%%%%%%%%%%%%%%%%%%%%%%%%%%%%%%%%%%%%%%%%%%%

\section{Survey data}\label{sec:data}

%%%%%%%%%%%%%%%%%%%%%%%%%%%%%%%%%%%%%%%%%%%%%%%%%%%%%%%%
%%%%%%%%%%%%%%%%%%%%%%%%%%%%%%%%%%%%%%%%%%%%%%%%%%%%%%%%

The National Optical Astronomy Observatory (NOAO) Deep Wide-field survey \citep[NDWFS;][]{januzziANDdey99} targeted the sky seen towards the constellation of Bo\"otes as one of its deep extragalactic fields.
Originally the NDWFS covered $\rm 9~deg^2$ in the optical and near-infrared B$_{\rm W},$ R, I and K bands. 
Since then, the Bo\"otes Field has been surveyed across the electromagnetic spectrum, including deep X-ray \citep{murray05}, mid-infrared \citep{jannuzi10} and far-infrared \citep{oliver12} photometric observations as well as extensive spectroscopic surveys \citep[e.g.][]{kochanek12}. The field therefore represents one of the richest multi-wavelength datasets among the wide deep extragalactic surveys and is complementary to the deep radio observations provided by LOFAR.

%%%%%%%%%%%%%%%%%%%%%%%%%%%%%%%%%%%%%%%%%%%%%%%%%%%%%%%%
\subsection{LOFAR 150 MHz }
    
\begin{figure}
\centering
\includegraphics[trim={0.7cm 0 0 1cm},clip,width=1.1\linewidth]{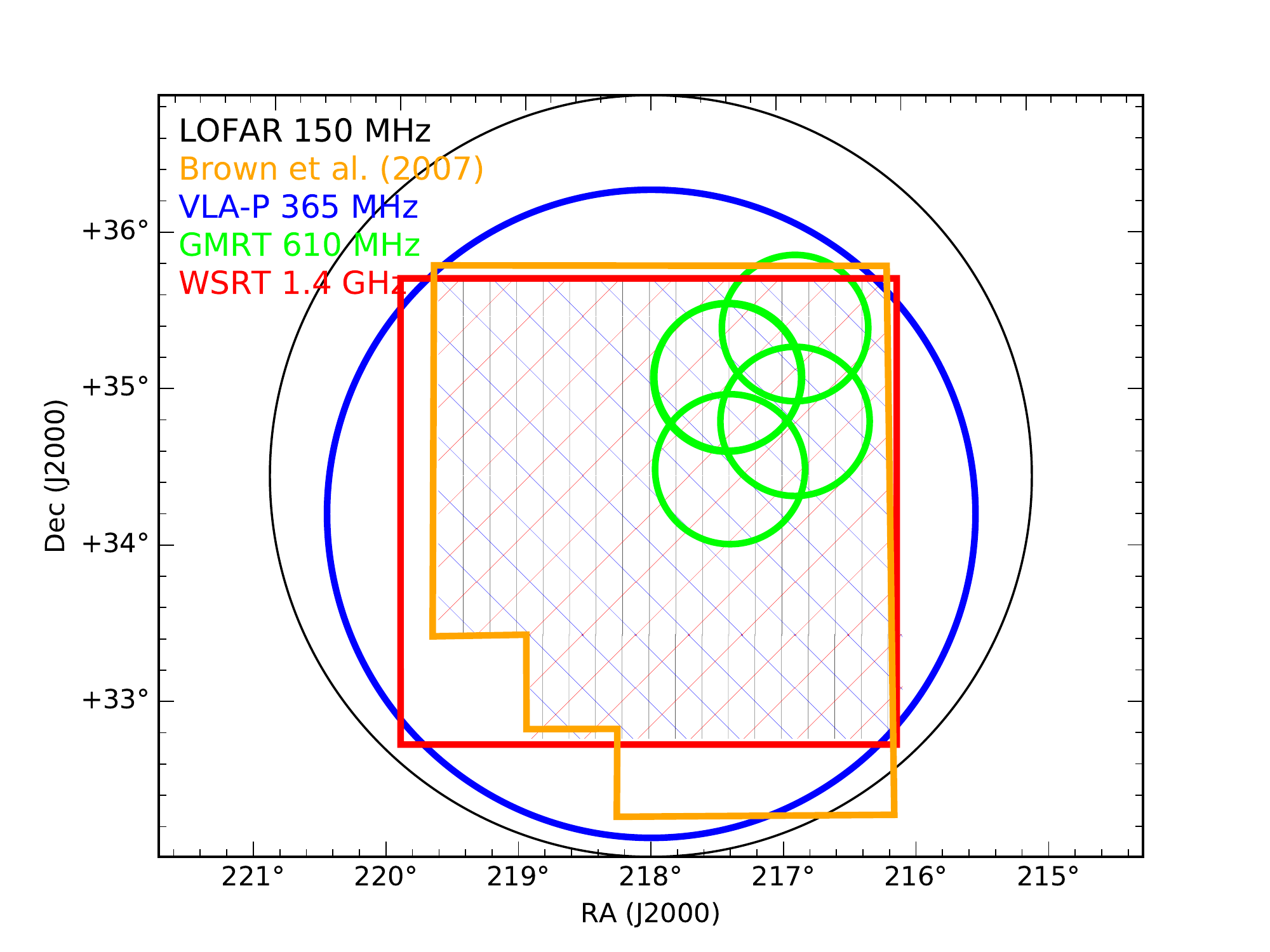}
\caption{Spatial coverage of the optical and radio photometry including the I-band selected photometry presented by \citet{brown07} and the four deep catalogues at 150, 325, 610, 1400 MHz, available for the Bo\"otes field. The dashed area is the area chosen for our study.}
\label{fig:radiocoverage}
\end{figure}

The first source catalogue used for our sample selection is based on 150 MHz radio observations of the Bo\"otes field \citep{williams16} taken with the LOFAR High Band Antennae (HBA). 
The calibration and imaging were achieved with the `Facet' calibration scheme presented by \cite{vanweeren16}, which corrects for direction dependent effects (DDEs) caused by the ionosphere and imperfect knowledge of LOFAR station beam shapes.
The resulting image, as presented by \cite{williams16}, covers 19 deg$^2$ with an rms noise of $\sigma_{\rm rms} \sim 120 - 150~ \mu$Jy~beam$^{-1}$ in the central region of the field and a resolution of $5.6 \times 7.4 \arcsec$.
These values represent up to more than one order of magnitude improvement compared to the images existing at this wavelength \citep{intema11, williams13}. 
The source extraction for the catalogue was done using the Python Blob Detection and Source Measurement software,  \citep[\textsc{PyBDSM}:][]{mohan15}, which performs Gaussian fitting to decompose radio interferometry images, grouping the Gaussians together into individual sources where appropriate.
The LOFAR 150 MHz radio source catalogue contains 6276 sources detected with a peak flux density threshold of 5$\sigma_{\rm rms}$ within the coverage region shown in Fig. \ref{fig:radiocoverage}.
The LOFAR 150 MHz luminosity distributions for our total sample classified into SF galaxies and AGN are shown in the lower panel of Figure \ref{fig:zandlumdist}.

%%%%%%%%%%%%%%%%%%%%%%%%%%%%%%%%%%%%%%%%%%%%%%%%%%%%%%%%
\subsection{Radio photometry for the Bo\"otes field}\label{subsec:radiophotometry}

The construction of radio SEDs for the sources in the Bo\"otes field is one of the main aims of this paper.
Conveniently, the Bo\"otes field has been covered by several previous radio surveys.
Among these, the high sensitivity and the resolution of the VLA-P Survey at 325 MHz, GMRT observations at 608 MHz and the deep WSRT catalogue at 1.4 GHz, make them the best data sets to complement the LOFAR data in our study.

The VLA-P catalogue \citep{coppejans15} used in this investigation, was drawn from a 324.5 MHz image of a radius of 2.05$^{\circ}$ in the NOAO Bo\"otes field using Karl G. Jansky Very Large Array (VLA) P-band observations. 
The image covered a single pointing and has a resolution of 5.6$\times$5.1 arcsec with a central noise of 0.2 mJy beam$^{-1}$ increasing to 0.8 mJy beam$^{-1}$ at the edge of the image.
The source extraction for the catalogue was done using the PyBDSM software, \citep[\textsc{PyBDSM}:][]{mohan15}.
The source detection threshold for the construction of the catalogue for this study is 3 $\sigma_{\rm rms}$. This is a reliable detection limit since our study includes only sources from the catalogue which have LOFAR detected counterparts.
As discussed by \cite{coppejans15}, the VLA-P catalogue was matched to WENSS \citep{WENSS} to check the absolute flux density scale and the primary beam correction. 

The GMRT image at 608 MHz is a mosaic constructed from four pointings covering 1.95 deg$^2$ of the Bo\"otes field (project code 28\_064).  
The mosaic has a resolution of $5 \times 5$ arcsec and a noise level of $\sigma_{\rm rms}\sim 40-70~\mu$Jy beam$^{-1}$.
Primary flux density calibration was done with 3C286 using the wideband low-frequency flux density standard of \cite{scaife12}.
The source extraction for the catalogue was also performed using the PyBDSM software, \citep[\textsc{PyBDSM}:][]{mohan15}.

The 1.4 GHz data \citep{devries02} are drawn from the deep (16$\times$12 hr) Westerbork Synthesis Radio Telescope (WSRT) observations of the approximately 6.68 deg$^2$ Bo\"otes Deep Field. 
The image covers 42 discrete pointings and has a limiting sensitivity of $\sigma_{\rm rms} \sim 28\, \mu$Jy beam$^{-1}$ and a resolution of 13$\times$27 arcsec. 
The source extraction in the public catalogue was done using automated routines described in detail by \cite{WENSS}, which consist basically of Gaussian fitting to islands of detected brightness in the radio map.
The source detection threshold for the inclusion of the flux densities in the catalogue is 5 times the local rms, resulting in the full catalogue containing 3172 sources.
Since the Bo\"otes field has been covered by previous radio surveys at this wavelength such as the NRAO VLA Sky Survey \citep[NVSS;][]{NVSS} and the Faint Images of The Radio Sky at 20 cm Survey \citep[FIRST;][]{FIRST}, these data were used to calibrate the survey flux densities and positions. 

Assuming a spectral index of $\sim$-0.7, the LOFAR measurements offer sensitivity values comparable to the 1.4 GHz WSRT map presented by \cite{devries02} (rms: 28 $\mu$Jy beam$^{-1}$) and resolution values comparable to the 365 MHz VLA-P map published by \cite{coppejans15} ($5.6 \times 5.1 \arcsec$), which are the best radio data available at the respective frequencies to date.

\begin{table}
\begin{minipage}{8cm}
\caption{Multiwavelength coverage of the  LOFAR+I-band selected sample in the Bo\"otes field. The column of `Detections' specifies which fraction of the final sample has counterparts in the respective bands.}

\hspace{-2cm}
\begin{center}
\begin{tabular}{ c| c| c | c  }
\hline
& Band & $\nu[Hz], \lambda[m]$ 

&Detections [\%]\\
\hline \hline
\rowcolor{lightgray}\multirow{1}{*}{Selection} & LOFAR & 150 MHz & 100\% \\
\rowcolor{lightgray} &I & 806 nm & 100\% \\
 
\hline
\multirow{2}{*}{UV}& NUV & 300 nm &  93\%\\
&u & 365 nm & 99\% \\
\hline
\multirow{2}{*}{optical} &B$_{\rm W}$& 445 nm &100\%\\
&R& 658 nm&100\%\\
\midrule
\multirow{5}{*}{NIR} &z& 900 nm & 99\%\\
&Y& 1020 nm & 99\%\\
&J& 1222 nm &100\%\\
&H& 1630 nm &100\%\\
&K &2190 nm&100\%\\
\hline
\multirow{5}{*}{MIR} & IRAC1 & 3.6 $\mu$m & 100\% \\
&IRAC2&4.5 $\mu$m&100\%\\
&IRAC3&5.8 $\mu$m&100\%\\
&IRAC4&8.0 $\mu$m&100\%\\
&MIPS24&24.0 $\mu$m&100\%\\

\midrule
\multirow{3}{*}{FIR} & SPIRE & 250 $\mu$m& 83\%   \\
&SPIRE &350$\mu$m & 79\% \\
&SPIRE& 500 $\mu$m& 67\%\\
\midrule
\multirow{3}{*}{RADIO}&WSRT& 1.4 GHz&67\% (88\%\footnote{\label{fluxcut}After the flux cut applied for the curvature analysis (section \ref{subsec:alphahighandlow})})\footnote{\label{forced}Forced photometry included for non-detections}\\
&GMRT & 610 MHz&23\%\footnote{Due to spatial partial coverage, see Fig.~\ref{fig:radiocoverage} }\\
&VLA-P & 325 MHz&63\% (88\%$^{\ref{fluxcut}}$)$^{\ref{forced}}$\\
\hline
\end{tabular}
\end{center}
\label{table:multi-wavelength}
\end{minipage}
\end{table}

% --------------------------------------------------------
%  FIGURE COVERAGE : (2. radio coverage)
%---------------------------------------------------------
% ----------------------------------------------------------

%%%%%%%%%%%%%%%%%%%%%%%%%%%%%%%%%%%%%%%%%%%%%%%%%%%%%%%%
\subsection{Optical and Infrared Photometry}

The second source-catalogue used for our sample selection is the combined I-band-selected psf-matched photometry catalogue presented by \cite{brown07}. The photometric bands included in the catalogue are presented in Table \ref{table:multi-wavelength} covering a wide range of wavelengths, spanning from 0.15 to 24 $\mu$m. 

This catalogue includes the original NDWFS observations in the optical and near- infrared; B$_{\rm W},\,$ R, I and K bands. 
With an absolute positional uncertainty of $< 0.1$ arcsec, the I-band images reach depths of 24.9 AB magnitude ($5 \sigma_{\rm rms}$ within 2 arcsec diameter aperture).

Mid-infrared (MIR) counterparts are drawn from the \textit{Spitzer} Deep Wide-field Survey \citep[SDWFS;][]{ashby09}, using the InfraRed Array Camera (IRAC) instrument on the \textit{Spitzer} Space Telescope, providing images at 3.6, 4.5, 5.8, and 8.0 $\mu$m.
Infrared 24 $\mu$m photometry from \textit{Spitzer} was provided by the Multiband Imaging Photometer (MIPS) AGN and Galaxy Evolution Survey (MAGES; \cite{jannuzi10}), while deep J, H, K$_{s}$ photometry was drawn from the NOAO Extremely Wide-field Infrared Imager \citep[NEWFIRM; ][]{autry03} survey.
A fraction of the total Bo\"otes field is covered by the z-Bo\"otes survey \citep{cool07}, providing photometry in the z-band for 7.62 deg$^2$ of the field, with some gaps in the coverage due to the 10 arcmin gaps in the instrument CCDs.
Y and U$_{\rm spec}$ bands photometry is provided using images from the Large Binocular Camera (LBC) mounted on the Large Binocular Telescope (LBT).
Finally, NUV photometry (1800-2750 \r{A}) has been included from the GALEX/GR6 surveys \citep{martin03, bianchi14}. 

As described in detail by \cite{brown07}, this psf-matched catalogue wass constructed by regridding and smoothing the individual survey images corresponding to the  u-, B$_{\rm w}$-,R-, I-, z-, Y-, J-, H- and K-bands to a common scale so that the sources' point spread functions (PSF) are Moffat profiles. Fluxes are extracted from these images for all the sources with I-band detections using SExtractor \citep{SExtractor}, while for the remaining bands (IRAC and MIPS bands) aperture fluxes were extracted. Regions surrounding very extended galaxies and saturated stars were excluded.

In total, the combined catalogue provides photometry for $\sim 830$k sources, including galaxies and AGN with I-band magnitudes $I < 24$.

% --------------------------------------------------------
%  FIGURE Z AND LUM DISTRIUTIONS
%---------------------------------------------------------
\begin{figure}
\centering
\includegraphics[trim={0.7cm 0 0 0.5cm},clip,width=1.0\linewidth]{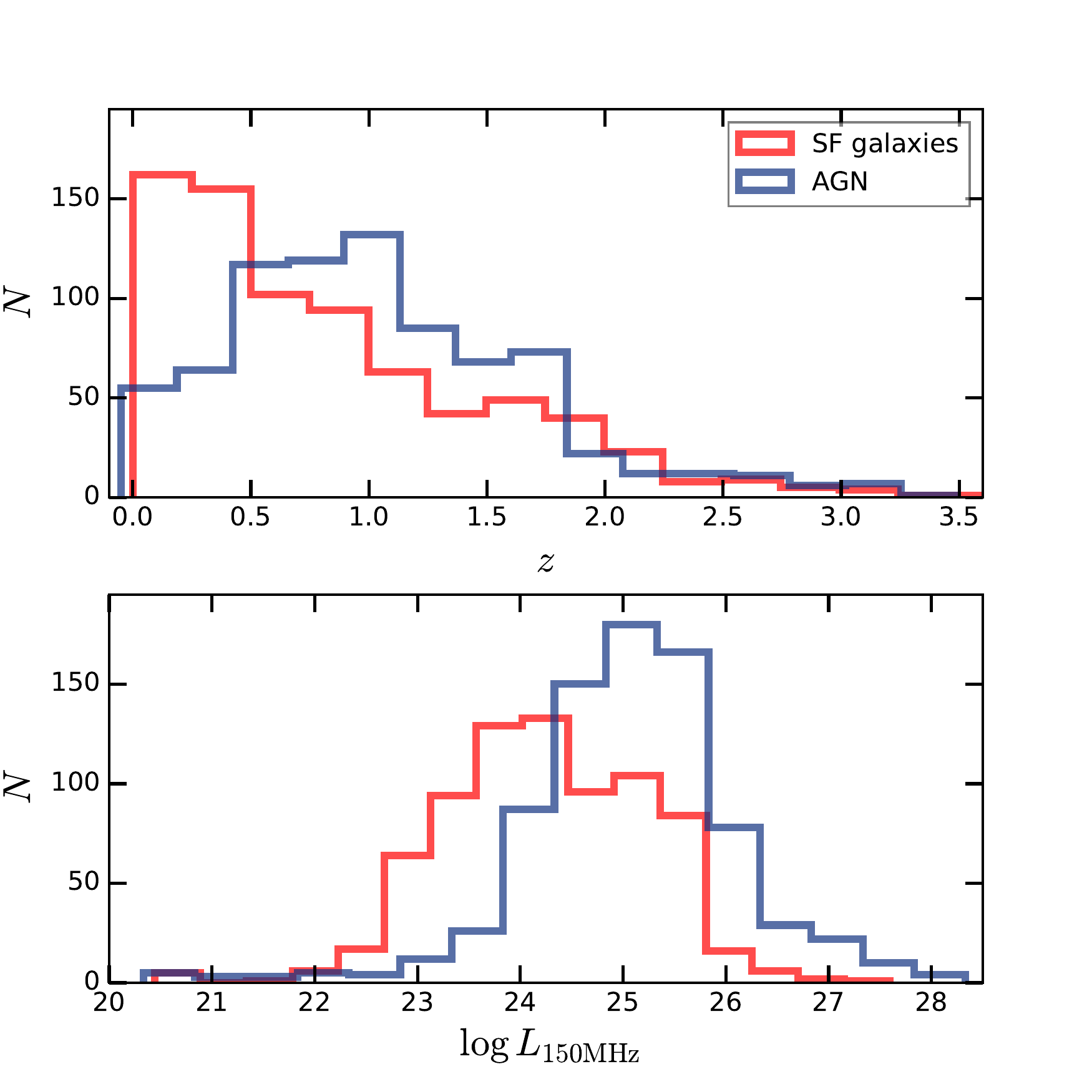}
\caption{Redshift and LOFAR 150 MHz luminosity distributions for the SF galaxies (red) and AGN (blue) populations. The classification of the total sample into SF galaxies and AGN is explained in detail in section \ref{subsec:classification}.}
\label{fig:zandlumdist}
\end{figure}
% ----------------------------------------------------------
%%%%%%%%%%%%%%%%%%%%%%%%%%%%%%%%%%%%%%%%%%%%%%%%%%%%%%%%
%%%%%%%%%%%%%%%%%%%%%%%%%%%%%%%%%%%%%%%%%%%%%%%%%%%%%%%%
\subsection{Far-infrared Photometry: HerMes DR3}

Far-infrared photometry is a key ingredient for this study, as it allows us to constrain total infrared luminosities and trace SFRs of the sources.
To increase the infrared spectral coverage of the final sample, we added SPIRE data at 250, 350 and 500 $\mu$m from the \textit{Herschel} Multi-tiered Extragalactic Survey \citep[HerMes;][]{oliver12}. 
One of the greatest challenges in measuring fluxes at such long wavelength is the high confusion noise, a result of the large SPIRE beam sizes.
Since one effect of confusion is that it increases the positional uncertainty of sources \citep[e.g., ][]{hogg01}, cross-identification with other wavelengths becomes very challenging. 
To deal with this issue, we use the third data release (DR3) cross-identification catalogues, which are selected with positional priors at 24 $\mu$m and obtained with the technique described by \cite{roseboom10}. 
This technique consists basically of performing cross-identifications in map-space so as to minimise source blending effects by using a combination of linear inversion and model selection techniques. 
In this way they produce reliable cross-identification catalogues based on Spitzer MIPS 24µm source positions, giving significantly greater accuracy in the flux density compared to other traditional source recovery methods and so recovering a much larger fraction of faint SPIRE sources.

The FIR photometry was included in the multi-wavelength data by cross-correlating the source selection catalogue described above with the \textit{Hermes} DR3 catalogue, using the task \textsc{TSKYMATCH2} from the \textsc{STILTS} software package \citep[\textsc{TOPCAT} implementation,][]{taylor06}. 
Using the optical position of the sources in the \cite{brown07} catalogue, \textsc{TSKYMATCH2} returned the SPIRE fluxes associated to all sources with MIPS 24$~\mu$m to optical separations of r$_{24-opt} < 3$ arcsec (less than half the FWHM of the MIPS 24$~\mu$m beam).

The percentage of sources with counterparts at different wavelengths are listed in Table \ref{table:multi-wavelength}.

\subsection{Redshifts}

The spectroscopic AGN and Galaxy Evolution Survey (AGES) has covered 7.7 deg$^2$ of the Bo\"otes Field providing spectroscopic redshifts for 23 745 galaxies and AGN. 
However, given that the majority of sources at $z>1$ in the AGES catalogue are QSOs, it is crucial to estimate robust photometric redshifts, as this study focuses on both AGN and star forming galaxies.
Spectroscopic redshifts are available for around 45 and 35 per cent of our total selection of SF galaxies and AGN respectively (the selection and classification is described in sections \ref{sec:selection} and \ref{sec:sedfitting}) . 
Photometric redshifts are estimated as described below for the remaining fraction of the sources. 

% --------------------------------------------------------
%  FIGURE COMPLETENESS LOFAR + I-band selection
%---------------------------------------------------------
\begin{figure}
\centering
\includegraphics[trim={0.6cm 0.6cm 0 0},clip,width=1.0\linewidth]{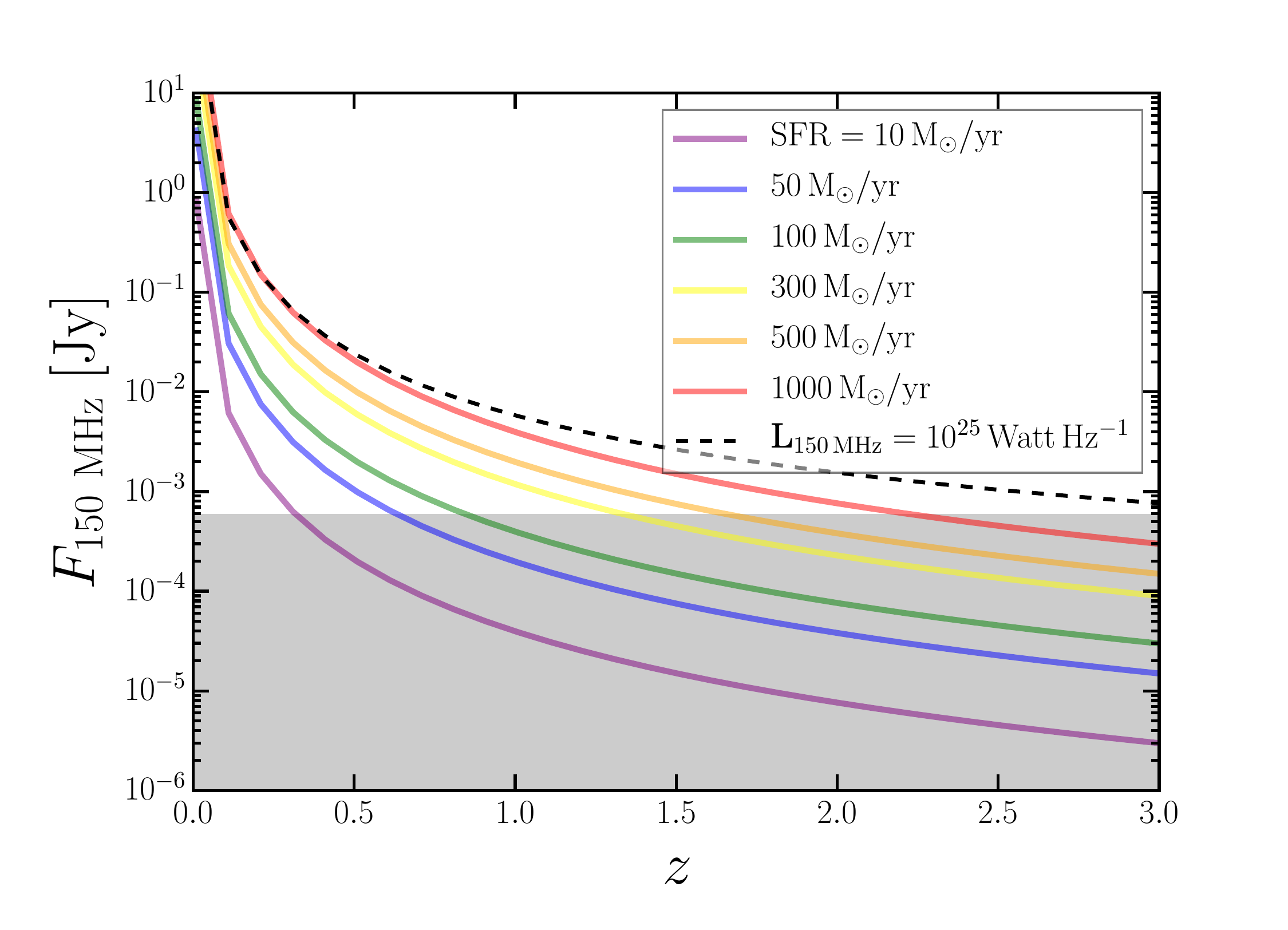}
\caption{Test for completeness of the LOFAR selection. The solid lines depict the redshift evolution of the expected radio fluxes for different SF galaxies and are colour-coded by SFR. The 150 MHz luminosities where calculated based on the IR-radio correlation, assuming $q_{1.4\,\rm GHz}=2.3$ and a spectral index of $\alpha=-0.7$ for the extrapolation from 1.4 GHz to 150 MHz. The dashed line corresponds to the 150 MHz flux evolution as a function of redshift for AGN of $\rm L_{150\rm MHz}> 10^{25} W\, Hz^{-1}$. The grey shaded area displays the flux limit implied in our radio selection, which correspond to 5 $\sigma_{\rm rms}.$}
\label{fig:completeness}
\end{figure}
% ----------------------------------------------------------
Photometric redshifts are provided by the optimised catalogue described by Duncan et al. (in prep.), produced using the Bo\"otes photometry of \cite{brown07}, presented above. 
The innovative method used for this redshift catalogue consists of combining three different $z_{\rm phot}$-estimation methods, by using the \textsc{EAZY} photometric redshift software \citep{EAZY} customized with three different template sets: one set of stellar only templates \citep[EAZY default library;][]{EAZY} and two sets including AGN and QSO contributions \citep[SWIRE;][]{polletta07} and \citep[Atlas of Galaxy SEDs;][]{brown14}. 
This methodology has been motivated by the fact that comparisons of $z_{\rm phot}$-estimation techniques have shown that for a suite of multiple z-estimates, the ensemble average estimates offer significant statistical improvements in redshift accuracy compared to the estimates of any single set of predictions \citep{dahlen13, carrasco14}.
These three individual $z_{\rm phot}$ estimates were then combined using a Hierarchical Bayesian combination method \citep{dahlen13}, as an alternative to a straight addition of the probability distributions of the three $z_{\rm phot}$ estimates. 
The main advantage of this method is that it determines the consensus probability \textit{P}$_{z_{\rm phot}}$ for each object, given the possibility that the individual measured probability distributions may be wrong.
These results were also optimised using zero-point offsets calculated from the spectroscopic redshift sample.
The redshift distributions for the total sample classified into SF galaxies and AGN are shown in the upper panel of Figure \ref{fig:zandlumdist}.

%%%%%%%%%%%%%%%%%%%%%%%%%%%%%%%%%%%%%%%%%%%%%%%%%%%%%%%%%%%%%%%%%%%%%%%%%%%%%%%%%%%%%%%%%%%%%%%%%%%%%%%%%%%%%%%%%%%%%%%%%%%

%%%%%%%%%%%%%%%%%%%%%%%%%%%%%%%%%%%%%%%%%%%%%%%%%%%%%%%%
%%%%%%%%%%%%%%%%%%%%%%%%%%%%%%%%%%%%%%%%%%%%%%%%%%%%%%%%

\section{Sample Selection} \label{sec:selection}
    
%%%%%%%%%%%%%%%%%%%%%%%%%%%%%%%%%%%%%%%%%%%%%%%%%%%%%%%%
%%%%%%%%%%%%%%%%%%%%%%%%%%%%%%%%%%%%%%%%%%%%%%%%%%%%%%%%    
    
The selection strategy for the sample of galaxies and AGN in the Bo\"otes Field involved several steps. 
First, sources detected both in LOFAR 150 MHz and I-band were selected to build the primary source catalogue, which is additionally complemented by multi-wavelength ancillary data. 
Second, we constrained this catalogue to sources with positions within the region of overlap among the 150 MHz, 325 MHz and 1.4 GHz radio maps (see Section \ref{subsec:radiophotometry} and Fig.~\ref{fig:radiocoverage}). 
Finally, we classified these sources into SF galaxies and AGN-dominated sources using the \textsc{AGNfitter} algorithm \citep{calistrorivera16} and restrict the posterior analysis on the sub-samples separately.

%%%%%%%%%%%%%%%%%%%%%%%%%%%%%%%%%%%%%%%%%%%%%%%%%%%%%%%%    
\subsection{LOFAR + I-band selection}

The identification of optical counterparts to the LOFAR detections from the \cite{brown07} catalogue is described in detail by Williams et al., in prep. 
We use the likelihood ratio (LR) method \citep{richter75} to quantify the probability of an I-band detected galaxy being the true counterpart of the radio emission observed in the LOFAR map. 
The fully developed method used is described by \cite{tasse08}.
From the 3317 LOFAR-detected sources, which lie within the 9 deg$^2$ boundary of the \cite{brown07} catalogue, 2326 (70 per cent) are found to have unique optical counterparts.

Fig.~\ref{fig:completeness} shows the implications of our radio selection for the nature of our sample. 
Lines correspond to the expected radio fluxes at 150 MHz, based on the IR-radio correlation, assuming $q_{1.4\,\rm GHz}=2.3$ and a spectral index of $\alpha=-0.7$ for the extrapolation from 1.4 GHz to 150 MHz. 
We choose to use a value of $q_{1.4\,\rm GHz}=2.3$ based on the observed median value of $q_{1.4}$ found for our sample (see Fig. \ref{fig:q-hist}).
The flux cut from our radio selection is represented by the grey shaded area.
Due to our selection, at $z<1.3$ our sample appears to be complete for SF galaxies of $SFR>300$  M$_{\odot}$/yr (yellow line).
At redshifts $1.5<z<2.3$ only galaxies with $SFR>1000~M_{\odot}$/yr are represented in a complete way according to the LOFAR selection.
We expect to be complete for AGN of $\rm L_{150\rm MHz}> 10^{25} W\, Hz^{-1}$ at all redshifts.
We thus conclude that at redshifts $z>1.5$ our sample is limited to highly SF galaxies (starbursts) and radio-selected AGN.

% --------------------------------------------------------
%  FIGURE AGN SED
%-------------------------
\begin{figure*}
\centering
\includegraphics[width=1.0\linewidth]{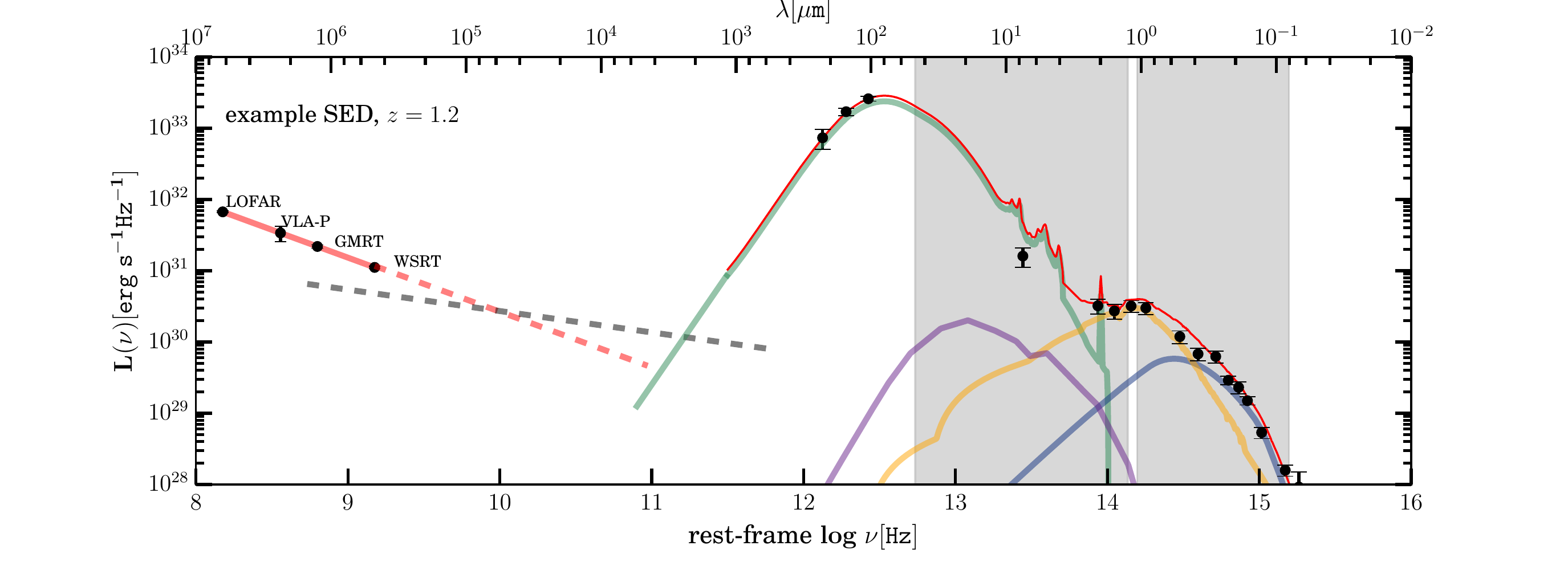}
\caption{Example SED of dusty SF galaxy from our sample at z=1.2. The multi-wavelength coverage of our total sample is shown here, from the low-frequency radio till the UV regime. The radio data points represent the radio SED covered in this study: 150, 325, 610 MHz and 1.4 GHz. The FIR-UV photometry is decomposed into physical components of the host galaxy and the AGN using the code \textsc{AGNfitter}. The integrated luminosities used in our classification scheme were computed by integrating the SED within the areas represented here as the grey-shaded rectangles. The left rectangle covers the integration limits for the main IR components: the galactic cold dust emission (green line) and the hot dust emission from the AGN torus (purple line). The right rectangle is the integration area for the optical main components: the stellar emission (orange line) and the accretion disk emission (blue line).The thin red line represents the total fitted SED. The dashed red and grey lines correspond to the synchroton and thermal contributions to the radio emission, respectively.}
\label{fig:SED}
\end{figure*}
%----------------------------------------------------------
%%%%%%%%%%%%%%%%%%%%%%%%%%%%%%%%%%%%%%%%%%%%%%%%%%%%%%%%
\subsection{Selection for radio SED construction}\label{subsec: selectionradio}

The radio SEDs constructed from the data described in Section \ref{subsec:radiophotometry} are built up mostly with three data points: LOFAR 150 MHz, VLA-P 365 MHz and WSRT 1400 MHz (Fig.~\ref{fig:radiocoverage}), complemented with a partial coverage of GMRT observations at 610 MHz. 
From the 2326 LOFAR sources with optical counterparts, 1955 sources lie within the region of overlap among the 150 MHz, 325 MHz and 1.4 GHz radio maps required for our study.
To study the shape of the radio continuum we need reliable spectral index measurements along the total frequency coverage. 
As spectral index measurements are prone to be easily affected by systematic offsets between different bands, special care should be exercised in having an homogeneous flux scales and in dealing with non-detections. 
We dedicate this section to this purpose, review all systematics intrinsic to each data set and corroborate the homogeneity of the flux scaling.

\subsubsection{Flux scales}

The combination or comparison of radio maps at different frequencies requires them to undergo a standard calibration scaling.
For this study we adjust the fluxes of the four different bands to the scale that is most accurate at low radio frequencies, that of \cite{scaife12}.
Since the main uncertainties of using a standard calibration scale are at frequencies lower that 1 GHz we focus our investigation on the LOFAR and VLA-P data sets.
However, we note that the scaling of the high-frequency range covered by the GMRT and WSRT datasets was verified to also be consistent with the scale by \citet{scaife12}. 

To investigate the reliability of LOFAR fluxes  at 150 MHz, \citet{williams16} have compared high signal-to-noise sources to three other data sets with known calibrations in the Bo\"otes Field: NVSS (at 1.4 GHz), WENSS \citep{WENSS} (at 365 MHz) and VLSSr \citep{VLSSR} (at 74 MHz), where the WENSS fluxes had been scaled a priori by a factor of 0.9 to be consistent with the \cite{scaife12} scale.
The inference of the correction factor for the LOFAR fluxes consists of calculating the spectral indices between the  lower  (74  MHz)  and  higher  (325,  1400  MHz)  frequencies and predicting the LOFAR flux density at the central frequency (150 MHz).
The mean flux density ratio between the predicted and uncorrected LOFAR flux was found to be $1.01 \pm 0.1$. Finally, as described by \citet{williams16}, this factor was used to adjust the LOFAR flux densities to the \cite{scaife12} flux scale.

The reliability of the VLA-P flux-scale was tested using two independent approaches which yielded consistent results.
First, we investigated the calibrator source (3C216) that lies in the Bo\"otes field. We calculated the ratio the 325 MHz flux measured during the calibration of the map (which assumes the flux scale of Perley\&Butler 2010) and the 325 MHz flux predicted by \citet{scaife12}. A correction factor of 0.91 was needed to adjust the calibrated fluxes to the standard scale of \cite{scaife12}. 
In a second approach, we compared high signal-to-noise VLA-P sources with their counterpart fluxes from the WSRT, GMRT and the LOFAR catalogues. The spectral indices between the lower (150 MHz) and higher frequencies (610 MHz, 1.4 GHz) were calculated and we could predict the VLA-P flux density at the central frequency (325 MHz).
The mean flux density ratio between the predicted and uncorrected VLA-P flux was $0.91 \pm 0.02$, in agreement with the data-independent method presented above.

\subsubsection{Non-detections}\label{subsubsec:nondetections}

Although the sensitivity of the LOFAR image is equivalent to the WSRT sensitivity under the assumption of a spectral index of $\alpha \sim 0.7$, 33 per cent of the LOFAR selected sources are not detected in WSRT. 
This issue is also relevant for the VLA-P data, which are less sensitive ($\sigma_{\rm rms} \sim  0.2 ~\mJy~\rm beam^{-1}$) than the equivalent LOFAR image at this frequency for the same assumed $\alpha$ ($ \sigma_{\rm rms} \sim 0.074~\rm mJy~ beam^{-1} $). We find that 37 per cent of the LOFAR selected sources are not detected in VLA-P.

To estimate the fluxes of the sources undetected in WSRT and VLA-P we use forced photometry technique.
This process consists of extracting aperture fluxes from the radio maps at the location of known LOFAR selected sources. 
To calibrate this process we first applied the forced photometry on catalogued sources.
We found that aperture fluxes were sensitive to artifacts of the imaging process (e.g. negative fluxes) and to reduce this effect, we used apertures smaller than the WSRT and VLA-P beam sizes.
Finally, we used our results on catalogued sources to derive a correction factor, which was applied on the undetected sources.
The flux uncertainties on the forced photometry were calculated by repeating the same flux extraction process as above using the corresponding rms-maps.

Although the aperture fluxes have been determined to be reliable, the results on radio continuum that are based on forced photometry (see Section \ref{sec:radiocontinuum}), are carefully differentiated from those of SEDs fully sampled with detected sources, using forced photometry fluxes just as upper limits.
Moreover, for more specific tests in the following sections we apply conservative flux cuts to minimize the fraction of non-detections in our data.

%%%%%%%%%%%%%%%%%%%%%%%%%%%%%%%%%%%%%%%%%%%%%%%%%%%%%%%%

\section{Analysis of AGN and SF galaxies with SED-fitting}\label{sec:sedfitting}

The FIR to UV SEDs of the sources in our sample were decomposed into different physical contributions through fitting the multi-wavelength photometry using the SED-fitting algorithm \textsc{AGNfitter}.
Some examples of the fitting output are included in Appendix \ref{app:SED}.
An advantage of using \textsc{AGNfitter} is that it is based on a Markov Chain Monte Carlo (MCMC) technique and infers the probability density functions (PDFs) of the physical parameters.
This way it provides a robust calculation of their uncertainties and improving the recognition of correlations and degeneracies among them.

The total active galaxy model in \textsc{AGNfitter} consists of the superposition of the host galaxy emission and the nuclear AGN emission. 
The host galaxy emission is modelled as a combination of a stellar component and the reprocessed emission of cold/warm dust in starburst regions. 
At nuclear scales, the AGN emission is modelled as a combination of an accretion disk component (Big Blue Bump, BBB) and a hot dust 'torus' component.

A large set of relevant physical parameters for the galaxy (SFR, $M_{*}$) and AGN (N$_{\rm H}$-torus, $L_{\rm bol}$) are calculated, some of which are detailed below. 
However, for full details we refer the reader to \citet{calistrorivera16}.
From the 1955 LOFAR-I-band selected sources within the region of overlap between the multi-frequency radio data (section \ref{subsec: selectionradio}), 1542 ($\sim 79$ per cent) can be classified as 758 SF galaxies and 784 AGN.

% --------------------------------------------------------
%  FIGURE \textsc{AGNfitter} SELECTION
%---------------------------------------------------------
\begin{figure}
\includegraphics[trim={0.2cm 0.3cm 0cm 0.25cm},clip,width=1.05\linewidth]{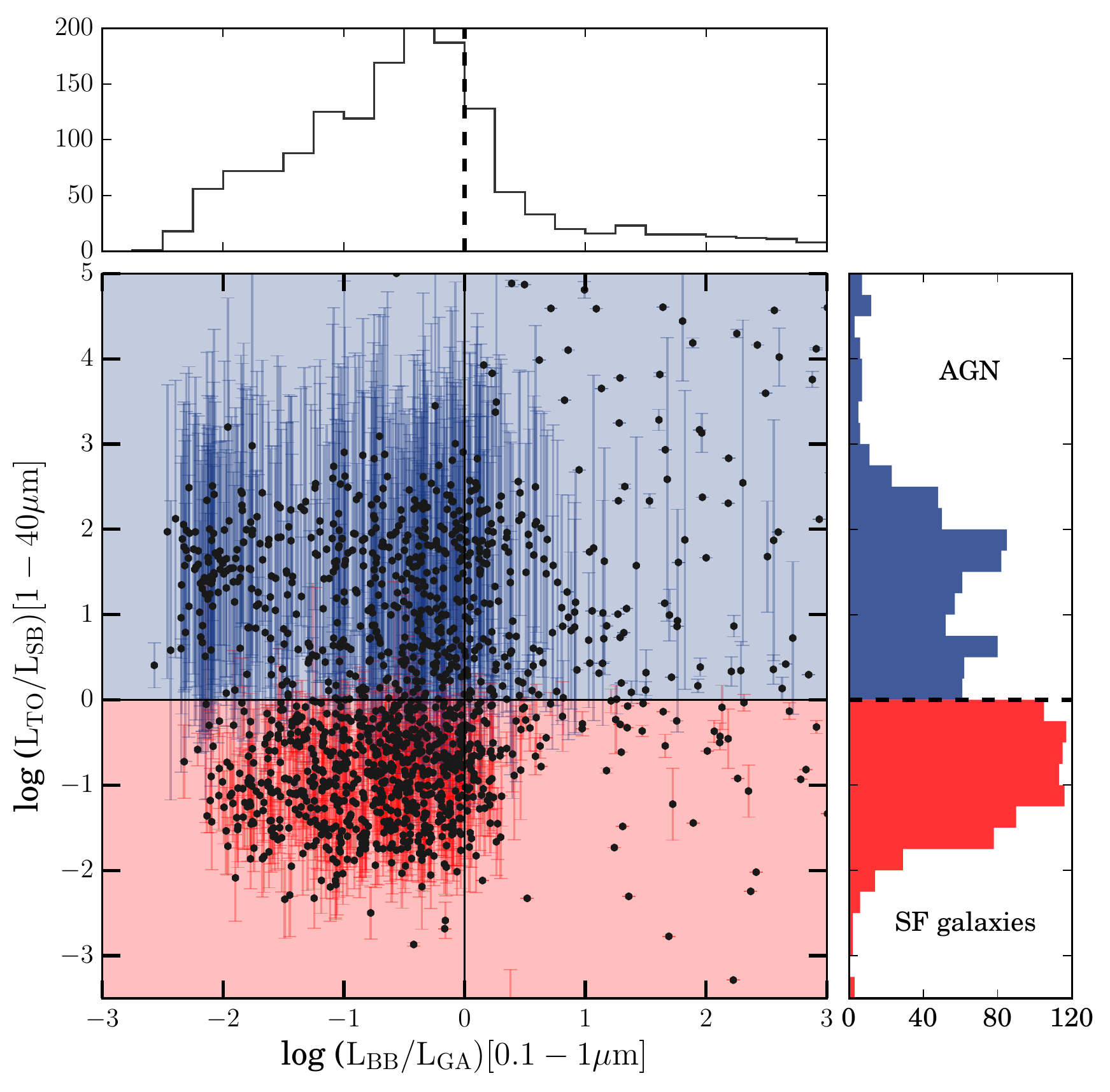}
\caption{Classification into AGN and SF galaxies using ratios of integrated luminosities of galactic versus AGN components. Integrated luminosities were computed from component templates fitted with \textsc{AGNfitter}. The red-shaded area shows the region populated by sources which are considered SF galaxies, while the blue-shaded areas are populated by sources classified as AGN. The data points are shown in black and their error bars are shown in red or blue, depending if their classified as SF galaxies or AGN, respectively.}% The colours of the error bars are chosen to show that although the uncertainties might be large, they spread mostly within their classification region.}
\label{fig:classification}
\end{figure}
% ---------------------------------------------------------

%%%%%%%%%%%%%%%%%%%%%%%%%%%%%%%%%%%%%%%%%%%%%%%%%%%%%%%%
\subsection{Stellar masses and star-formation rates}

Stellar masses and SFRs ($\textup{SFR}_{\rm opt}$) were estimated by fitting the observed data to \cite{bruzual03} templates, assuming a \citet{chabrier03} initial mass function (IMF) and an exponentially declining star formation history modulated by the time scale parameter $\tau$.
The distribution of stellar masses derived for our sample has a median value of $3.01^{+5.01}_{-2.50}\times 10^{10} \rm M_{\odot}$, where the error bars correspond to the scatter given by the 16th and 84th percentiles. 
The templates were corrected for dust absorption assuming the \cite{calzetti94} reddening law.
Additional SFRs from the IR-emission were calculated using integrated infrared luminosities of cold dust templates ($\textup{SFR}_{\rm IR}$) \citep[][]{dale02, chary01}.   
To derive the SFRs from the fitted cold dust emission we use the calibrations presented by \citet{murphy11}, which are an updated version from those by \citet{kennicutt98}.
These estimates are described in detail in \citet{calistrorivera16}.

%%%%%%%%%%%%%%%%%%%%%%%%%%%%%%%%%%%%%%%%%%%%%%%%%%%%%%%%
\subsection{Classification into AGN and SF galaxies} \label{subsec:classification}

% --------------------------------------------------------
%  FIGURE AGNFITTER SELECTION
%---------------------------------------------------------
\begin{figure}
\centering
\includegraphics[trim={0.2cm 0.3cm 0cm -3.2cm},clip,width=1.05\linewidth]{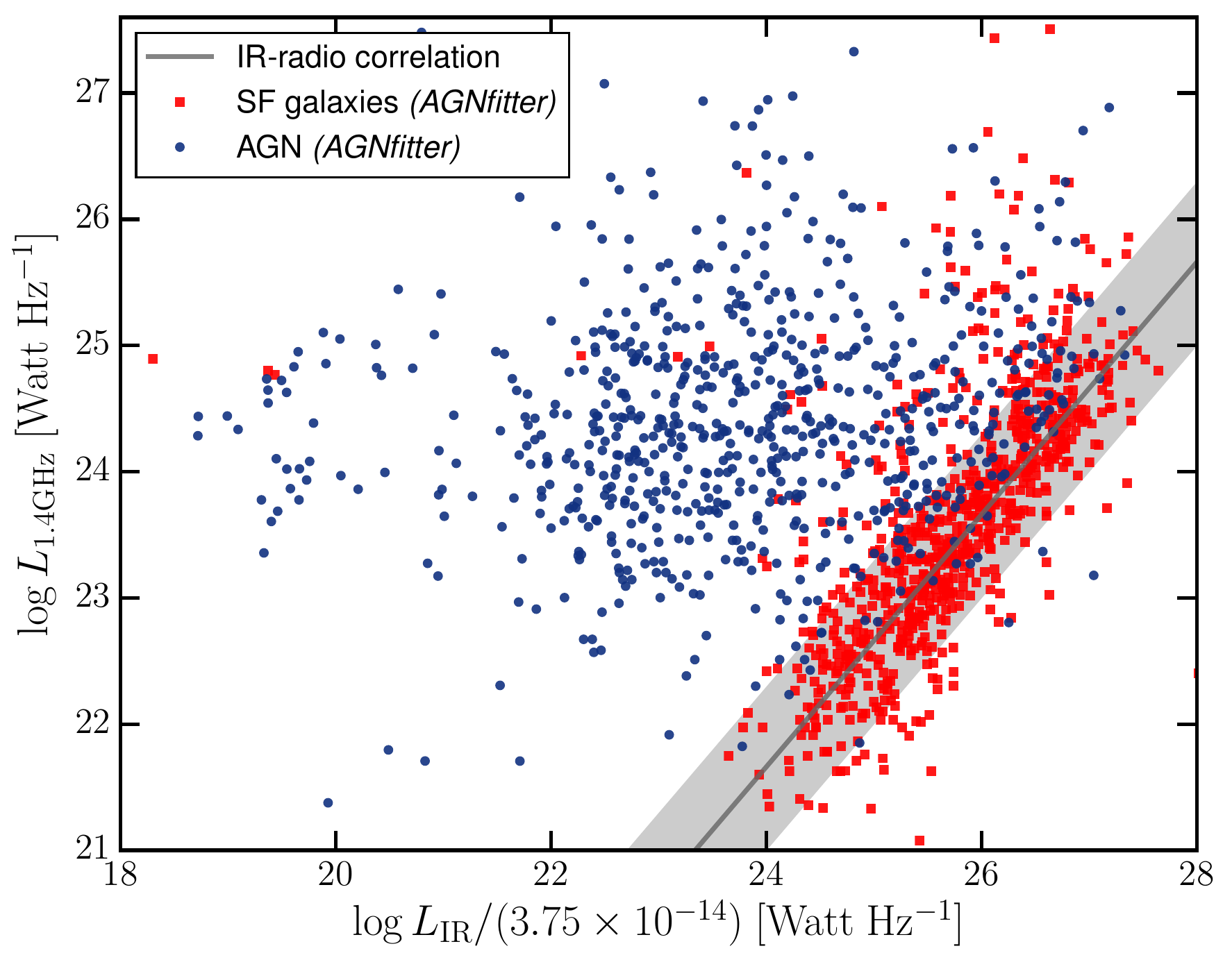}
\caption{Comparison of the radio-independent classification by \textsc{AGNfitter} to the IR-radio correlation. IR luminosities $L_{IR}$ are inferred as the luminosities of the cold dust emission component in the sources' SEDs integrated within the range 8-1000 $\mu$m, while radio luminosities are k-corrected observations at 1.4 GHz, assuming a spectral index of $\alpha=-0.7$. The solid line represents the canonical value of the correlation $q=2.3$, while the shaded area show a scatter of $\sigma=0.5$. The red and blue data points represent the values for galaxies classified as SF galaxies and AGN respectively.}
\label{fig:classificationvsFRC}
\end{figure}
% ----------------------------------------------------------

The classification strategy consisted of comparing the disentangled contributions of the AGN and the host galaxy in both the optical-UV and the infrared regimes.
It is important to note that this classification may still be contaminated by low excitation radio AGN (LERGs), which lack AGN signatures in their non-radio SED.
In section \ref{subsec:lergs} we will describe how we corrected for these potential mis-classifications.

In the optical-UV, the AGN contribution arises from the BBB luminosity $L_{\rm BB}$, which is then compared to the galaxy stellar emission $L_{\rm GA}$, integrating both inside the same frequency range 0.1 $ < \lambda < 10 \, \mu$m.
In the MIR, the AGN luminosity contribution is produced mainly by the torus $L_{\rm TO}$ and is compared to the cold dust emission in starburst regions of the host galaxy $L_{\rm SB}$, as well integrating both components within the same frequency range 1 $< \lambda < 40 \, \mu$m.
A representation of the SED decomposition and the integration ranges is shown in Fig.~\ref{fig:SED}.

As presented in Fig.~\ref{fig:classification}, in order to classify the total sample into SF galaxies and AGN we define the following scheme:

If the ratio of AGN torus to galaxy cold dust luminosities is smaller than one ($\log (L_{\rm TO}/L_{\rm SB}) < 0$), we consider the source's emission in the IR is dominated by star formation and we classify it as a SF-galaxy. 
The optical/UV contributions do not change the classification significantly, since the results for the 94 per cent of the sample which satisfy $\log(L_{\rm TO}/L_{\rm SB}) < 0$, also satisfy $\log(L_{\rm BB}/L_{\rm GA})<0$. 
In the cases with  $\log(L_{\rm BB}/L_{\rm GA})>0$ (only around 6 per cent of the sample), the direct emission of the AGN accretion disk appears to dominate over the stellar emission. 
However, since their FIR emission is dominated by the dust component in star-forming regions, these sources may represent the fraction of highly obscured star forming galaxies \citep[e.g.][]{casey14} and are thus classified as star-forming galaxies.

If the ratio of AGN torus to galaxy cold dust luminosities is greater than one  ($\log(L_{\rm TO}/L_{\rm SB}) > 0$) the source is considered an AGN according to our scheme. In the optical, the classification as AGN is less stringent and the source may host an AGN both in the case were $ \log(L_{\rm BB}/L_{\rm GA})>0$, where the accretion disk direct emission dominates the optical and if $\log(L_{\rm BB}/L_{\rm GA})<0$, where the source may present obscuration at nuclear scales and is considered an obscured AGN. 

To test this classification, we take advantage of the availability of the radio data and use the infrared-radio correlation \citep[IRC; e.g.][]{yun01}, which is a property observed almost exclusively by SF galaxies and radio-quiet AGN.
Fig.~\ref{fig:classificationvsFRC} shows an excellent agreement between the radio independent classification method based on SED-fitting compared to the IRC results.
This illustrates the capability of our method in classifying the sources into AGN and galaxies.
Prior to our correction for contamination by LERGs our samples consists of 810 SF galaxies and 732 AGN, both populating the redshift range from 0.05 to 2.5.

%%%%%%%%%%%%%%%%%%%%%%%%%%%%%%%%%%%%%%%%%%%%%%%%%%%%%%%%
\subsection{Contamination by Low Excitation Radio Galaxies }\label{subsec:lergs}

A consequence of classifying AGN and galaxies based on their SEDs from the FIR-UV is that the population of LERGs are prone to be misclassified as galaxies, due to the lack of AGN signature in their non-radio SED.
This mis-classification can be partially corrected by the assumption that LERGs would populate the area in the IRC typically covered by AGN, despite being classified as SF galaxies.
In Fig.~\ref{fig:classificationvsFRC} LERG candidates are the red data points which are outliers from the IRC area ($q\sim2.3\pm0.5$, as estimated below in section \ref{sec:IRC}).

We applied a conservative correction for LERGs contamination and reclassified the LERG candidates into the AGN class by choosing the values below the 2.5 sigma region of the IRC value to be the division line ($q= 2.3 - 2.5\sigma =0.8$, as observed for our sample).
This correction finds a total of 52 LERGs in our sample, which represents around 7 per cent of the total AGN population in our data.
To test our correction we investigate other properties of the 52 sources identified as LERGs. 
We find that although LERGs have IR luminosities dominated by SF processes by definition $\log(L_{\rm TO}/L_{\rm SB})<0$, the dominance is clearly weak ($\log(L_{\rm TO}/L_{\rm SB})\sim -0.50 \pm 0.51$) compared to the values for SF galaxies ($\log(L_{\rm TO}/L_{\rm SB})\sim -1.08 \pm 1.04$, were the errorbars represent the scatter given by the 14th and 86th percentiles).
Since these sources present a tendency towards the values of the HERGs classified as AGN ($\sim 1.4 \pm 1.2$), these results support the reliability of our correction.
For a detailed discussion of the properties of mass-selected LERGs and HERGs populations within Bo\"otes see Williams et al. (in prep).

We note that this correction applies for sources where AGN and star-formation activity are mutually exclusive processes, but this is not generally expected since AGN activity may occur in strongly star-forming galaxies \citep{stevens03, rees16}. 
Moreover, since the IRC is a central topic of this work, the classification of our sources needs to be as independent of it as possible to avoid biases in the results.
Through the chosen limits, we consider our correction is conservative enough for our purposes.

Our final sample consists of 758 SF galaxies and 784 AGN, which are well covered in the redshift range $z=[0.05,1.7]$, while the redshift range $z=[1.7,2.5]$ is slightly affected by incompleteness due to the flux limit of the data.

%%%%%%%%%%%%%%%%%%%%%%%%%%%%%%%%%%%%%%%%%%%%%%%%%%%%%%%%
\subsection{Comparison of the classification strategy with literature values}

A classification based solely on the SED-fitting output yields that the relative source fractions of SF galaxies and AGN is 49 and 51 per cent respectively, given our LOFAR-I-band selected sample.
A direct comparison to other observational and theoretical studies on this fraction is a complex task, since it is highly dependent on the sensitivity of the surveys \cite[e.g.][]{jarvis&rawlings04, appleton04, ibar08, simpson12}.
For instance, a similar LOFAR-selected sample presented by \cite{hardcastle16} shows a number density split of SF galaxies to AGN of approximately 30/70.
The lower SF galaxies fraction found can be explained due to the shallower radio data used in that study but also due to their different selection criteria (based on the FIR-radio correlation only). 
However, by using their selection criteria in our sample we still recover our original fraction of $\sim$50/50.
Similarly, a study of radio sources selected at 3 GHz in the COSMOS field \citep{delvecchio17} finds a SF galaxies/AGN ratio of $\sim 60/40$ based on an equivalent SED-fitting classification.

Investigations of the SF galaxy fraction in 1.4 GHz radio surveys \citep[e.g.][]{seymour08, bonzini13, smolcic08} have found that while SF galaxies show predominance at low flux densities, the AGN population start dominating at 100 $\mu$Jy. 
Assuming a spectral index value of $\sim -0.7$ this turning point corresponds to $\sim 600 \, \mu$Jy at 150 MHz, which is close to the detection limit of our sources.
Based on these assumptions, the observed 49/51 ratio in our study is thus in agreement with the ratio expected for the flux regime around the dominance turning point.

%%%%%%%%%%%%%%%%%%%%%%%%%%%%%%%%%%%%%%%%%%%%%%%%%%%%%%%%
%%%%%%%%%%%%%%%%%%%%%%%%%%%%%%%%%%%%%%%%%%%%%%%%%%%%%%%%

\section{The radio SEDs of SF galaxies and AGN}\label{sec:radiocontinuum}

%%%%%%%%%%%%%%%%%%%%%%%%%%%%%%%%%%%%%%%%%%%%%%%%%%%%%%%%
%%%%%%%%%%%%%%%%%%%%%%%%%%%%%%%%%%%%%%%%%%%%%%%%%%%%%%%%

We characterize the radio SED for galaxies and AGN by calculating the distributions of different spectral indices, which we define as:
\begin{equation}
\alpha^{\nu_2}_{\nu_1} =  \dfrac{\log (S_{\nu_1}/S_{\nu_2})}{\log(\nu_2/\nu_1) }, 
\label{eq:alpha}
\end{equation}

For the total sample we calculate the most general spectral index $\alpha^{1400}_{150}$ using only the two extreme points of the total frequency range studied, from 150 MHz to 1.4 GHz. 

More detailed spectral properties such as curvature are studied by calculating adjacent frequency pairs, $\alpha^{1400}_{325}$ and $\alpha^{325}_{150}$ in Section \ref{subsec:2points}.
To reliably measure spectral indices over narrow frequency ranges requires higher signal to noise data.
We therefore use a subsample with LOFAR fluxes above 2 mJy in order to minimize the effect of non-detections on these results.
Finally, the spectral study is refined for a fraction of the total sample which has GMRT spatial coverage at 610 MHz in Section \ref{subsec:3points}. 
For this sub-sample ($25$ per cent of the total) the distributions of three different spectral indices  are calculated for adjacent frequency pairs $\alpha^{1400}_{610},\alpha^{610}_{325}$ and $\alpha^{325}_{150}$.

%%%%%%%%%%%%%%%%%%%%%%%%%%%%%%%%%%%%%%%%%%%%%%%%%%%%%%%%
\subsection{Spectral index $\alpha^{1400}_{150}$ -- all sources}

% --------------------------------------------------------
%  FIGURE SPECTRAL indices STARBURSTS: ALPHAHIST
%---------------------------------------------------------
 \begin{figure}
      \includegraphics[trim={0 2.25cm 1cm 0.5cm},clip,width=1\linewidth]{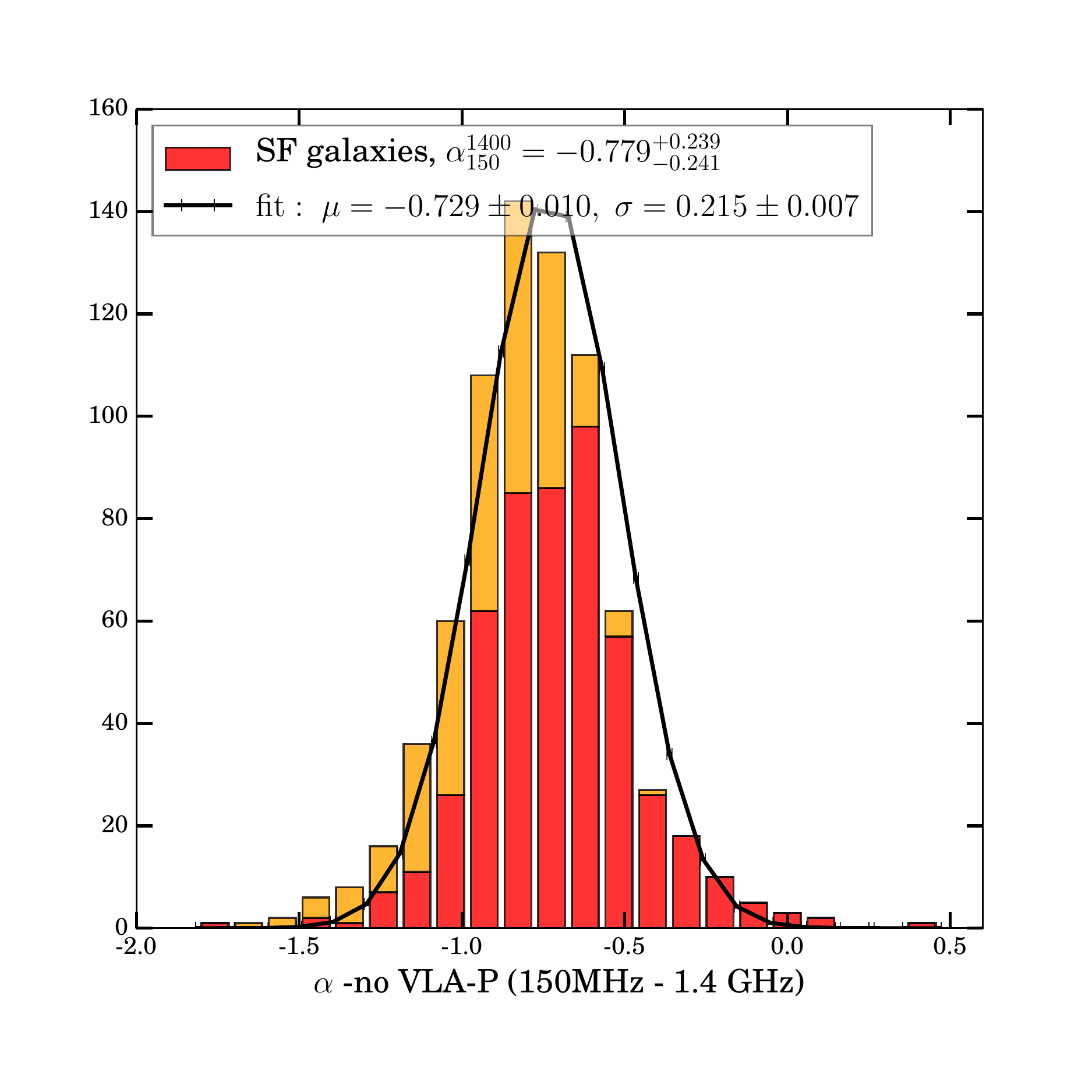}
      \includegraphics[trim={0 0 1cm 1cm},clip,width=1\linewidth]{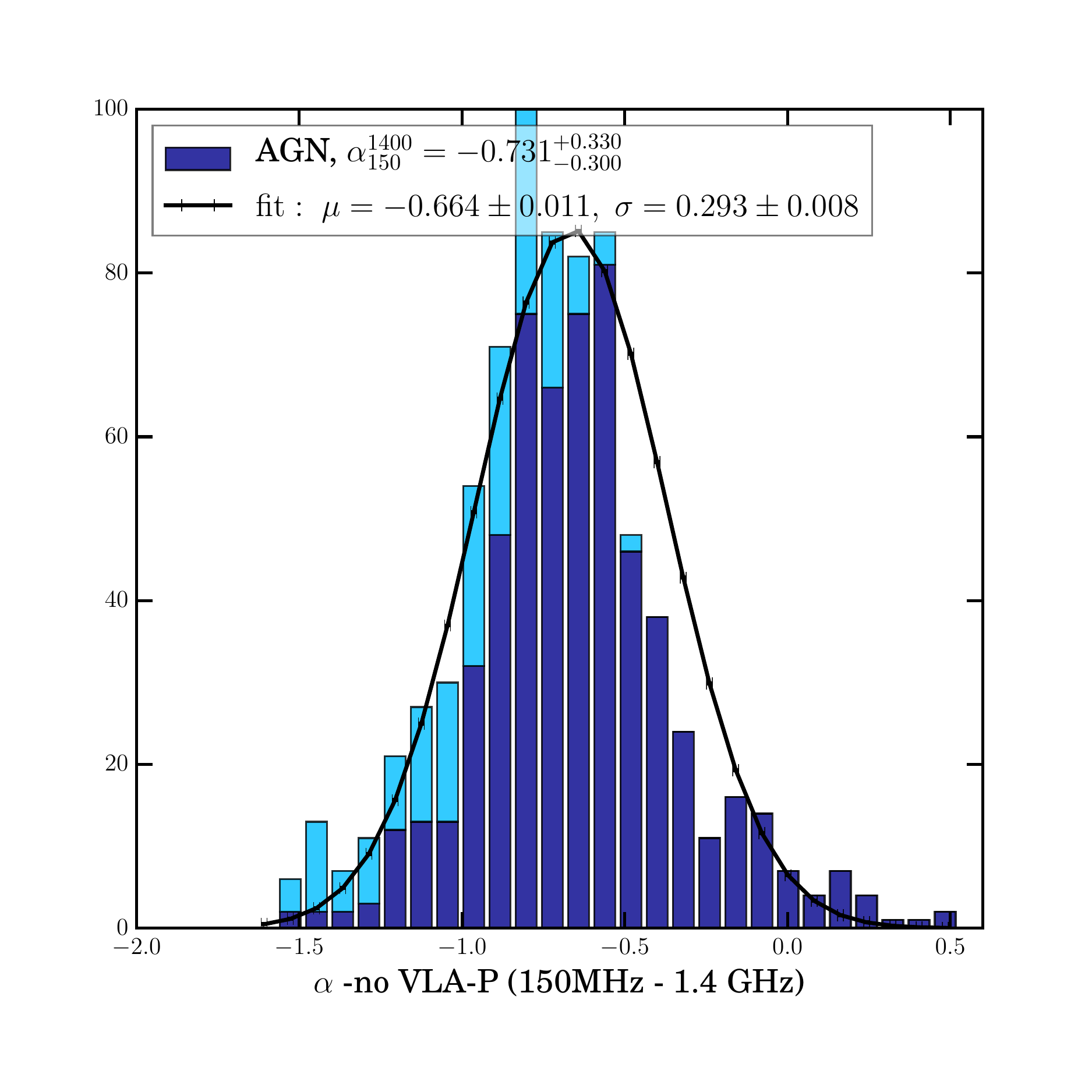}
    \caption{Spectral index $\alpha^{1400}_{150}$ for SF galaxies (\textit{upper panel}) and AGN-dominated galaxies (\textit{lower panel}).  Values corresponding to the 16th, 50th and 84th percentiles are calculated for the total sample. Detected and undetected sources in 1.4 GHz are depicted in red and orange respectively for SF galaxies, and dark blue and sky blue for AGN. The \textit{solid line} displays the fitted Gaussian with mean $\mu$ and intrinsic scatter $\sigma$. } 
    \label{fig:alpha_noVLAP}
  \end{figure}

% ----------------------------------------------------------

The distribution of the spectral index $\alpha^{1400}_{150}$ in Fig.~\ref{fig:alpha_noVLAP} includes the complete LOFAR-I-band selected sample. 
Note that we use forced photometry in the cases of non-detections in 325 MHz or 1.4 GHz.
We obtain the median  and scatter values for the distributions of SF galaxies and AGN:  $\alpha^{1400}_{150} =-0.78^{+0.24}_{-0.24}$ and $\alpha^{1400}_{150} =-0.73^{+0.33}_{-0.30}$ respectively, derived according to the 16th, 50th and 84th percentiles of the total samples.

For a proper interpretation of the width of the distribution, whether it corresponds to measurement errors or intrinsic spread of the spectral index distribution, we fit a Gaussian distribution to the total samples.
Using an MCMC algorithm \citep[emcee:][]{fm13} we infer the parameters $\mu$ and $\sigma$ (mean value and intrinsic spread) that characterise the intrinsic distribution of the parameters, independently of the measurement errors.
In order to avoid that extreme outliers have a strong effect on the fit we restrict the fit on 99.7 percent of the population distributed around the median value.
We choose this very conservative cut in order not to alter the shape of the distributions, which are characterised by longer tails than a normal distribution.
The logarithm of the likelihood function obtained through the fitting of these parameters, assuming the measurement errors are Gaussian, is given by
\begin{equation}
ln~\mathcal{L} = C - \dfrac{1}{2} \sum \left( ln(\sigma^2 + e_i^2) + \dfrac{(x_i - \mu)^2}{\sigma^2 + e_i^2}\right),
\label{eq:posteriorgaussian}
\end{equation}
where $\sigma$ (spread of the intrinsic distribution) and $e_i$ (measurement errors) are coupled.
Strictly speaking, the distribution described by Eq. \ref{eq:posteriorgaussian} is not a Gaussian, but a weighted sum of Gaussians with varying widths.
To visualize the shape of the intrinsic distribution, Gaussian distributions with the $\mu$ and intrinsic $\sigma$ inferred above are over-plotted on the histograms of Fig. \ref{fig:alpha_noVLAP}.\footnote{The slight apparent mismatch between the fit and the data (especially for the AGN sample) is explained by the presence of the small bulk of outlier sources at the high frequency tail of the spectral index distribution, by undetected sources (orange / sky blue bars) being weighted less than detections (red /dark blue bars)  due to their larger errors (following Eq. \ref{eq:posteriorgaussian}) and by the fact that a Gaussian fit is only an approximate description of the distribution.}

% --------------------------------------------------------
%  FIGURE SPECTRAL indices SB and AGN: ALPHAHIST
%---------------------------------------------------------
 \begin{figure*}
    \includegraphics[trim={2.5cm 2cm 0 3cm},clip,width=0.8\linewidth]{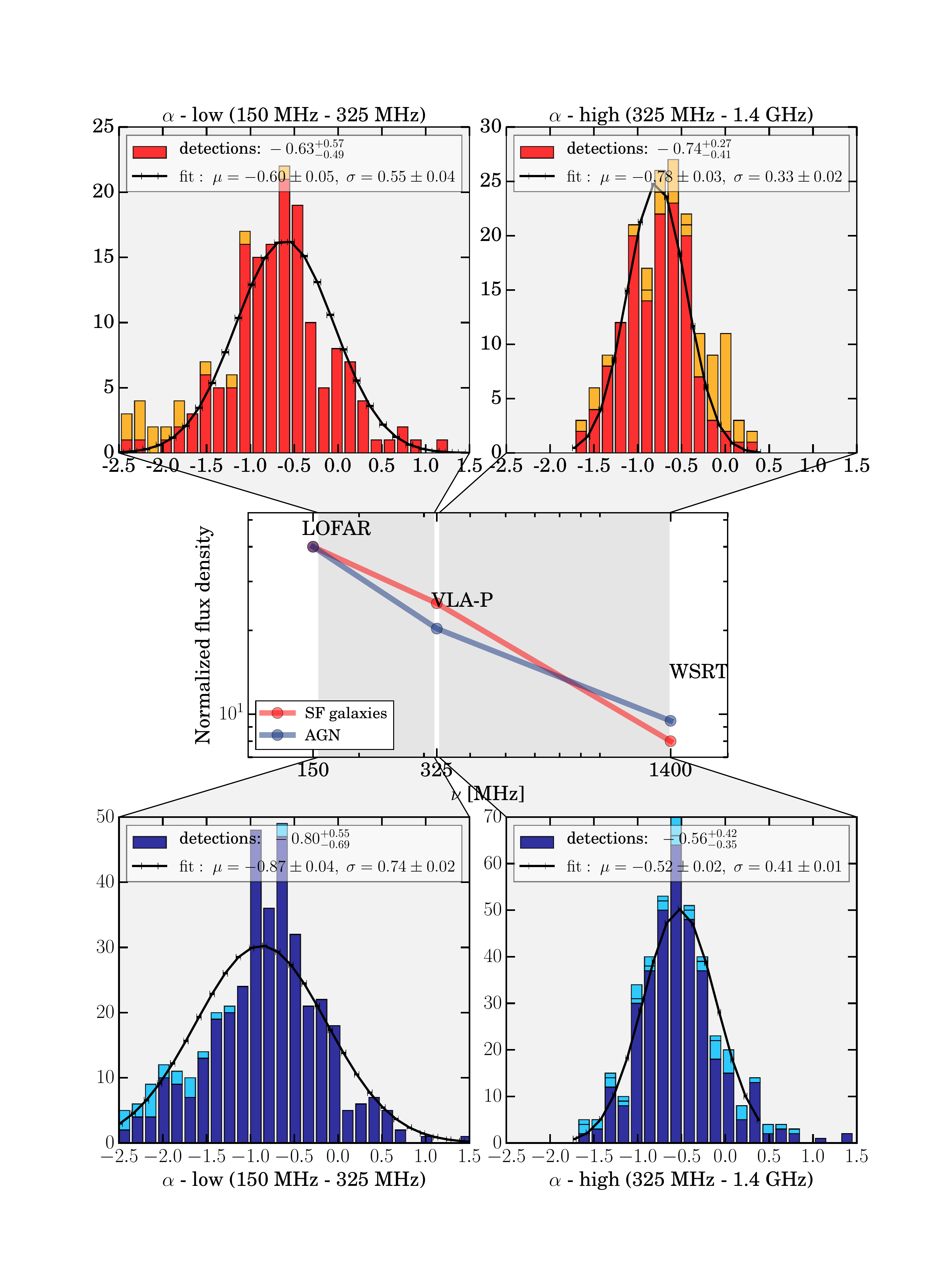}
    \caption{Spectral indices $\alpha^{1400}_{325}$ and $\alpha^{325}_{150}$ for SF galaxies (\textit{upper panel}) and AGN-dominated galaxies (\textit{lower panel}). Detected and undetected sources in 1.4 GHz are depicted in red and orange respectively for SF galaxies, and dark blue and sky blue for AGN. Median and scatter values corresponding to the 16th, 50th and 84th percentiles are calculated for all detected sources in VLA-P and WSRT ($\sim 90$ per cent of the total). The \textit{solid line} displays the fitted Gaussian with mean $\mu$ and intrinsic scatter $\sigma$. A schematic representation of the observed curvature is also shown in the central panel for SF galaxies (red line) and AGN (blue line).}
    \label{fig:alphaSBandAGN}
  \end{figure*}

%---------------------------------------------------------

The distribution of the SF galaxy sample is best fitted with a mean value $\mu=-0.729\pm0.010$ and intrinsic scatter of $\sigma = 0.215\pm0.007$, while the spectral index distribution for AGN presents $\mu=-0.664\pm0.0112$ and $\sigma = 0.294\pm0.008$. 
Although the median values of the SF-galaxy population are slightly higher than the AGN population, and the fitted mean values show a similar trend, both populations are consistent within the error bars.
To test the difference between both distributions for significance we used the non-parametric two-sample Kolmogorov-Smirnov (KS) test and 
found that the distributions are statistically different at a confidence level greater than 99.9 per cent for all sources and even greater when considering only detections.
In conclusion we find that using the total spectral index $\alpha^{1400}_{150}$ to characterize the radio SEDs of galaxies, a statistically significant difference between starburst-dominated and the AGN-dominated sample is found, where the total sample of SF galaxies show a slightly steeper spectra than AGN with a difference in their median values of $\delta \alpha^{1400}_{150} =-0.048 \pm 0.011$. 
The resulting values agree with previous calculations of the low-frequency spectral slope in the literature  \citep[starbursts:][]{condon92, ibar08, ivison10a, marvil14}, \citep[AGN:][]{singh13}. 

% --------------------------------------------------------
%  FIGURE SPECTRAL indices AGN: ALPHAHIST GMRT
%---------------------------------------------------------
 \begin{figure*} 
    \includegraphics[trim={4cm 2cm 0 3cm},clip,width=0.9\linewidth]{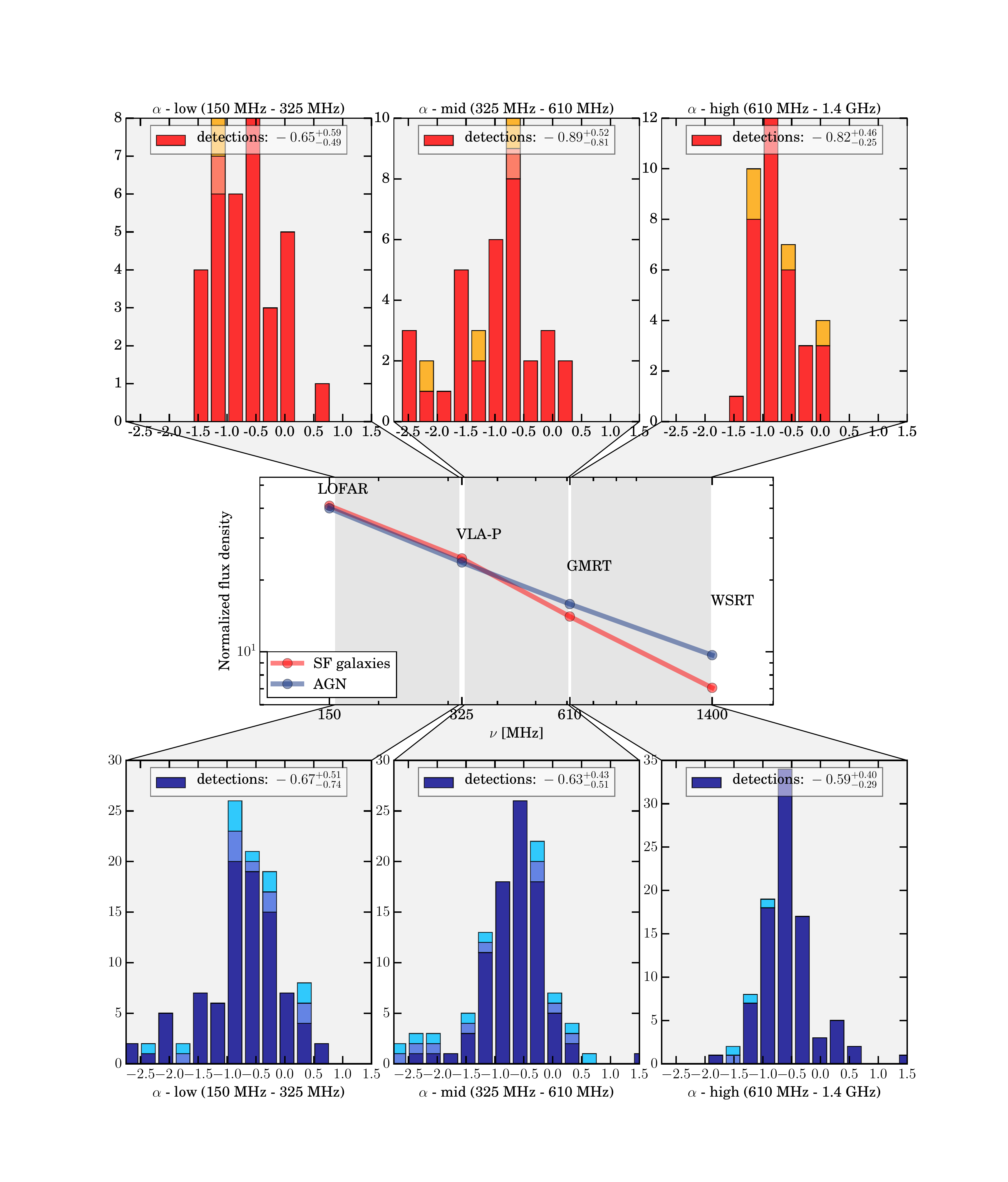}

    \caption{Spectral indices $\alpha^{1400}_{610}$,$\alpha^{610}_{325}$ and $\alpha^{325}_{150}$ for SF galaxies (\textit{upper panel}) and AGN-dominated galaxies (\textit{lower panel}). These distributions corresponds to the fraction of our sample with GMRT coverage (see Fig.~\ref{fig:radiocoverage}), which makes up $\sim 25$ per cent of the total sample. Detected and undetected sources in 1.4 GHz are depicted in red and orange respectively for SF galaxies, and dark blue and sky blue for AGN. Values corresponding to the 16th, 50th and 84th percentiles are calculated for all detected sources in VLA-P and WSRT ($\sim 95$ per cent of the total). A schematic representation of the observed curvature is also shown in the central panel for SF galaxies (red line) and AGN (blue line). }
    \label{fig:alphaGMRT}
  \end{figure*}

% --------------------------------------------------------

%%%%%%%%%%%%%%%%%%%%%%%%%%%%%%%%%%%%%%%%%%%%%%%%%%%%%%%%
\subsection{Spectral indices $\alpha^{325}_{150}$ and $\alpha^{1400}_{325}$ }\label{subsec:2points}

Higher order spectral properties such as curvature and higher order features can be studied by adding the central P-band observations at 325 MHz to the data set.

\subsubsection{Biased curvature observations due to non-detections}\label{subsubsec:bias}

Before including the central data point, we need to model our expectations according to the characteristics of the available data.
Following the distribution of $\alpha^{1400}_{150}$ presented above, a single slope spectrum around $-0.73$ and $-0.67$ for galaxies and AGN would imply observing similar values for the spectral indices $\alpha^{1400}_{325}$ and $\alpha^{325}_{150}$, when adding the central data point at 325 MHz.
However, measurement uncertainties can easily alter these observations and produce biased conclusions on curvature.
An important bias is the detection level of the central band, which can severely compromise the interpretation of steep spectra undetected in this band, shifting the median to flatter spectra towards lower frequencies.
We simulated the behaviour of our sample of LOFAR selected sources assuming that their spectral indices follow the ones observed in the histograms of Fig. \ref{fig:alpha_noVLAP}, and proceed to mimic similar systematics to those of our sample to investigate the expected values after the addition of the 325 MHz data point.
The distributions resulting from the simulations show a clear offset in these values compared to the initial distribution; demonstrating the biasing effect of using upper limits or excluding undetected sources without a proper treatment.

We would like to point out that this is important for the interpretation of results on radio continuum studies constructed from data of diverse sensitivity similar to ours \citep[see e.g.:][]{marvil14, clemens10}.
To overcome these biases we proceed with a cut in LOFAR fluxes and select sources above 2 mJy, reducing the percentage of undetected sources in VLA-P from 41 per cent to 14 per cent for SF galaxies and 27 per cent to 10 per cent for AGN, leaving us still with a large enough sample for a statistical study with 189 SF galaxies and 421 AGN.

\subsubsection{Results on flux-selected sources with S$_{150}> 2.0 \mJy$}\label{subsec:alphahighandlow}

Histograms of the low and high frequency spectral indices for SF galaxies and AGN are presented in Fig.~\ref{fig:alphaSBandAGN}.
We quote the median values and scatter based on the 16th and 84th percentiles of the distributions and fit a Gaussian to the distribution of sources with VLA-P detections (red and dark blue areas for galaxies and AGN respectively).

The upper panel of Fig.~\ref{fig:alphaSBandAGN} shows the median and scatter values of the spectral index distributions for SF galaxies: $\alpha^{325}_{150}= -0.63^{+0.57}_{-0.49}$ and $\alpha^{1400}_{325}= -0.74^{+0.27}_{-0.41}$.
A simple comparison of the median values of the two different spectral index distributions (similarly, using mean values from the fitted Gaussian curves) shows in general a slight flattening towards lower frequencies.
Using the non-parametric two-sample KS-test, we find that the difference between the low- and high frequency distributions of SF galaxies is highly significant at a level greater than 99.99 per cent (p-value$<10^{-11}$).
Our statistical study thus concludes that the spectral continuum for our SF galaxy sample is consistent with a curved spectrum, showing a slight flattening towards lower frequencies.

Similarly, the lower panels of Fig.~\ref{fig:alphaSBandAGN} present the median and scatter values for the population of AGN: $\alpha^{325}_{150}= -0.80^{+0.55}_{-0.69}$ and $\alpha^{1400}_{325}= -0.56^{+0.42}_{-0.35}$.
In contrast to the SF galaxies population, the medians of the distributions and fitted mean values in the lower panels of Fig.~\ref{fig:alphaSBandAGN} show a steepening towards lower frequencies with a difference in the mean values of the Gaussian fits of $\Delta \alpha \sim -0.35\pm 0.05$. Also here, we test this observation for statistical significance using the two-sample KS-test. 
We find that the sample of galaxies hosting AGN present curvature in their radio SED with a confidence level greater than 99.99 per cent (p-value$<10^{-11}$).
Our study thus concludes that radio SEDs of AGN in our sample show a statistically significant steepening in their radio continuum going to lower frequencies.

These results will be discussed in a physical context in section \ref{sec:discussioncurvature}.
We would like to remark that the significant differences found between the spectral curvature of SF galaxies and AGN in Fig.~\ref{fig:alphaSBandAGN} confirm that the observed curvature is a real spectral feature and not due to calibration issues in our sample.

%%%%%%%%%%%%%%%%%%%%%%%%%%%%%%%%%%%%%%%%%%%%%%%%%%%%%%%%
\subsection{Spectral indices $\alpha^{325}_{150}$, $\alpha^{610}_{325}$ and $\alpha^{1400}_{610}$}\label{subsec:3points}

We take advantage of the partial availability of GMRT data at 610 MHz and construct low-frequency radio SEDs with four data points for the fraction of sources that lie inside the GMRT coverage (green region in Fig.~\ref{fig:radiocoverage}).
The GMRT subsample constitutes 24 per cent of the total sample of LOFAR-I-band selected sources with LOFAR fluxes $S_{150}>2 $ mJy, which implies a total of 198 sources.
Only 11 per cent have VLA-P non-detections and are replaced by forced-photometry.
Taking into account detected sources this sub-sample consists of 41 SF galaxies and 101 AGN.

As expected from a solely spatial cut, the spectral index distributions present similar median values to the total field, with general spectral indices for SF galaxies of $\alpha^{1400}_{150}= -0.80\pm0.27$, and for AGN of $\alpha^{1400}_{150}= -0.70\pm0.31$.
Adding one more data point to the radio SEDs, we calculate the three spectral indices $\alpha^{325}_{150}$, $\alpha^{610}_{325}$ and $\alpha^{1400}_{610}$ following Eq.~\ref{eq:alpha}. Constructing the total SED, as sketched in the central panel of Fig.~\ref{fig:alphaGMRT}, we are able to recover spectral index ratios which suggest a consistent behaviour (within the larger statistical errors) with our previous finding in Fig.~\ref{fig:alphaSBandAGN}.

% --------------------------------------------------------
%  FIGURE SPECTRAL INDEX TOT VS REDSHFT
%---------------------------------------------------------
 \begin{figure}
 \centering
 \includegraphics[width=\linewidth]{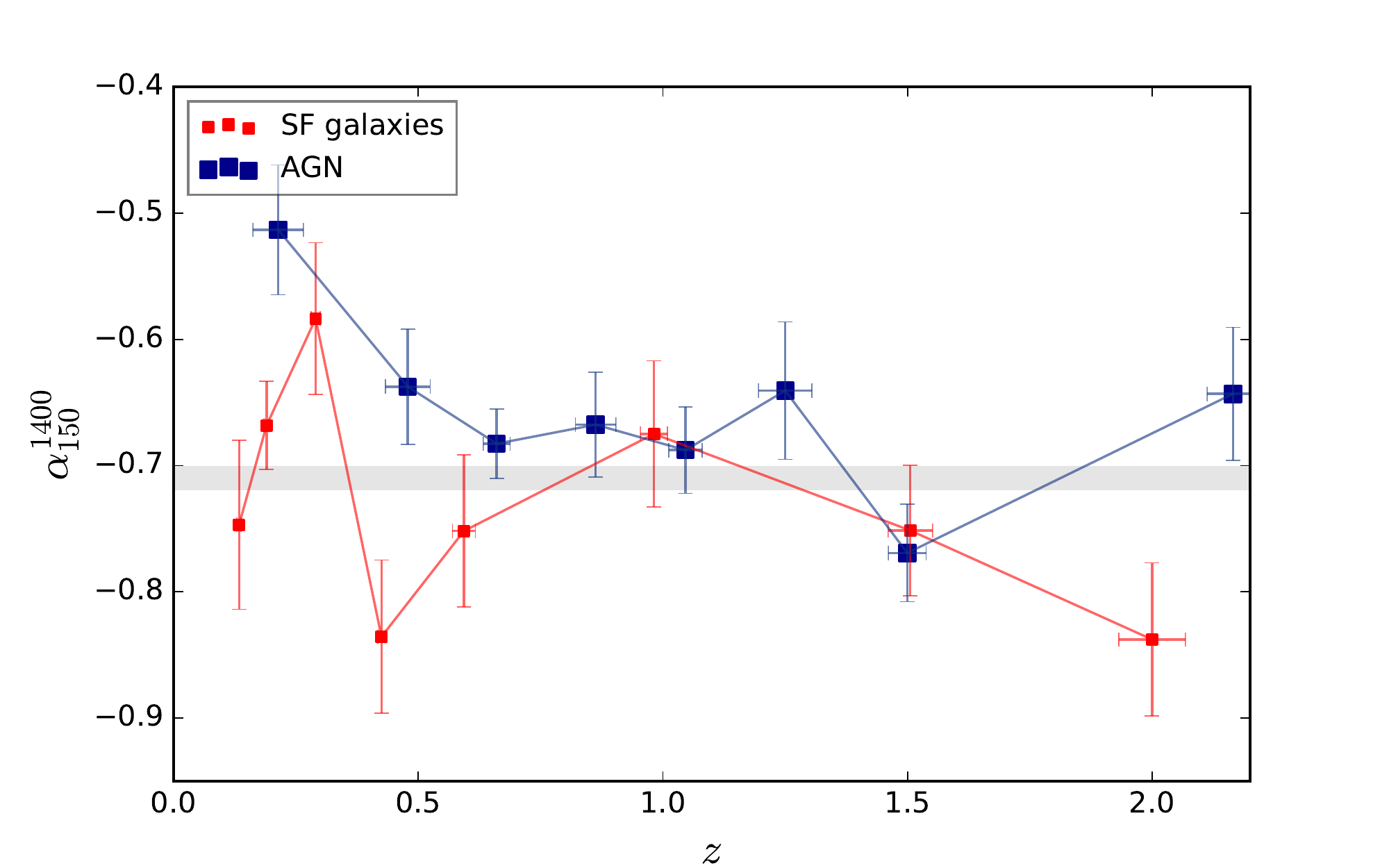}
 \caption{Evolution of spectral index $\alpha^{1400}_{150}$ as a function of redshift for SF galaxies (red) and AGN (blue). The data points are the mean values of $\alpha^{1400}_{150}$, grouped in bins of equal density. The error bars represent bootstrap errors.}
 \label{fig:alphatot-z}
  \end{figure}

% --------------------------------------------------------
The upper and lower panels of Fig.~\ref{fig:alphaGMRT} show three spectral index distributions for low, medium and high frequency pairs for both SF galaxies and AGN, respectively.
While the distributions' median values for SF galaxies ($\alpha^{1400}_{610}=-0.82, \alpha^{610}_{325}=-0.89$ and $\alpha^{325}_{150}=-0.65$) suggest a slight tendency to have flatter spectra towards low frequencies, the distributions' medians for AGN ($\alpha^{1400}_{610}=-0.59, \alpha^{610}_{325}=-0.63$ and $\alpha^{325}_{150}=-0.67$) are consistent with a steepening of the continuum going towards lower frequencies.
A quick examination to these characteristic values suggest these results are similar to the results of section \ref{subsec:alphahighandlow}, finding curvature as sketched in the middle panel of Fig.~\ref{fig:alphaSBandAGN}.

To test this systematic curvature for statistical significance, a KS-test on the difference of the spectral index distributions was performed. 
Comparing all three spectral index distributions at adjacent frequency pairs we find that the curvature is not statistically significant, since the KS-test could not reject the null-hypothesis due to the small statistics of the sub-sample with GMRT coverage (40 SF galaxies and 102 AGN).
However, a KS-test on the distributions of the spectral indices at the extremes of the frequency coverage, $\alpha^{1400}_{610}$ and $\alpha^{325}_{150}$, shows that a difference between these distributions is statistical significant for the AGN population at a level of 98 per cent, consistent with curvature and confirming the results of the previous section. 
For the SF-galaxy population this is not the case. This may be due to the small sample used for the study in this section (40 sources), which weakens the statistics.

%%%%%%%%%%%%%%%%%%%%%%%%%%%%%%%%%%%%%%%%%%%%%%%%%%%%%%%%
\subsection{Redshift evolution of the spectral curvature}\label{subsec:curvature-z}

\begin{figure}
\centering
\includegraphics[width=\linewidth]{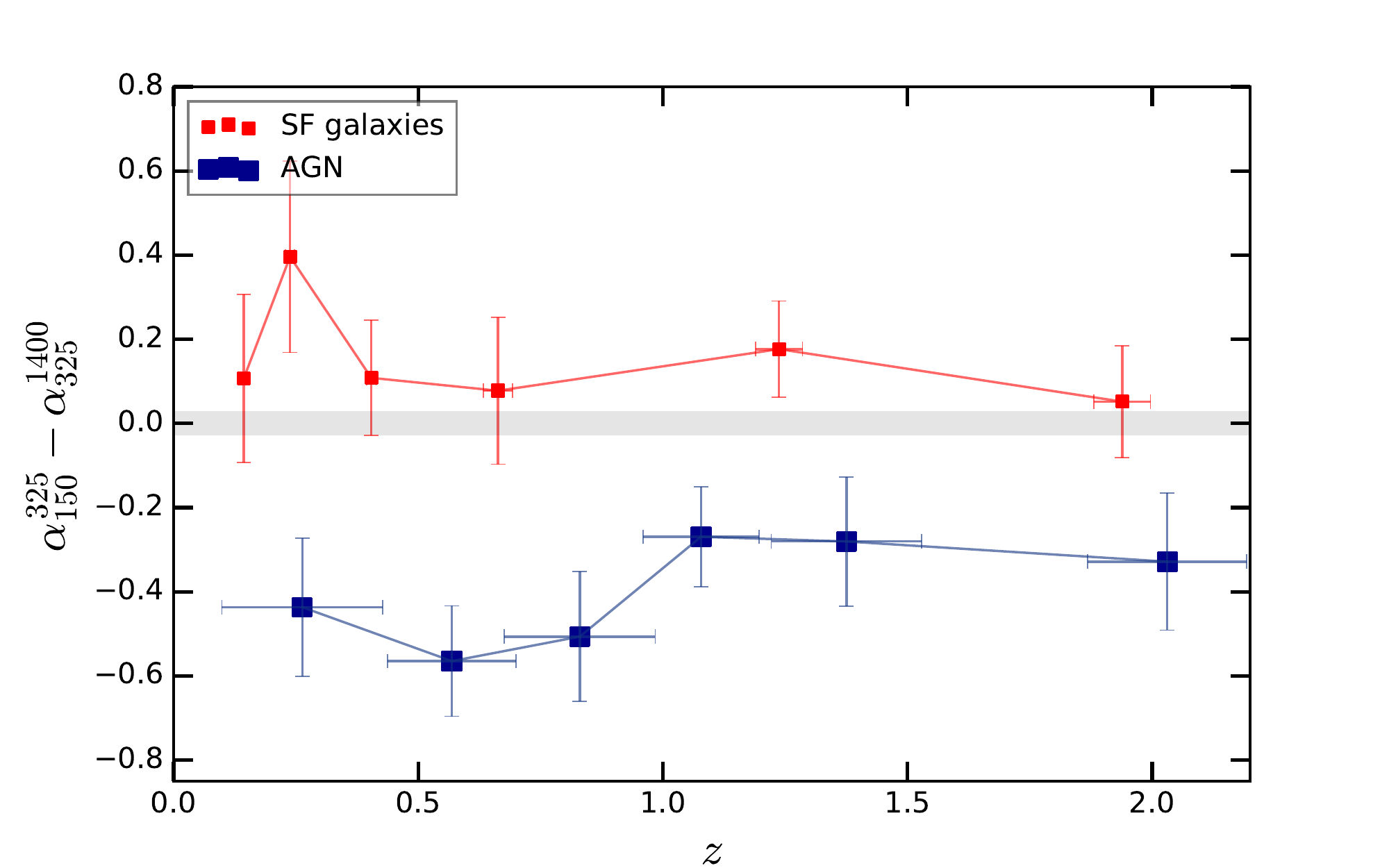}
\caption{Evolution of the spectral curvature parameter ($\alpha^{325}_{150} -\alpha^{1400}_{325}$) as a function of redshift for SF galaxies (red) and AGN (blue). The data points are the mean values of the curvature parameter, binned in groups of equal density in redshift. The error bars represent bootstrap errors.}
\label{fig:curvature-z}
\end{figure}

In Figs.~\ref{fig:alphatot-z} and \ref{fig:curvature-z} we study the redshift evolution of the total spectral index $\alpha^{1400}_{150}$  and the curvature parameter  $\alpha^{325}_{150} - \alpha^{1400}_{325}$, respectively.
For this study we use the samples with detections in both VLAP and WSRT ($\sim$90 per cent of the sample with S$_{150}> 2.0 \mJy$ ), which correspond to the characteristic values estimated from the distributions in Fig. \ref{fig:alphaSBandAGN}.

Investigating the change in total slope as a function of redshift is important to validate the approximation of a single power law as the k-correction for radio luminosities.
This assumption holds as long as the SED slope remains constant at the rest-frame frequencies for low redshift (150 MHz-1.4 GHz at $z\sim 0$) and high redshift (450 MHz-4.2 GHz at $z=2$) sources.
Fig.~\ref{fig:alphatot-z} shows that evolution of the slope as a function of redshift cannot be clearly recognized for either populations but shows a large scatter around the canonical value of $\alpha^{1400}_{150} =-0.7$.
Specifically for AGN, some data points in Fig.~\ref{fig:alphatot-z} may suggest an unclear trend as a function of redshift. However, using a linear fitting to the data points we find a redshift evolution very close to 0 (slope= $-0.05\pm 0.03$). 
We also tested the robustness of this finding using different binning schemes and the result remains consistent.
This result is expected since the rest-frame frequency range covered by our sample should be completely dominated by synchrotron emission and the strong flattening from thermal emission should not be visible below frequencies around 10 GHz \citep[e.g.][]{gioia82}. 

% --------------------------------------------------------
%  FIGURE CURVATURE - REDSHIFT
%---------------------------------------------------------
% ----------------------------------------------------------

A non-evolving mean spectral slope disfavours a scenario where the synchrotron spectral shape would be significantly affected by large scale properties of the galaxy that present a strong redshift evolution, such as integrated luminosities or SFRs. 
On the contrary, the absence of such an evolution suggests that the synchrotron emission would be determined by other rather local properties affecting the CR injection mechanisms that are not coupled to an evolution in redshift.
Examples of such processes are CR acceleration mechanisms and energy losses confined to sub-kpc structure around individual supernova remnants, galactic magnetic fields, and the immediate surrounding interstellar medium. 
We suspect these local properties establish the shape of the radio SED, while  differences in global properties may primarily affect the normalisation of the spectrum.
This result is consistent with the findings of \citet{magnelli15} and \citet{ivison10b}, who found no significant evolution of the spectral index $\alpha^{\rm 1.4GHz}_{\rm 610MHz}$ with redshift across $0<z<3$ for SF galaxies.

Fig.~\ref{fig:curvature-z} shows the redshift evolution of the spectral curvature, parametrized as the difference between spectral indices $\alpha^{325}_{150}$ and $\alpha^{1400}_{325}$, binned in six different redshift bins of equal sample sizes.
The most remarkable observation is that a contrast between the curvature of both populations is clearly present at all redshifts. 
While SF galaxies show continuously positive values of  $\alpha^{325}_{150} - \alpha^{1400}_{325}$, AGN present negative values, consistent with Figs. \ref{fig:alphaSBandAGN} and \ref{fig:alphaGMRT}.

We find no clear evolution of the curvature with redshift for both SF galaxies and AGN populations, as shown in Fig.~\ref{fig:curvature-z}.
We quantify this by fitting straight lines of different slopes for both binned populations. 
We find that the slope for the SF galaxies population is consistent with zero ($-0.055 \pm 0.103$), while the slope for AGN is found to be $0.147\pm 0.096$.
We test these results for significance by calculating the  coefficient of determination statistics $R^2$, which describes the proportion of the variance in the curvature parameter that is predictable from the redshift evolution given by the fit. 
From the results of the coefficient of determination we find that the fits for the SF galaxies and AGN populations account for less than 13 and 40 per cent of the scatter respectively.
Given the very small slope values, the relatively large errors and the low values of $R^2$, we conclude that we find no significant redshift evolution of the spectral curvature.

%\begin{figure}
%\centering
%    \includegraphics[trim={1.1cm 1.5cm 0 1.5cm}, clip, width=1.1\linewidth]{figures/radiosed.pdf}
%\caption{Radio continuum spectra for the SF galaxies (red) and AGN populations (blue). The spectra are sampled by binning the sources into different redshift bins, calculating the median expected luminosities from spectral indexes within each bin and plotting them at the corresponding rest-frame frequencies. The data points correspond to the median expected luminosities of the sources inside each bin and the error bars describe bootstrap errors. We fit the data points using two different descriptions: First, the best-fit of a power law using one single spectral index (dashed line) describes the spectrum of SF galaxies as $ \log(L^{SB}_{\nu}) \propto 0.69^{+0.44}_{-0.03} \times \log(\nu) $, and the spectrum of AGN is described as $\log(L^{SB}_{\nu}) \propto 0.74^{+0.50}_{-0.03} \times \log(\nu) $ . Using a non-linear function of the frequency results in  $\log(L^{\rm AGN}_{ \nu}) \propto -0.77^{+0.55}_{-0.49} \times \log{\nu}  +0.19 ^{+0.18}_{-0.19} \times \log^{2}{\nu}+ 0.04 ^{+0.02}_{-0.02} \times \log^{3}{\nu} $ for SF galaxies and   $\log(L^{\rm SB}_{\nu}) \propto 0.73^{+0.50}_{-0.51} \times \log{\nu} +  0.12 ^{+0.18}_{-0.18} \times \log^{2}{\nu} + 0.02 ^{+0.02}_{-0.02} \times \log^{3}{\nu}$ for AGN. }
%\label{fig:sampledSED}
%\end{figure}

Finally, we explore the redshift dependence of the total sample if undetected sources were also included. 
Although non-detections produce no significant differences to the total characterization of the curvature due to their low fraction (10 percent), we find that they imprint a weak redshift dependence on the results.
This effect occurs since luminosities of undetected fluxes are calculated as the product of a narrow distribution of fluxes close to the noise limit and the squared luminosity distance which imprints a strong redshift dependence.
This observation is especially important for samples of large non-detection fractions, as we will discuss in section \ref{sec:IRC}.

%%%%%%%%%%%%%%%%%%%%%%%%%%%%%%%%%%%%%%%%%%%%%%%%%%%%%%%%
\subsection{Spectral curvature dependence on other properties}

To investigate the origin of the spectral curvature for galaxies and AGN, we studied its dependence on the different physical properties inferred by SED-fitting.
For the SF galaxy sample, no dependence on IR luminosity was found and consequently on $\textup{SFR}_{\rm IR}$. 
Similarly we investigated the evolution of spectral curvature as a function of specific star formation rate sSFR, estimated by the SED fitting routine, and found no significant dependence.

For the AGN sample, a significant evolution was found as a function of the integrated luminosity of the torus component $L_{\rm TO}$, which represents a proxy for the total AGN power.
As shown in Fig.~\ref{fig:curvature-Ltor}, this suggests that more powerful AGN seem to present weaker curvature, i.e. flatter spectra.
We quantify this by fitting straight lines of different slopes to the z-evolution of the binned AGN population presented in Fig.~\ref{fig:curvature-Ltor}. 
We estimate the slope to be $0.168 \pm 0.019$, where the fit accounts for 93 per cent of the scatter according to the coefficient of determination of this fit. Since the spectral curvature of AGN does not present a redshift dependence as significant as this luminosity dependence (Fig.~\ref{fig:curvature-z}), our results suggest that AGN-luminosity is the driver of this spectral curvature trend.
We discuss these observations in section \ref{sec:discussioncurvature}.

%%%%%%%%%%%%%%%%%%%%%%%%%%%%%%%%%%%%%%%%%%%%%%%%%%%%%%%%
\subsection{Comparison with previous spectral index studies}\label{sec:discussioncurvature}

In this section radio SEDs from 150 MHz to 1.4 GHz have been constructed for the total population of LOFAR-selected sources with fluxes above 2 mJy.
Our results suggest the presence of oppositely directed curvature in the radio continuum of star-forming compared to AGN-dominated galaxies.
While SF galaxies present a slight flattening towards lower frequencies suggesting positive curvature, AGN radio SEDs reveal a systematic steepening towards lower frequencies, implying negative curvature.
Moreover, no redshift evolution of the spectral curvature was found for both the SF galaxy and AGN populations.
Comparing our results to previous findings in the literature is not a trivial task due to the unparalleled depth of our observations, particularly at low frequencies, which allows us to study the radio SED properties of sources to higher redshifts and lower luminosities than previous studies.

\subsubsection{SF galaxies}

% --------------------------------------------------------
%  FIGURE AGN CURVATURE VS LTOR
%---------------------------------------------------------
\begin{figure}
\centering
\includegraphics[width=1.1\linewidth]{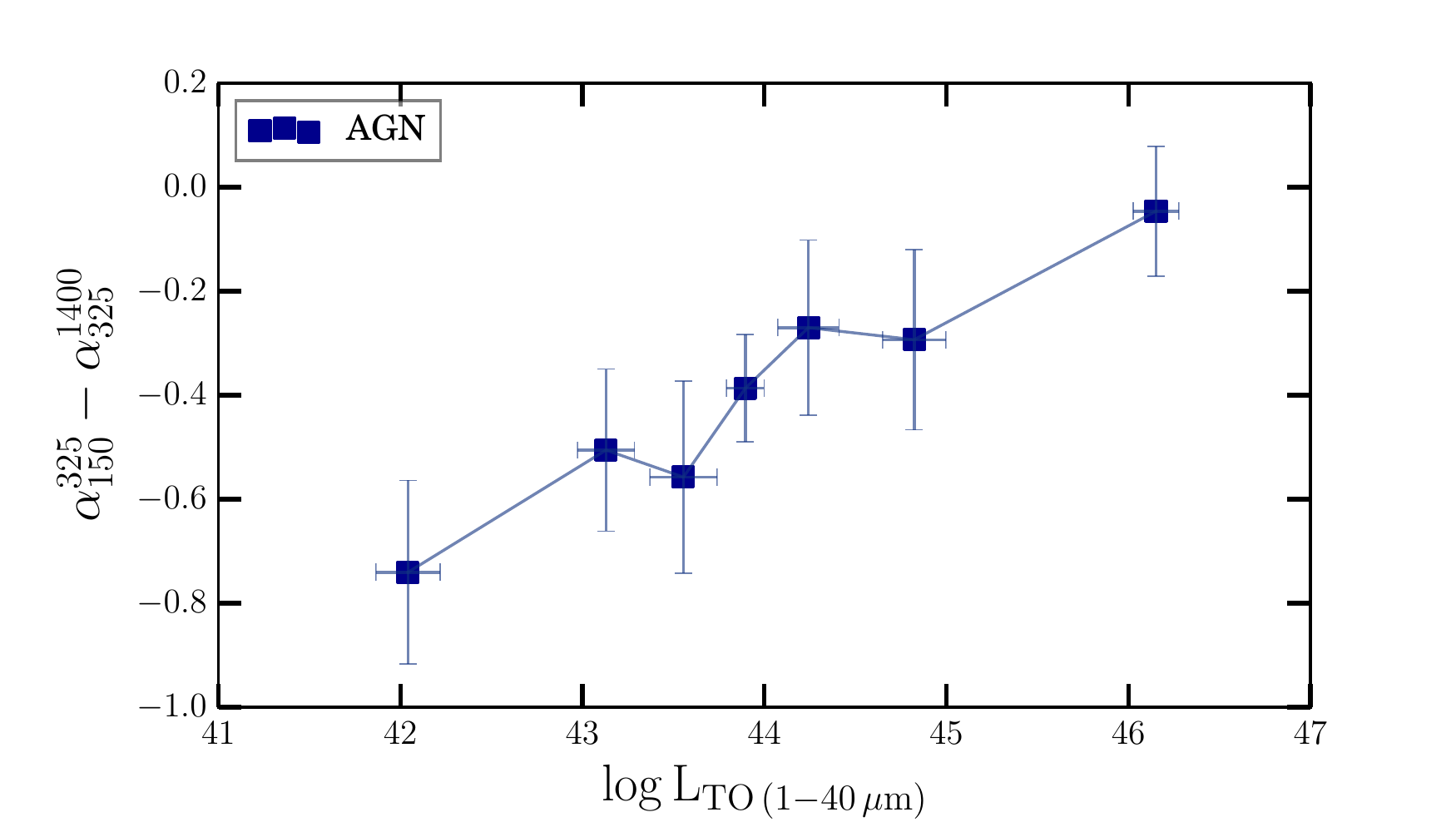}
\caption{Curvature observed in AGN plotted against the torus luminosity $L_{\rm TO}$, integrated in the wavelength range 1-40 $\mu$m. The data points are the mean values of the curvature parameter, binned in groups of equal density in $L_{\rm TO}$. The error bars represent bootstrap errors.}
\label{fig:curvature-Ltor}
\end{figure}
%----------------------------------------------------------

%%% literature on curvature of STARBURSTS
Several studies of the radio SEDs of nearby SF galaxies agree with our results that positive curvature is not an unusual spectral property at low frequencies \citep[e.g.][]{marvil14, murphy13, williamsandbower10, clemens10, condon92, israelmahoney90}, although few differences are to be noted.
For instance, the average spectrum of \cite{marvil14} nearby sources is curved following a change of $\Delta \alpha = 0.2$ per logarithmic frequency decade towards lower frequencies, which is a stronger flattening than that found in this study ($\Delta \alpha = 0.1$). 
One possible explanation is that \cite{marvil14} include a large fraction of non detections at the central wavelengths ($>60$ per cent for two of the four frequencies sampled), which is an important source of biased curvature if the results of our simulation in Section \ref{subsubsec:bias} applies here as well.
Similarly, \cite{clemens10} studied 20 nearby LIRGS and ULIRGS from 244 to 8.4 GHz, finding complex curvature in their source as well as claiming not only flattening but a strong turn over at around 1 GHz, which when extrapolated towards lower frequencies, implies a positive spectral index at frequencies $<1$ GHz. 
With the availability of LOFAR data at 150 MHz we can confirm this is not the case for most of the SF galaxies present in our field but the curvature is rather small, presenting a slightly flatter slope but still of negative sign.

%%%% SFR dependence
Spectral flattening in SF galaxies hints towards physical conditions which, integrated over the total galaxy, appear to be a common property in our SF galaxies sample and are expected to remain constant at all redshifts.
These conditions can be intrinsic to the production of the cosmic ray electrons (CRE) responsible for the synchrotron emission component \citep{basu12} or environmental, connected to the extreme interstellar medium (ISM) in star-forming regions into which the CREs are injected \citep{lacki13}.
%% Freefree
Consistent with the second hypothesis, \cite{murphy13} find increasing flattening of radio spectral index of galaxies with increasing specific star formation rate (sSFR).
We find such a dependence neither for sSFR nor for SFR.
% Although we find a SFR dependence on our total sample of sources, when binning our sample into redshift bins to break the $SFR-z$ degeneracy, we find that such a trend disappears at all bins, including sources of $z<0.3$ which are more similar to the \cite{murphy13} sample.
% In this case, the absorbing material that might be responsible for the flattening may not be directly related to SFR.
Similarly, \cite{basu15b} find that the flattening of the spectral index of nearby SF galaxies is unlikely to be caused by thermal free-free absorption but instead suggest a dependence on gas surface density.
Due to the lack of gas surface density tracers in our sample, we can not test for this.

\subsubsection{AGN}

%%% literature on curvature of AGN
Other attempts have been made in the construction of radio SED for radio AGN, but these have been limited to small samples and/or single spectral index studies \citep[e.g.][]{laing80, singh13, mauch13, kharb16,hardcastle16}.
Understanding the physical processes which drive integrated radio spectra of AGN is not an easy task due to the multiple physical conditions contributing to its total emission, as multi-component structures and orientation effects, and hence studies have focused on spectral index maps ($\alpha$ maps) of spatially resolved sources \citep{vardoulaki15, harwood15}.
For instance, in the case of FRI and FRII sources, the spectral shape can be dramatically different between jet, lobes to hot spots due to spectral ageing  \citep{harwood15}. 
While hot-spots are sources of constant CRE supply and present flat spectral indices \citep[][]{laing80}, lobes consist of earlier-ejected material which has lost energy, producing considerably steeper radio spectral indices.
The overall spectrum thus depends on the history and environment of the radio source.

An important question to understand the integrated radio spectra is which of the physical components dominate the total emission, at different frequency ranges, and across different orders of magnitude in luminosity?
A basic consensus among the observations is that radio AGN present generally flatter spectra in the GHz regime than SF galaxies \citep[][]{murphy13}, similar to our finding in Section \ref{sec:radiocontinuum} and, mainly for sources at $z>0.7$ and frequencies $\nu>325 \MHz$, in Fig.~\ref{fig:alphaSBandAGN}.
As hot-spots can be the dominant source in total flux density in luminous sources \citep{jenkins77}, this could suggest that these frequencies are dominated by hotspot emission.
Consistent with our observation in Fig.~\ref{fig:curvature-Ltor}, \cite{laing80} finds a luminosity dependence of the integrated spectral shapes for a sample of radio sources selected at 178 and 2700 MHz, arguing that more powerful sources have flatter spectra than weaker sources, while weaker sources have spectra which steepen at low frequencies \citep[but see also ][for different results]{rees16}.
While their study is limited to a much smaller sample and luminosity range due to their shallower data, we are able to confirm the luminosity dependence, ruling out a redshift dependence.
The steepening at low frequencies observed for our weaker AGN sources could be explained by steep-spectrum components dominating in this frequency regime, while flat-spectrum components become relatively more important at higher frequencies, causing the spectral index to flatten. 
These observations are in good agreement with the results presented by \citet{whittam16}.
They report spectral flattening for radio-selected galaxies at the high-frequency end (15.7 GHz) and suggest this may be due to the cores of \citet{fanaroff74} class I sources (FRI) becoming dominant at these high frequencies.

However, considering high flux uncertainties is especially relevant for AGN due to the different combinations of resolution and source extraction methods used, as they are extended sources. 
In contrast, this is not a relevant issue for SF galaxies, which are mostly compact sources.
These uncertainties might weaken the strength of our conclusions on AGN and better radio data and larger samples are needed to confirm this finding.

%%%%%%%%%%%%%%%%%%%%%%%%%%%%%%%%%%%%%%%%%%%%%%%%%%%%%%%%%%%%%%
%%%%%%%%%%%%%%%%%%%%%%%%%%%%%%%%%%%%%%%%%%%%%%%%%%%%%%%%%%%%%%

\section{The infrared-radio correlation (IRC)}\label{sec:IRC}

%%%%%%%%%%%%%%%%%%%%%%%%%%%%%%%%%%%%%%%%%%%%%%%%%%%%%%%%%%%%%%%
%%%%%%%%%%%%%%%%%%%%%%%%%%%%%%%%%%%%%%%%%%%%%%%%%%%%%%%%%%%%%%%

A remarkable property intrinsic to the radio SED of SF galaxies is the tight correlation with infrared luminosity, which has been observed at a few radio bands across five orders of magnitude in luminosity \citep{yun01}.
In this section we will characterize the IRC for the 1.4 GHz and the LOFAR data at 150 MHz and investigate its evolution with redshift for our sample. 
For this study we include all detected and undetected sources in both the 1.4 GHz WSRT and the \textit{Herschel}-SPIRE data.
As explained in detail in section \ref{subsubsec:nondetections}, undetected fluxes are estimated using forced photometry measurements on the positions of LOFAR-detected sources.
In this section we discard AGN and consider only sources selected as SF galaxies from the total LOFAR-I-band sample, which comprises $\sim$810 galaxies.

The IRC is parametrized by the widely used value $q_{\rm irc}$, which is defined as the ratio
\begin{equation}
q_{irc} = \log \left( \dfrac{L_{\rm IR}/(3.75\times 10^{12} \rm~ Hz)}{L_{\rm radio}} \right),
\label{eq:IRCdef}
\end{equation}
where  $L_{\rm IR}$ is the total rest-frame infrared luminosity (in $\rm ~erg~s^{-1}$) of the cold dust SED template integrated along the rest-frame frequency range $8-1000\,\mu$m \citep[equivalent to ][]{helou85, bell03, ivison10a, sargent10b, bourne11, murphy11}. L$_{\rm radio}$ is the luminosity at the radio frequency to be studied, in this case 1.4 GHz or 150 MHz in $\rm erg~s^{-1}~Hz^{-1}$. 
The factor $3.75\times 10^{12} \rm~ Hz$ is the frequency corresponding to $80~\mu$m, used in the definition to make $q_{irc}$ a dimensionless quantity.

% --------------------------------------------------------
%  FIGURE SPECTRAL indices STARBURSTS: QHIST
%---------------------------------------------------------
 \begin{figure}
    \centering
      \includegraphics[trim={0.5cm 0.75cm 0.9cm 0},clip,width=0.8\linewidth]{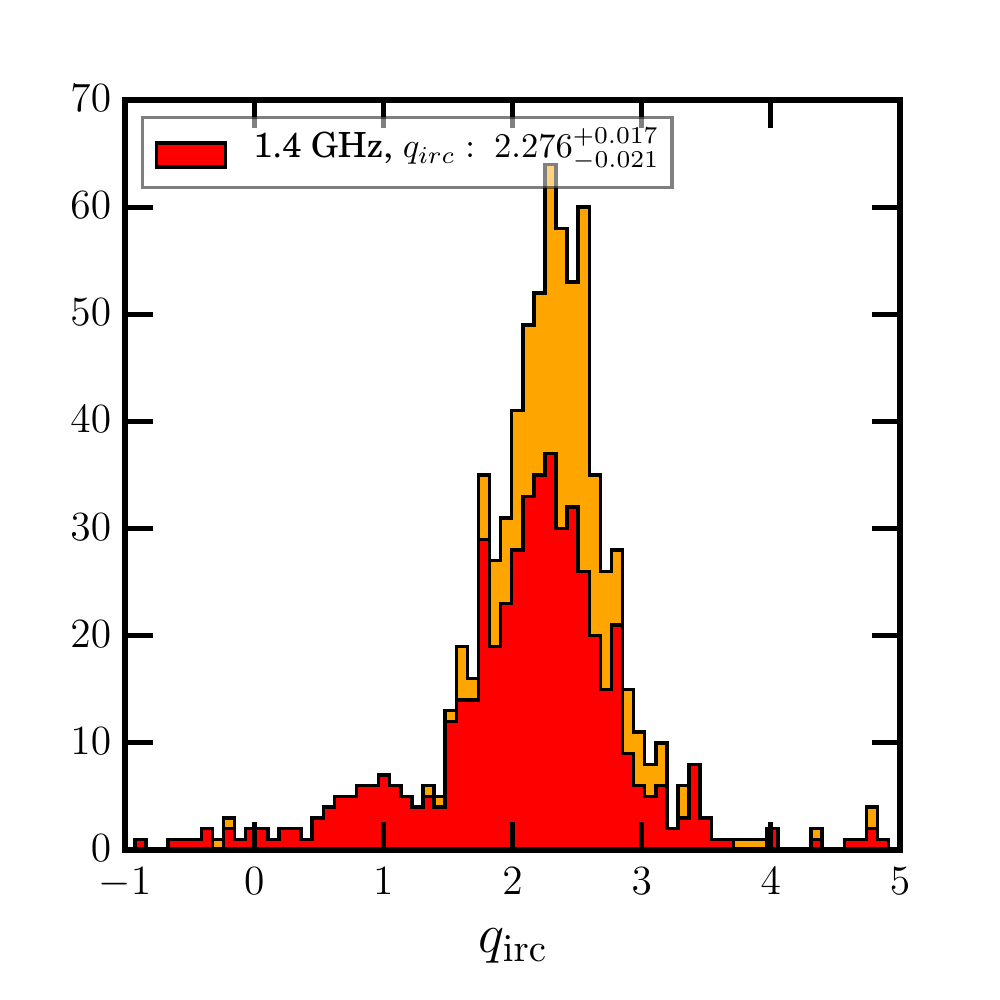}
      \includegraphics[trim={0.5cm 0 0.9cm 0.75cm},clip,width=0.8\linewidth]{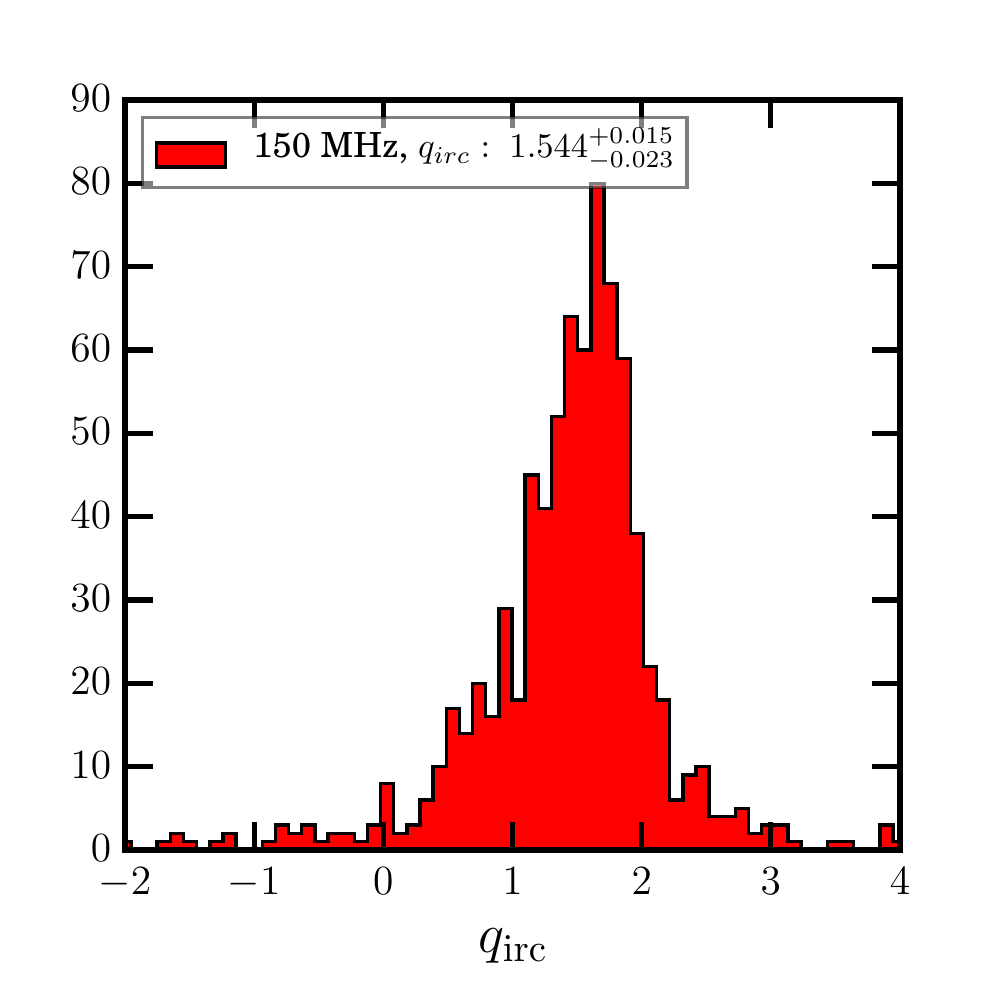}
    \caption{Histograms representing the distributions of $q_{irc}$ for the subsample of SF galaxies only.  The $q_{irc}$ values are calculated} using radio luminosities at 1.4 GHz (\textit{upper panel}) and at 150 MHz (\textit{lower panel}). Red bars represent detections at the corresponding radio frequencies, while orange bars represent forced-photometry values at 1.4 GHz.  We quote the median values of $q_{irc}$ in the legends, where the error bars represent the error on the median.  Note that the small number of outlier galaxies with low $q_{irc}$ values in both panels (misclassified by our IR-based method) were carefully excluded for the redshift-evolution study in Fig. \ref{fig:FRC1.4}. 
    \label{fig:q-hist}
  \end{figure}

% ----------------------------------------------------------

Radio rest-frame luminosities are computed using k-corrections inferred assuming $S_{\nu}\propto \nu^{\alpha}$, where $\alpha$ is the mean value observed for the SF galaxies distribution $\alpha=-0.73$ (see Fig.~\ref{fig:alpha_noVLAP}). 
We use a fixed $\alpha$ value for this study, since the available data does not sample the radio SED curvature accurately enough for a useful k-correction.

In contrast to most previous studies, our unique infrared coverage and the large dust models library allow us to robustly compute the total infrared luminosities. 
This approach has several advantages over using bolometric corrections based on monochromatic luminosities and single models as discussed by \citet{smith14}.
Moreover our MCMC approach in SED fitting allows us to infer robust uncertainties for the luminosities and consequently for q$_{irc}$.

% --------------------------------------------------------
%  FIGURE FRC of 1.4 GHz VS REDSHIFT - ALL
%---------------------------------------------------------
\begin{figure*}
\centering
\includegraphics[width=0.8\linewidth]{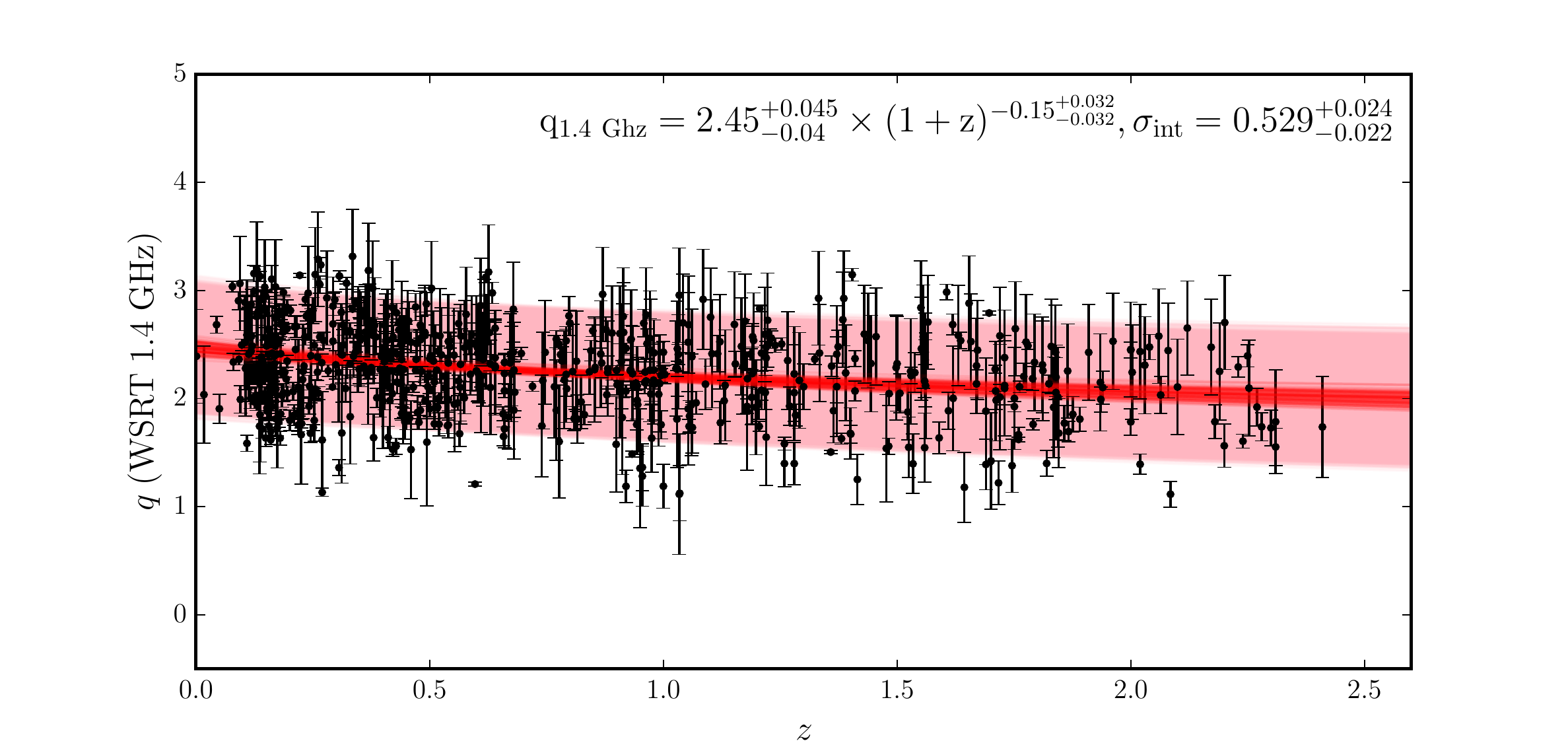}
\caption{q-value for the IRC corresponding to radio luminosities at 1.4 GHz plotted against redshift. Black error bars depict the observed values within the 2$\sigma$ region of the $q$-distribution in Fig.~\ref{fig:q-hist}, while red lines correspond to the fitted q-values inferred by the equation in the legend, taking into account uncertainties on the parameters calculated through MCMC sampling. The pink shaded area correspond to the fitted intrinsic scatter of the correlation.}
\label{fig:FRC1.4}
\end{figure*}
%----------------------------------------------------------

\subsection{The IRC at 1.4 GHz and its redshift evolution}

While 1.4 GHz luminosities of SF galaxies clearly follow the IRC (Fig.~\ref{fig:classificationvsFRC}), in the upper panel of Fig.~\ref{fig:q-hist}  $q_{1.4}$ is scattered in a distribution with a median value of $q_{1.4}= 2.28^{+0.017}_{-0.021}$, where the error bars represent the error on the median.
Fig.~\ref{fig:q-hist} reveals a small number of outlier galaxies exhibiting a radio excess, which is the consequence of contamination as part of our IR-based classification. Note that these were carefully excluded for the redshift-evolution study in the next section.
This $q_{1.4}$ distributions agrees well with the literature within the errors, although it is slightly lower than most of the local values observed for local SF galaxies (e.g. using FIR emission: $q_{1.4}=2.34\pm0.26$ \cite{yun01}, $q_{1.4}=2.3$ \cite{helou85}, using total IR emission as in our case: $q_{1.4}=2.64\pm0.26$  \cite{bell03} ).
In this section we investigate which parameter drives this offset and is able to explain the scatter observed.

In Fig.~\ref{fig:FRC1.4} we test the evolution of $q_{1.4}$ (1.4 GHz) as a function of redshift. 
To quantify a redshift evolution we fit the function: $q_{1.4}= C \times (1+z)^{\gamma} $, where $C$ is a constant and $\gamma$ the variable driving the evolution.
We include only 86 per cent of sources, which lie within the $2\sigma$ region of the distribution in Fig.~\ref{fig:q-hist}, so that the fit is not driven by outliers.
To account for measurement uncertainties in both variables and intrinsic scatter, the linear regression to infer the parameters is based on the likelihood function
\begin{equation}
    \mathcal{L} = \sum \left( \sigma_{\rm tot} + \dfrac{(q_{\rm obs} - (C\times (1+z)^{\gamma}) )^{2}} {\sigma_{\rm tot}^{2}} \right), 
\end{equation}
where 
\begin{equation}
\sigma_{\rm tot}  = \sqrt{\sigma^{2}_{\rm int} + \Delta q^{2}_{obs}}
\end{equation}

The fitting results for q$_{1.4}$ in  Fig.~\ref{fig:FRC1.4} suggest an evolution with redshift given by
\begin{equation}
   q_{1.4}(z) = (2.45\pm0.04) \times (1+z)^{-(0.15\pm0.03)}
   \label{eq:1.4z}
\end{equation}
with an intrinsic scatter of $\sigma_{\rm int} = 0.53$.

As expected, after the fit for redshift evolution the value $q_{1.4}$ at $z=0$ is higher than the median value of the total sample by $\delta=0.11$.
To infer the proportion of the variance in the distribution of Fig.~\ref{fig:q-hist}, which is explained by the $z$-evolution, we infer the coefficient of determination $R^2$, defined in this case as,
\begin{equation}
R^2 = \dfrac{\sum_{i} (q_{\rm obs-i} - q_{\rm1.4}(z_{i}))^2} {\sum_{i} (q_{\rm obs-i} - \langle q_{\rm obs}\rangle)^2}   
\end{equation}
where q$_{\rm obs-i}$ are the observed q-values and q$_{\rm 1.4}(z)$ are values calculated following Eq.~\ref{eq:1.4z}.
We calculate that the redshift evolution accounts for $\sim 6$ per cent of the variance in the total distribution, while most of the intrinsic scatter is still not explicable.

%%%%%%%%%%%%%%%%%%%%%%%%%%%%%%%%%%%%%%%%%%%%%%%%%%%%%%%%%%%%%%%

A dependence of the q$_{\rm 1.4}-z$ evolution on the $\alpha-z$ evolution around 1.4 GHz is unlikely, since this would imply a clear evolution of $\alpha_{\rm tot}$ in Fig.~\ref{fig:alphatot-z}, which is not observed.
Moreover, since the spectral curvature discussed in our study is observed only at lower frequencies this should not affect the k-correction for the 1.4 GHz luminosities. 
To investigate whether this redshift dependency is rather a luminosity dependence of the $q_{1.4}$ we separated the total sample into luminosity bins of similar sample sizes and were able to confirm a redshift evolution inside most of the luminosity bins, discarding such a degeneracy (see plots in Appendix \ref{app:IRClum} for details).
Finally, we test whether the z-evolution of $q_{1.4}$ is related to a changing ratio of AGN and SF contribution to the IR emission ($L_{\rm TO}/L_{\rm SB}$).
We find no correlation between these quantities indicating that the redshift evolution is not due to AGN contamination.

We discuss possible explanations to the offset observed between the mean $q_{1.4}$ of our sample ($q_{1.4}\sim 2.28\pm0.02$) and the local mean IRC found by \cite[][ $q_{1.4}\sim 2.64\pm0.26$)]{bell03}. 
An important factor to take into account in IRC studies is its sensitivity to selection biases \citep{sargent10b}.
A common selection bias in radio-selected samples arises from samples being incomplete at IR data for sources with faint radio fluxes \citep{ibar08,sargent10b}. This might not affect our sample, since the FIR emission expected for the 150 MHz fluxes in our sample (based on the IRC) is considerably above the flux limit of the XID SPIRE fluxes \citep{roseboom10}. Non-detections comprise less than 17 per cent and are compensated using forced photometry. 
Moreover, as discussed above, no direct dependence of the z-evolution of $q_{1.4}$ on radio luminosity could be found (Appendix). 
We note that by using a radio-selected sample we might be losing the FIR-dominated sources which drop out at the radio at the high redshift end of our sample and could shift the distribution to lower $q_{1.4}$ values. However, \cite{casey12} remarked that the XID SPIRE fluxes are complete in relation to a pure SPIRE selected sample and that the fraction of sources that could drop out in the radio at $z<2$ is negligible. Since $>90$ per cent of the sources in our sample have redshifts $z<2$, the possible bias against faint radio sources is not significant.

The mean $q_{1.4}$ value of our total distribution is lower ($\delta q_{1.4}\sim 0.36$) than the local mean IRC presented by \citet{bell03}.  
However, from the fitted redshift-evolution in Equation \ref{eq:1.4z} the predicted $q_{1.4}$ value corresponding to the local universe would be $q_{1.4}\sim 2.45$ which is well within the error on the mean quoted by \cite[][ $q_{1.4}\sim 2.64\pm0.26$)]{bell03}. 
The tension between our observations and those by \citet{bell03} is therefore reduced.
The offset between the mean $q_{1.4}$ values can be primarily explained by the redshift evolution and the different redshift ranges probed in the two studies.

% --------------------------------------------------------
%  FIGURE FRC of 150 MHz
%---------------------------------------------------------
\begin{figure*}
\centering
\includegraphics[width=0.8\linewidth]{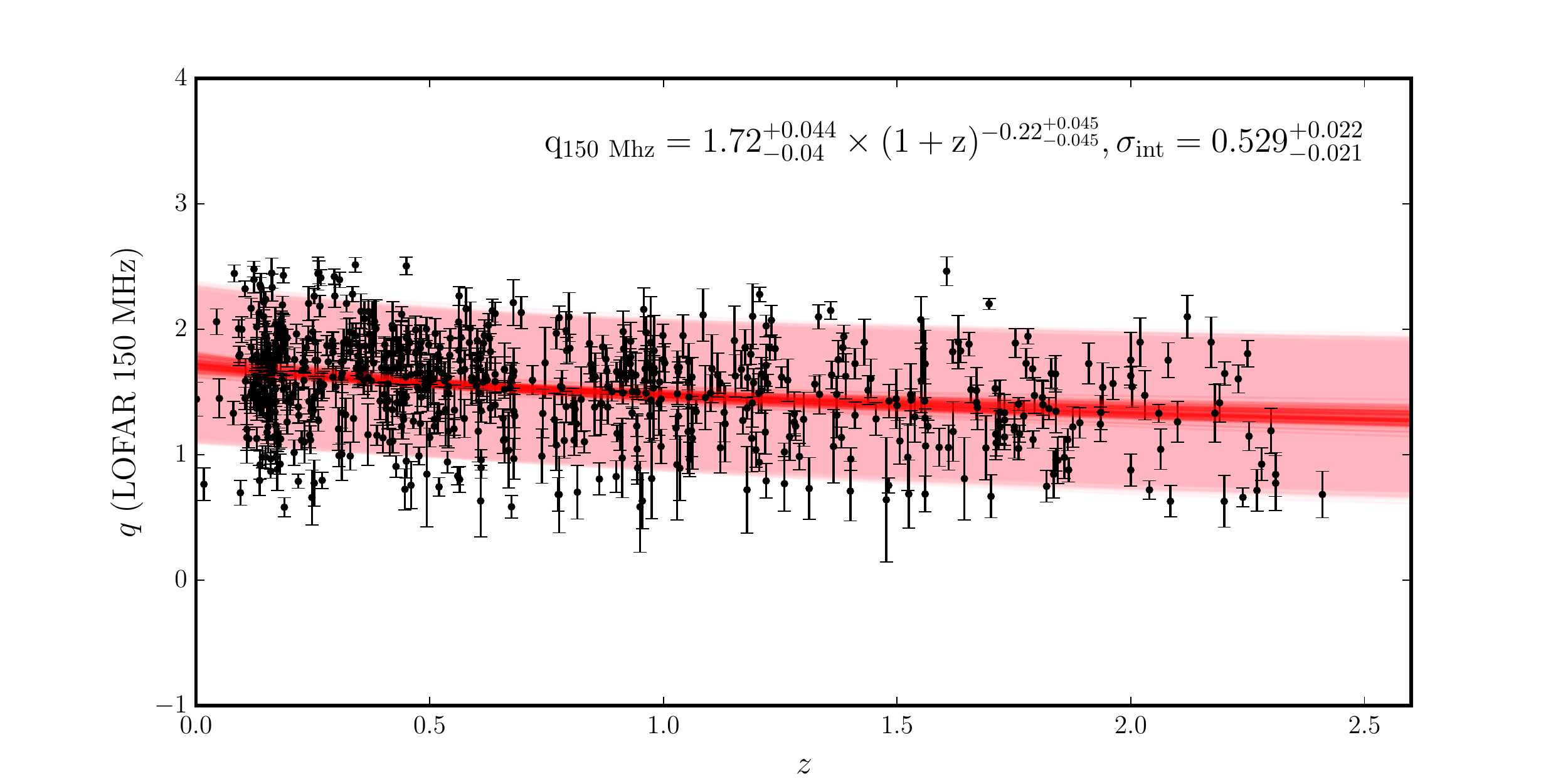}
\caption{q-value for the IRC corresponding to radio luminosities at 150 MHz plotted against redshift. Black error bars depict the observed values within the 2$\sigma$ region of the $q$-distribution in Fig.~\ref{fig:q-hist}, while red lines correspond to the fitted q-values inferred by the equation in the legend, taking into account uncertainties on the parameters calculated through MCMC sampling. The pink shaded area correspond to the fitted intrinsic scatter of the correlation.}
\label{fig:FRClofar}
\end{figure*}

% ----------------------------------------------------------
The picture of decreasing $q_{1.4}$ values as a function of redshift is being supported by a number of studies based on FIR and radio data sets of variate properties.
Although some theoretical studies predict the opposite trend \citep{schober16} and few observational studies claim to find no observable trend \citep[][]{pannella15}, a long list of recent \textit{Herschel}-FIR and \textit{Spitzer}-IR based studies have observed a redshift evolution of  $q_{1.4}$ \citep[e.g.,][]{ivison10a, bourne11, casey12, jarvis10, smith14, magnelli15, delhaize17}. 
However, some of these studies have chosen to favour a no-trend scenario due to the modest evolution compared to the large size of their measurement errors or other observed dependences. 
It is reassuring that our results are consistent with studies that take into account the treatment of biases due to selection as presented by \citet[][]{ivison10a}, which compare different selections to test their results and \citet[][]{ magnelli15}, which is based on a mass-selected galaxy sample. 
They find an evolution that goes as $q_{1.4} \propto (1+z)^{-0.12}$ and $q_{1.4} \propto (1+z)^{-0.15}$ respectively.
Similarly, \citet{delhaize17} find a redshift evolution of $q_{1.4} \propto (1+z)^{-0.19}$  using highly sensitive 3GHz VLA observations and \textit{Herschel}-FIR data and accounting for radio and IR non-detections through survival analysis.

\subsection{The IRC at 150 MHz and its redshift evolution}\label{subsec:q150}

One main focus of this study is to investigate for the first time the IRC for radio emission at low frequencies, which is crucial for LOFAR observations and relevant as well for future SKA data.
Similarly to the previous section, we calculate the value of the IRC parameter for the 150 MHz data and present the distribution of $q_{150}$ in the lower panel of Fig.~\ref{fig:q-hist}.
The distribution of $q_{150}$ presents a median value and error on the median of $q_{150}= 1.544^{+0.015}_{-0.023}$.

It is customary to calculate the value of $q_{irc}$ for different radio frequencies interpolated from the well calibrated value $q_{1.4}$ at 1.4 GHz \citep[e.g.,][]{yun01, murphy11} and under the assumption of a single power law radio emission.
Such a value would be calculated as
\begin{equation}
  q_{150, exp} = q_{1.4} + \log \left(\dfrac{1400}{150} \right)^{\alpha}
\end{equation}
Assuming a unique power law value from 1400 to 150 MHz of $\alpha=-0.73$, we obtain an expected q-value of $q_{150, exp} = 1.59$. 
This value is well in agreement with the distribution observed in the lower panel Fig.~\ref{fig:q-hist} ($q_{150, obs} = 1.54$) given the errors.

Also here we consider sources with $q$-values within the 2$\sigma$ region of the distribution in Fig.~\ref{fig:q-hist} and investigate the evolution of $q_{150}$ with redshift.
We find the relation
\begin{equation}
   q_{150}(z)= (1.72\pm0.04 )\times (1+z)^{-(0.22\pm0.05)}
   \label{150z}
\end{equation}
with an intrinsic scatter of $\sigma_{\rm int} = 0.53$.
Proceeding with a similar analysis as for $q_{1.4}$, we calculate the $R^2$ values for this regression and obtain that the redshift evolution accounts for $\sim 8.4$ per cent of the variance in the total distribution, while most of the intrinsic scatter is still not accounted for.

One remarkable property of the redshift evolution of $q_{150}$ is that its redshift dependence, $q_{150} \propto (1+z)^{-0.22\pm0.05}$, is slightly stronger ($\delta \gamma =0.07 $) than that observed for $q_{1.4}$, with $q_{1.4} \propto (1+z)^{-0.15\pm0.04}$.
We explore the connection between the stronger redshift dependence of $q_{150}$ and our results on the spectral curvature described in section \ref{subsec:curvature-z}.

% --------------------------------------------------------
%  FIGURE SFR of 150 MHz
%---------------------------------------------------------
\begin{figure*}
\centering
\includegraphics[width=0.8\linewidth]{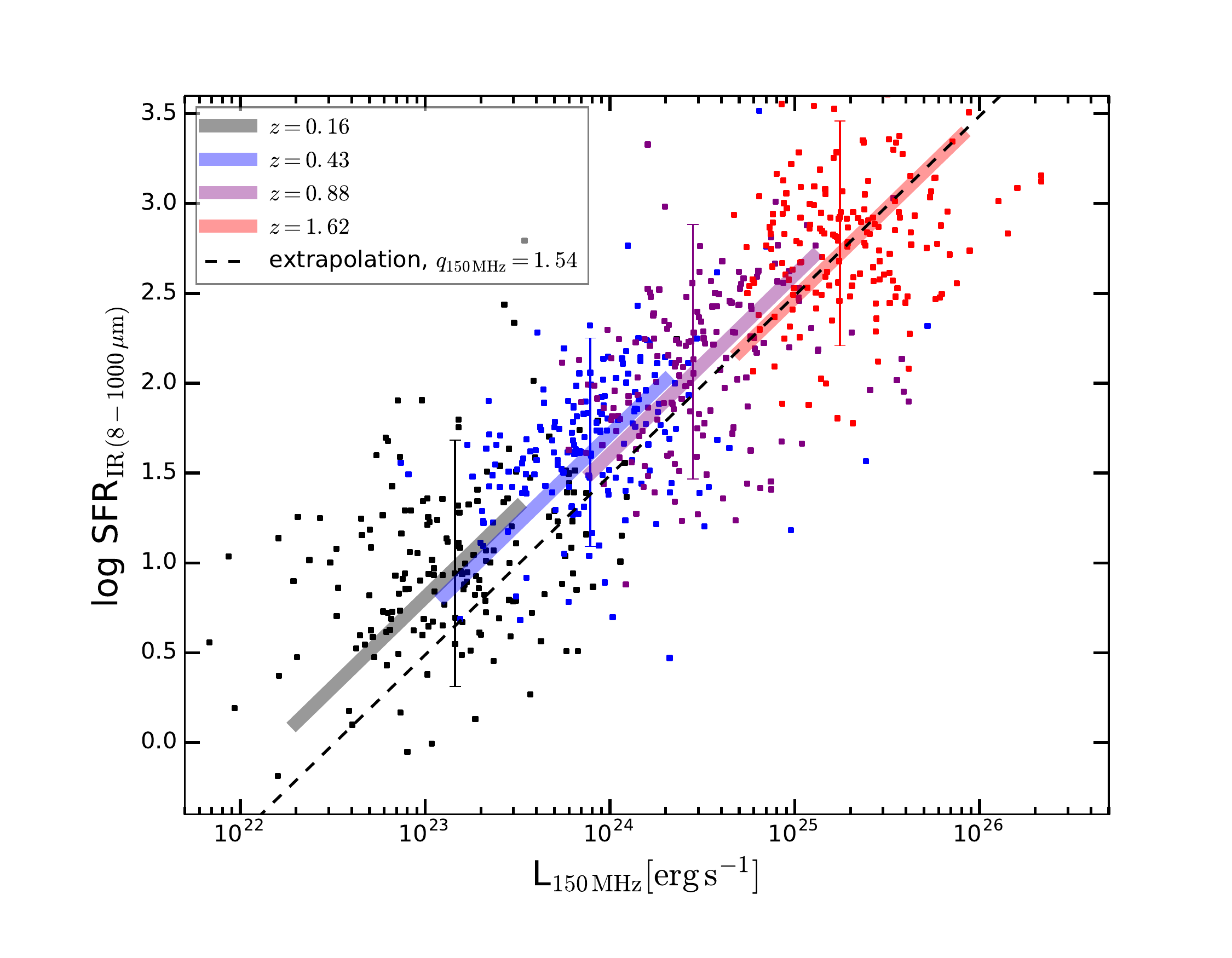}
\caption{SFR inferred from the total infrared luminosity as a function of radio luminosity at 150 MHz. Scatter points are the observed median SFR values of all SF-sources, colour-coded by redshift bin. Redshifts were binned in groups of equal density, where median values of each bins are labeled in the legend. The transparent lines following the same colour-coding correspond to the calculated SFR values using Eq.\ref{eq:SFR150}.The dotted line shows the SFR values, when a non-evolving q-value is assumed.
}
\label{fig:SFR150}
\end{figure*}
% ----------------------------------------------------------
In Fig. \ref{fig:curvature-z} we have shown that the spectral curvature is not a function of redshift when considering only the detected sources in 1.4 GHz (90 per cent of total). 
We added that the inclusion of the remaining  fraction of undetected sources (10 per cent) would imprint a weak redshift dependence on the results (section \ref{subsec:curvature-z}). 
However, in that case we clarified that this finding produced no significant differences to the total characterization of the curvature.
In contrast to section \ref{subsec:curvature-z}, for the present IRC study we use the total LOFAR-I-band selected sample for which 33 percent of sources have non-detections at 1.4 GHz (Table \ref{table:multi-wavelength}). These non-detections will have a stronger impact here than in the spectral curvature study. 

We quantify this impact by investigating the redshift evolution of subsamples of 1.4 GHz detections and non-detections exclusively. For the subsample of detected sources we find an evolution of $q_{1.4} \propto (1+z)^{-0.22\pm0.04}$, which is a similar result to the one obtained from $q_{150}$. The study comprising exclusively sources undetected in 1.4 GHz returns an evolution of $q_{1.4} \propto (1+z)^{-0.07\pm0.04}$. 
From these results we conclude that the weaker redshift evolution of $q_{1.4}$ compared to $q_{150}$ is an effect of non-detections in 1.4 GHz rather than a result of any evolution in the spectral curvature (see Fig. \ref{fig:curvature-z}, which illustrates the lack of redshift evolution).
We thus conclude that the redshift evolution in both IRC parameter distributions are consistent to a large extent.
These observations suggest that the strength of the redshift dependence found for $q_{1.4}$ can be considered a lower limit and that a stronger evolution is possible.

\section{The low frequency radio luminosity as a SFR tracer}\label{sec:SFRtracer}

Calibrating non-thermal radio continuum emission as a SFR tracer is done mostly based on the IRC due to the reliability of the IR emission as an extinction-free SFR diagnostic (see also Brown et al. (in prep.) and G\"urkan et al. (in prep.) for other studies on this at low redshifts).  
Since this is especially important for our sample due to the dusty environments observed in highly SF galaxies, we use the total IR data ($8-1000\, \mu \rm m$, see section \ref{sec:sedfitting} for details ) to compute the SFRs throughout the paper following:
\begin{equation}
\left(\dfrac{\rm SFR_{\rm IR}}{\rm M_{\odot}yr^{-1}}\right) = 3.88 \times 10^{-44} \left(\dfrac{\rm L_{\rm IR}}{{\rm ~erg~s}^{-1}}\right).
\label{eq:SFRir}
\end{equation} 
as presented as well by \cite{murphy11} and \cite{bell03}.
Combining Equations \ref{eq:SFRir}, \ref{eq:IRCdef} and \ref{150z} one can express the 150 MHz luminosity as a SFR as:
\begin{equation}
\left(\dfrac{\rm SFR_{\rm 150\,\MHz}(z)}{M_{\odot}yr^{-1}}\right) = 1.455 \times 10^{-24}  \times 10^{\rm q_{150}(z)} \times  \left(\dfrac{\rm L_{150}}{{\rm ~erg~s}^{-1}}\right),
\label{eq:SFR150}
\end{equation} 
where the factor $(1.455 \times 10^{-24}  \times 10^{\rm q_{150}(z)})$ is equivalent to $2.450\times 10^{-24}$, $ 2.289\times 10^{-24}$, $2.149\times 10^{-24}$ and $1.965\times 10^{-24}$ for redshifts $z$ around 0.1, 0.5, 1.0 and 2.0, respectively.

In Fig.~\ref{fig:SFR150} we show how the redshift evolution of the $q_{150}$-value suggests a more accurate SFR estimation compared to assuming a unique $q_{150}$ value.
We over-plot the observed median SFR$_{\rm IR}$ values of all SF-sources (scatter points) to the inferred SFR values using Eq.\ref{eq:SFR150} (transparent lines), both colour-coded by redshift bins.
In contrast we plot the SFR values, as assuming a unique $q_{150}$-value at all redshifts as a dotted black line.
As can be seen in Fig.~\ref{fig:SFR150}, values derived from Eq.\ref{eq:SFR150} give a better description of the data.

%%%%%%%%%%%%%%%%%%%%%%%%%%%%%%%%%%%%%%%%%%%%%%%%%%%%%%%%%%%%%%%%%%%%%%%%%%%%%%%%%%%%%%%%%%%%%%%%%%%%%%%%%%%%%%

\section{Summary}\label{sec:summary}

In this paper we have investigated two properties of radio selected SF galaxies and AGN, namely their radio spectral energy distributions and the IR-radio correlation (IR), and how these evolve across the bulk of cosmic history.
%In this paper we have investigated the radio spectral energy distributions of both star-forming galaxies and AGN across the bulk of cosmic history.
Our study is based on a radio and I-band selected sample from the Bo\"otes field and comprises $\sim 1500$ galaxies at redshifts of $0.05<z<2.5$.
Taking advantage of the rich multi-wavelength coverage available for the Bo\"otes field, we constructed panchromatic SEDs for our sample, incorporating UV/optical to far-infrared observations. The sources were then classified into star-formation and AGN-dominated galaxies using a multi-wavelength SED-fitting method \citep[\textsc{AGNfitter} code,][]{calistrorivera16}.

We have characterized the statistical radio spectral index and spectral curvature properties for our total samples of SF galaxies and AGN using deep radio data across a wide range in radio frequency (150, 325, 610 MHz and 1.4 GHz). 
Additionally, intrinsic physical properties of the galaxy and AGN were investigated by calibrating spectral properties with parameters inferred from multi-wavelength SED-fitting. 
We arrive at the following conclusions:
\begin{itemize}
\item SF galaxies exhibit an average radio spectral slope of $\alpha^{1400}_{150} =-0.73$. In comparison, AGN were found to exhibit a flatter spectral slope, $\alpha^{1400}_{150} =-0.67$. Although the difference is small compared to the width of the observed $\alpha^{1400}_{150}$ distributions we find it to be statistically significant. 
\item Observations of radio SEDs indicate a difference in the curvature of the radio continuum for SF galaxies compared to AGN-dominated galaxies.
SF galaxies exhibit a slightly curved radio spectrum which flattens at frequencies $<$ 325 MHz.
Conversely, AGN-dominated radio galaxies display a curved radio spectrum which shows a systematic steepening towards low frequencies below 1.4 GHz.
In both cases we find these results to be significant based two-sample KS-tests.
\item No evolution of the curvature with redshift is observed for both the SF galaxies and AGN samples.
\item In SF galaxies we found no dependence of the curvature on star-formation rate or specific star-formation rate.
\item A correlation between the radio spectral curvature and torus luminosity is observed for AGN, with low luminosity AGN exhibiting a higher degree of steepening than high luminosity sources.
\end{itemize}

In addition to studying the radio SEDs, the extensive mid- to far-IR data available for our sample allowed us to also study in detail the infrared radio correlation (IRC) for the star-forming galaxies:
%Infrared-radio correlation (IRC)
\begin{itemize}
\item An evolution with redshift of the IRC for 1.4 GHz radio luminosities is observed, following $q_{1.4}(z)= 2.45\pm0.04 \times (1+z)^{-0.15\pm0.03}$, where $q_{1.4}$ is the ratio between the total infrared luminosity ($L_{\rm{IR}}, 8-1000\, \mu m$) and the 1.4 GHz radio luminosity ($L_{\textup{1.4 GHz}}$). 
The observed evolution accounts only for $\sim 6$ per cent of the variance of the distribution. 
The large observed intrinsic scatter of $\sigma_{int}=0.53$ is still predominantly unexplained.
We discuss that the inclusion of radio non-detections in this sample may bias the result towards a weaker redshift dependence and thus conclude that this evolution can be considered a lower limit for the IRC at 1.4 GHz.
\item Similarly, an evolution of the IRC with redshift for 150 MHz radio luminosities is found following $q_{150}(z) = 1.72\pm0.04 \times (1+z)^{-0.22\pm0.05}$. 
A comparison of the observed $q_{150}$ values and those extrapolated from $q_{1.4}$, as a function of redshift, give similar results but with some scatter around $z\sim 1$.
\item After converting $L_{\rm IR}$ to $\rm SFR_{\rm IR}$ to calibrate the low frequency radio luminosity as a measure of SFR, we show that the observed IRC redshift evolution is potentially an important factor for the future use of radio-luminosity to estimate un-biased star-formation rates.

\end{itemize}

\section{Acknowledgements}
The authors thank the anonymous referee for the very careful reading of the manuscript and the many insightful comments and suggestions that improved the paper.
We thank Jacqueline Hodge, Frank Israel and Michael Brown for useful discussions and suggestions.
GCR, KJD and HJR acknowledge support from the European Research Council under the European Union’s Seventh Framework Programme (FP/2007- 2013) /ERC Advanced Grant NEW-CLUSTERS-321271.
WLW and MJH acknowledge support from the UK Science and Technology Facilities Council [ST/M001008/1]. 
PNB is grateful for support from the UK STFC via grant ST/M001229/1. 
MJJ acknowledges support from UK Science and Technology Facilities Council and the South African SKA Project
EKM acknowledges support from the Australian Research Council Centre of Excellence for All-sky Astrophysics (CAASTRO), through project number CE110001020. 
GJW Gratefully acknowledges support from The Leverhulme Trust.
This research made use of ASTROPY, a community-developed
core Python package for astronomy (Astropy Collaboration et al.
2013)  hosted  at \url{http://www.astropy.org/}.

%%%%%%%%%%%%%%%%%%%%%%%%%%%%%%%%%%%%%%%%%%%%%%%%%%%%%%%%%%%%%%%%%%%%%%%%%%%%%%%%%%%%%%%%%%%%%%%%%%%%%%%%%%%%%%%%%%%%%%%%%%%

\label{lastpage}

\bibliographystyle{mnras}
\bibliography{LOFAR-SBandAGN}

\begin{thebibliography}{}
\makeatletter
\relax
\def\mn@urlcharsother{\let\do\@makeother \do\$\do\&\do\#\do\^\do\_\do\%\do\~}
\def\mn@doi{\begingroup\mn@urlcharsother \@ifnextchar [ {\mn@doi@}
  {\mn@doi@[]}}
\def\mn@doi@[#1]#2{\def\@tempa{#1}\ifx\@tempa\@empty \href
  {http://dx.doi.org/#2} {doi:#2}\else \href {http://dx.doi.org/#2} {#1}\fi
  \endgroup}
\def\mn@eprint#1#2{\mn@eprint@#1:#2::\@nil}
\def\mn@eprint@arXiv#1{\href {http://arxiv.org/abs/#1} {{\tt arXiv:#1}}}
\def\mn@eprint@dblp#1{\href {http://dblp.uni-trier.de/rec/bibtex/#1.xml}
  {dblp:#1}}
\def\mn@eprint@#1:#2:#3:#4\@nil{\def\@tempa {#1}\def\@tempb {#2}\def\@tempc
  {#3}\ifx \@tempc \@empty \let \@tempc \@tempb \let \@tempb \@tempa \fi \ifx
  \@tempb \@empty \def\@tempb {arXiv}\fi \@ifundefined
  {mn@eprint@\@tempb}{\@tempb:\@tempc}{\expandafter \expandafter \csname
  mn@eprint@\@tempb\endcsname \expandafter{\@tempc}}}

\bibitem[\protect\citeauthoryear{{Appleton} et~al.,}{{Appleton}
  et~al.}{2004}]{appleton04}
{Appleton} P.~N.,  et~al., 2004, \mn@doi [\apjs] {10.1086/422425}, \href
  {http://adsabs.harvard.edu/abs/2004ApJS..154..147A} {154, 147}

\bibitem[\protect\citeauthoryear{{Ashby} et~al.,}{{Ashby}
  et~al.}{2009}]{ashby09}
{Ashby} M.~L.~N.,  et~al., 2009, \mn@doi [\apj] {10.1088/0004-637X/701/1/428},
  \href {http://adsabs.harvard.edu/abs/2009ApJ...701..428A} {701, 428}

\bibitem[\protect\citeauthoryear{{Autry} et~al.,}{{Autry}
  et~al.}{2003}]{autry03}
{Autry} R.~G.,  et~al., 2003, in {Iye} M.,  {Moorwood} A.~F.~M.,  eds,
  \procspie Vol. 4841, Instrument Design and Performance for Optical/Infrared
  Ground-based Telescopes. pp 525--539, \mn@doi{10.1117/12.460419}

\bibitem[\protect\citeauthoryear{{Basu}, {Roy}  \& {Mitra}}{{Basu}
  et~al.}{2012}]{basu12}
{Basu} A.,  {Roy} S.,   {Mitra} D.,  2012, \mn@doi [\apj]
  {10.1088/0004-637X/756/2/141}, \href
  {http://adsabs.harvard.edu/abs/2012ApJ...756..141B} {756, 141}

\bibitem[\protect\citeauthoryear{{Basu}, {Beck}, {Schmidt}  \& {Roy}}{{Basu}
  et~al.}{2015}]{basu15b}
{Basu} A.,  {Beck} R.,  {Schmidt} P.,   {Roy} S.,  2015, \mn@doi [\mnras]
  {10.1093/mnras/stv510}, \href
  {http://adsabs.harvard.edu/abs/2015MNRAS.449.3879B} {449, 3879}

\bibitem[\protect\citeauthoryear{{Becker}, {White}  \& {Helfand}}{{Becker}
  et~al.}{1995}]{FIRST}
{Becker} R.~H.,  {White} R.~L.,   {Helfand} D.~J.,  1995, \mn@doi [\apj]
  {10.1086/176166}, \href {http://adsabs.harvard.edu/abs/1995ApJ...450..559B}
  {450, 559}

\bibitem[\protect\citeauthoryear{{Bell}}{{Bell}}{2003}]{bell03}
{Bell} E.~F.,  2003, \mn@doi [\apj] {10.1086/367829}, \href
  {http://adsabs.harvard.edu/abs/2003ApJ...586..794B} {586, 794}

\bibitem[\protect\citeauthoryear{{Bertin} \& {Arnouts}}{{Bertin} \&
  {Arnouts}}{1996}]{SExtractor}
{Bertin} E.,  {Arnouts} S.,  1996, \mn@doi [\aaps] {10.1051/aas:1996164}, \href
  {http://adsabs.harvard.edu/abs/1996A%26AS..117..393B} {117, 393}

\bibitem[\protect\citeauthoryear{{Best} \& {Heckman}}{{Best} \&
  {Heckman}}{2012}]{bestANDheckman12}
{Best} P.~N.,  {Heckman} T.~M.,  2012, \mn@doi [\mnras]
  {10.1111/j.1365-2966.2012.20414.x}, \href
  {http://adsabs.harvard.edu/abs/2012MNRAS.421.1569B} {421, 1569}

\bibitem[\protect\citeauthoryear{{Bianchi}, {Conti}  \& {Shiao}}{{Bianchi}
  et~al.}{2014}]{bianchi14}
{Bianchi} L.,  {Conti} A.,   {Shiao} B.,  2014, \mn@doi [Advances in Space
  Research] {10.1016/j.asr.2013.07.045}, \href
  {http://adsabs.harvard.edu/abs/2014AdSpR..53..900B} {53, 900}

\bibitem[\protect\citeauthoryear{{Bonzini}, {Padovani}, {Mainieri},
  {Kellermann}, {Miller}, {Rosati}, {Tozzi}  \& {Vattakunnel}}{{Bonzini}
  et~al.}{2013}]{bonzini13}
{Bonzini} M.,  {Padovani} P.,  {Mainieri} V.,  {Kellermann} K.~I.,  {Miller}
  N.,  {Rosati} P.,  {Tozzi} P.,   {Vattakunnel} S.,  2013, \mn@doi [\mnras]
  {10.1093/mnras/stt1879}, \href
  {http://adsabs.harvard.edu/abs/2013MNRAS.436.3759B} {436, 3759}

\bibitem[\protect\citeauthoryear{{Bourne}, {Dunne}, {Ivison}, {Maddox},
  {Dickinson}  \& {Frayer}}{{Bourne} et~al.}{2011}]{bourne11}
{Bourne} N.,  {Dunne} L.,  {Ivison} R.~J.,  {Maddox} S.~J.,  {Dickinson} M.,
  {Frayer} D.~T.,  2011, \mn@doi [\mnras] {10.1111/j.1365-2966.2010.17517.x},
  \href {http://adsabs.harvard.edu/abs/2011MNRAS.410.1155B} {410, 1155}

\bibitem[\protect\citeauthoryear{{Brammer}, {van Dokkum}  \& {Coppi}}{{Brammer}
  et~al.}{2008}]{EAZY}
{Brammer} G.~B.,  {van Dokkum} P.~G.,   {Coppi} P.,  2008, \mn@doi [\apj]
  {10.1086/591786}, \href {http://adsabs.harvard.edu/abs/2008ApJ...686.1503B}
  {686, 1503}

\bibitem[\protect\citeauthoryear{{Brown}, {Dey}, {Jannuzi}, {Brand}, {Benson},
  {Brodwin}, {Croton}  \& {Eisenhardt}}{{Brown} et~al.}{2007}]{brown07}
{Brown} M.~J.~I.,  {Dey} A.,  {Jannuzi} B.~T.,  {Brand} K.,  {Benson} A.~J.,
  {Brodwin} M.,  {Croton} D.~J.,   {Eisenhardt} P.~R.,  2007, \mn@doi [\apj]
  {10.1086/509652}, \href {http://adsabs.harvard.edu/abs/2007ApJ...654..858B}
  {654, 858}

\bibitem[\protect\citeauthoryear{{Brown} et~al.,}{{Brown}
  et~al.}{2014}]{brown14}
{Brown} M.~J.~I.,  et~al., 2014, \mn@doi [\apjs] {10.1088/0067-0049/212/2/18},
  \href {http://adsabs.harvard.edu/abs/2014ApJS..212...18B} {212, 18}

\bibitem[\protect\citeauthoryear{{Bruzual} \& {Charlot}}{{Bruzual} \&
  {Charlot}}{2003}]{bruzual03}
{Bruzual} G.,  {Charlot} S.,  2003, \mn@doi [\mnras]
  {10.1046/j.1365-8711.2003.06897.x}, \href
  {http://adsabs.harvard.edu/abs/2003MNRAS.344.1000B} {344, 1000}

\bibitem[\protect\citeauthoryear{{Calistro Rivera}, {Lusso}, {Hennawi}  \&
  {Hogg}}{{Calistro Rivera} et~al.}{2016}]{calistrorivera16}
{Calistro Rivera} G.,  {Lusso} E.,  {Hennawi} J.~F.,   {Hogg} D.~W.,  2016,
  \mn@doi [\apj] {10.3847/1538-4357/833/1/98}, \href
  {http://adsabs.harvard.edu/abs/2016ApJ...833...98C} {833, 98}

\bibitem[\protect\citeauthoryear{{Calzetti}, {Kinney}  \&
  {Storchi-Bergmann}}{{Calzetti} et~al.}{1994}]{calzetti94}
{Calzetti} D.,  {Kinney} A.~L.,   {Storchi-Bergmann} T.,  1994, \mn@doi [\apj]
  {10.1086/174346}, \href {http://adsabs.harvard.edu/abs/1994ApJ...429..582C}
  {429, 582}

\bibitem[\protect\citeauthoryear{{Carrasco Kind} \& {Brunner}}{{Carrasco Kind}
  \& {Brunner}}{2014}]{carrasco14}
{Carrasco Kind} M.,  {Brunner} R.~J.,  2014, \mn@doi [\mnras]
  {10.1093/mnras/stu1098}, \href
  {http://adsabs.harvard.edu/abs/2014MNRAS.442.3380C} {442, 3380}

\bibitem[\protect\citeauthoryear{{Casey} et~al.,}{{Casey}
  et~al.}{2012}]{casey12}
{Casey} C.~M.,  et~al., 2012, \mn@doi [\apj] {10.1088/0004-637X/761/2/140},
  \href {http://adsabs.harvard.edu/abs/2012ApJ...761..140C} {761, 140}

\bibitem[\protect\citeauthoryear{{Casey}, {Narayanan}  \& {Cooray}}{{Casey}
  et~al.}{2014}]{casey14}
{Casey} C.~M.,  {Narayanan} D.,   {Cooray} A.,  2014, \mn@doi [\physrep]
  {10.1016/j.physrep.2014.02.009}, \href
  {http://adsabs.harvard.edu/abs/2014PhR...541...45C} {541, 45}

\bibitem[\protect\citeauthoryear{{Chabrier}}{{Chabrier}}{2003}]{chabrier03}
{Chabrier} G.,  2003, \mn@doi [\pasp] {10.1086/376392}, \href
  {http://adsabs.harvard.edu/abs/2003PASP..115..763C} {115, 763}

\bibitem[\protect\citeauthoryear{{Chary} \& {Elbaz}}{{Chary} \&
  {Elbaz}}{2001}]{chary01}
{Chary} R.,  {Elbaz} D.,  2001, \mn@doi [\apj] {10.1086/321609}, \href
  {http://adsabs.harvard.edu/abs/2001ApJ...556..562C} {556, 562}

\bibitem[\protect\citeauthoryear{{Clemens}, {Scaife}, {Vega}  \&
  {Bressan}}{{Clemens} et~al.}{2010}]{clemens10}
{Clemens} M.~S.,  {Scaife} A.,  {Vega} O.,   {Bressan} A.,  2010, \mn@doi
  [\mnras] {10.1111/j.1365-2966.2010.16534.x}, \href
  {http://adsabs.harvard.edu/abs/2010MNRAS.405..887C} {405, 887}

\bibitem[\protect\citeauthoryear{{Condon}}{{Condon}}{1992}]{condon92}
{Condon} J.~J.,  1992, \mn@doi [\araa] {10.1146/annurev.aa.30.090192.003043},
  \href {http://adsabs.harvard.edu/abs/1992ARA26A..30..575C} {30, 575}

\bibitem[\protect\citeauthoryear{{Condon}, {Cotton}, {Greisen}, {Yin},
  {Perley}, {Taylor}  \& {Broderick}}{{Condon} et~al.}{1998}]{NVSS}
{Condon} J.~J.,  {Cotton} W.~D.,  {Greisen} E.~W.,  {Yin} Q.~F.,  {Perley}
  R.~A.,  {Taylor} G.~B.,   {Broderick} J.~J.,  1998, \mn@doi [\aj]
  {10.1086/300337}, \href {http://adsabs.harvard.edu/abs/1998AJ....115.1693C}
  {115, 1693}

\bibitem[\protect\citeauthoryear{{Cool}}{{Cool}}{2007}]{cool07}
{Cool} R.~J.,  2007, \mn@doi [\apjs] {10.1086/511179}, \href
  {http://adsabs.harvard.edu/abs/2007ApJS..169...21C} {169, 21}

\bibitem[\protect\citeauthoryear{{Coppejans}, {Cseh}, {Williams}, {van Velzen}
  \& {Falcke}}{{Coppejans} et~al.}{2015}]{coppejans15}
{Coppejans} R.,  {Cseh} D.,  {Williams} W.~L.,  {van Velzen} S.,   {Falcke} H.,
   2015, \mn@doi [\mnras] {10.1093/mnras/stv681}, \href
  {http://adsabs.harvard.edu/abs/2015MNRAS.450.1477C} {450, 1477}

\bibitem[\protect\citeauthoryear{Dahlen et~al.,}{Dahlen
  et~al.}{2013}]{dahlen13}
Dahlen T.,  et~al., 2013, The Astrophysical Journal, 775, 93

\bibitem[\protect\citeauthoryear{{Dale} \& {Helou}}{{Dale} \&
  {Helou}}{2002}]{dale02}
{Dale} D.~A.,  {Helou} G.,  2002, \mn@doi [\apj] {10.1086/341632}, \href
  {http://adsabs.harvard.edu/abs/2002ApJ...576..159D} {576, 159}

\bibitem[\protect\citeauthoryear{{Delhaize} et~al.,}{{Delhaize}
  et~al.}{2017}]{delhaize17}
{Delhaize} J.,  et~al., 2017, preprint, \href
  {http://adsabs.harvard.edu/abs/2017arXiv170309723D} {} (\mn@eprint {arXiv}
  {1703.09723})

\bibitem[\protect\citeauthoryear{{Delvecchio} et~al.,}{{Delvecchio}
  et~al.}{2017}]{delvecchio17}
{Delvecchio} I.,  et~al., 2017, preprint, \href
  {http://adsabs.harvard.edu/abs/2017arXiv170309720D} {} (\mn@eprint {arXiv}
  {1703.09720})

\bibitem[\protect\citeauthoryear{{Dumas}, {Schinnerer}, {Tabatabaei}, {Beck},
  {Velusamy}  \& {Murphy}}{{Dumas} et~al.}{2011}]{dumas11}
{Dumas} G.,  {Schinnerer} E.,  {Tabatabaei} F.~S.,  {Beck} R.,  {Velusamy} T.,
   {Murphy} E.,  2011, \mn@doi [\aj] {10.1088/0004-6256/141/2/41}, \href
  {http://adsabs.harvard.edu/abs/2011AJ....141...41D} {141, 41}

\bibitem[\protect\citeauthoryear{{Fanaroff} \& {Riley}}{{Fanaroff} \&
  {Riley}}{1974}]{fanaroff74}
{Fanaroff} B.~L.,  {Riley} J.~M.,  1974, \mn@doi [\mnras]
  {10.1093/mnras/167.1.31P}, \href
  {http://adsabs.harvard.edu/abs/1974MNRAS.167P..31F} {167, 31P}

\bibitem[\protect\citeauthoryear{{Foreman-Mackey}, {Hogg}, {Lang}  \&
  {Goodman}}{{Foreman-Mackey} et~al.}{2013}]{fm13}
{Foreman-Mackey} D.,  {Hogg} D.~W.,  {Lang} D.,   {Goodman} J.,  2013, \mn@doi
  [\pasp] {10.1086/670067}, \href
  {http://adsabs.harvard.edu/abs/2013PASP..125..306F} {125, 306}

\bibitem[\protect\citeauthoryear{{Gioia}, {Gregorini}  \& {Klein}}{{Gioia}
  et~al.}{1982}]{gioia82}
{Gioia} I.~M.,  {Gregorini} L.,   {Klein} U.,  1982, \aap, \href
  {http://adsabs.harvard.edu/abs/1982A%26A...116..164G} {116, 164}

\bibitem[\protect\citeauthoryear{{Hardcastle}}{{Hardcastle}}{2009}]{hardcastle09}
{Hardcastle} M.~J.,  2009, in {Saikia} D.~J.,  {Green} D.~A.,  {Gupta} Y.,
  {Venturi} T.,  eds,  Astronomical Society of the Pacific Conference Series
  Vol. 407, The Low-Frequency Radio Universe. p.~121

\bibitem[\protect\citeauthoryear{{Hardcastle}, {Evans}  \&
  {Croston}}{{Hardcastle} et~al.}{2007}]{hardcastle07}
{Hardcastle} M.~J.,  {Evans} D.~A.,   {Croston} J.~H.,  2007, \mn@doi [\mnras]
  {10.1111/j.1365-2966.2007.11572.x}, \href
  {http://adsabs.harvard.edu/abs/2007MNRAS.376.1849H} {376, 1849}

\bibitem[\protect\citeauthoryear{{Hardcastle} et~al.,}{{Hardcastle}
  et~al.}{2016}]{hardcastle16}
{Hardcastle} M.~J.,  et~al., 2016, preprint, \href
  {http://adsabs.harvard.edu/abs/2016arXiv160609437H} {} (\mn@eprint {arXiv}
  {1606.09437})

\bibitem[\protect\citeauthoryear{{Harwood}, {Hardcastle}  \&
  {Croston}}{{Harwood} et~al.}{2015}]{harwood15}
{Harwood} J.~J.,  {Hardcastle} M.~J.,   {Croston} J.~H.,  2015, \mn@doi
  [\mnras] {10.1093/mnras/stv2194}, \href
  {http://adsabs.harvard.edu/abs/2015MNRAS.454.3403H} {454, 3403}

\bibitem[\protect\citeauthoryear{{Helou}, {Soifer}  \&
  {Rowan-Robinson}}{{Helou} et~al.}{1985}]{helou85}
{Helou} G.,  {Soifer} B.~T.,   {Rowan-Robinson} M.,  1985, \mn@doi [\apjl]
  {10.1086/184556}, \href {http://adsabs.harvard.edu/abs/1985ApJ...298L...7H}
  {298, L7}

\bibitem[\protect\citeauthoryear{{Hogg}}{{Hogg}}{2001}]{hogg01}
{Hogg} D.~W.,  2001, \mn@doi [\aj] {10.1086/318736}, \href
  {http://adsabs.harvard.edu/abs/2001AJ....121.1207H} {121, 1207}

\bibitem[\protect\citeauthoryear{{Ibar} et~al.,}{{Ibar} et~al.}{2008}]{ibar08}
{Ibar} E.,  et~al., 2008, \mn@doi [\mnras] {10.1111/j.1365-2966.2008.13077.x},
  \href {http://adsabs.harvard.edu/abs/2008MNRAS.386..953I} {386, 953}

\bibitem[\protect\citeauthoryear{{Ibar}, {Ivison}, {Biggs}, {Lal}, {Best}  \&
  {Green}}{{Ibar} et~al.}{2009}]{ibar09}
{Ibar} E.,  {Ivison} R.~J.,  {Biggs} A.~D.,  {Lal} D.~V.,  {Best} P.~N.,
  {Green} D.~A.,  2009, \mn@doi [\mnras] {10.1111/j.1365-2966.2009.14866.x},
  \href {http://adsabs.harvard.edu/abs/2009MNRAS.397..281I} {397, 281}

\bibitem[\protect\citeauthoryear{{Ibar}, {Ivison}, {Best}, {Coppin}, {Pope},
  {Smail}  \& {Dunlop}}{{Ibar} et~al.}{2010}]{ibar10}
{Ibar} E.,  {Ivison} R.~J.,  {Best} P.~N.,  {Coppin} K.,  {Pope} A.,  {Smail}
  I.,   {Dunlop} J.~S.,  2010, \mn@doi [\mnras]
  {10.1111/j.1745-3933.2009.00786.x}, \href
  {http://adsabs.harvard.edu/abs/2010MNRAS.401L..53I} {401, L53}

\bibitem[\protect\citeauthoryear{{Intema}, {van Weeren}, {R{\"o}ttgering}  \&
  {Lal}}{{Intema} et~al.}{2011}]{intema11}
{Intema} H.~T.,  {van Weeren} R.~J.,  {R{\"o}ttgering} H.~J.~A.,   {Lal} D.~V.,
   2011, \mn@doi [\aap] {10.1051/0004-6361/201014253}, \href
  {http://adsabs.harvard.edu/abs/2011A26A...535A..38I} {535, A38}

\bibitem[\protect\citeauthoryear{{Israel} \& {Mahoney}}{{Israel} \&
  {Mahoney}}{1990}]{israelmahoney90}
{Israel} F.~P.,  {Mahoney} M.~J.,  1990, \mn@doi [\apj] {10.1086/168513}, \href
  {http://adsabs.harvard.edu/abs/1990ApJ...352...30I} {352, 30}

\bibitem[\protect\citeauthoryear{{Ivison} et~al.,}{{Ivison}
  et~al.}{2010a}]{ivison10b}
{Ivison} R.~J.,  et~al., 2010a, \mn@doi [\mnras]
  {10.1111/j.1365-2966.2009.15918.x}, \href
  {http://adsabs.harvard.edu/abs/2010MNRAS.402..245I} {402, 245}

\bibitem[\protect\citeauthoryear{{Ivison}, {Magnelli}  \& {Ibar}}{{Ivison}
  et~al.}{2010b}]{ivison10a}
{Ivison} R.~J.,  {Magnelli} B.,   {Ibar} E. e.~a.,  2010b, \mn@doi [\aap]
  {10.1051/0004-6361/201014552}, \href
  {http://adsabs.harvard.edu/abs/2010A%26A...518L..31I} {518, L31}

\bibitem[\protect\citeauthoryear{{Jannuzi} \& {Dey}}{{Jannuzi} \&
  {Dey}}{1999}]{januzziANDdey99}
{Jannuzi} B.~T.,  {Dey} A.,  1999, in {Weymann} R.,  {Storrie-Lombardi} L.,
  {Sawicki} M.,   {Brunner} R.,  eds,  Astronomical Society of the Pacific
  Conference Series Vol. 191, Photometric Redshifts and the Detection of High
  Redshift Galaxies. p.~111

\bibitem[\protect\citeauthoryear{{Jannuzi} et~al.,}{{Jannuzi}
  et~al.}{2010}]{jannuzi10}
{Jannuzi} B.,  et~al., 2010, in American Astronomical Society Meeting Abstracts
  \#215. p.~513

\bibitem[\protect\citeauthoryear{{Jarvis} \& {Rawlings}}{{Jarvis} \&
  {Rawlings}}{2004}]{jarvis&rawlings04}
{Jarvis} M.~J.,  {Rawlings} S.,  2004, \mn@doi [\nar]
  {10.1016/j.newar.2004.09.006}, \href
  {http://adsabs.harvard.edu/abs/2004NewAR..48.1173J} {48, 1173}

\bibitem[\protect\citeauthoryear{{Jarvis} et~al.,}{{Jarvis}
  et~al.}{2010}]{jarvis10}
{Jarvis} M.~J.,  et~al., 2010, \mn@doi [\mnras]
  {10.1111/j.1365-2966.2010.17772.x}, \href
  {http://adsabs.harvard.edu/abs/2010MNRAS.409...92J} {409, 92}

\bibitem[\protect\citeauthoryear{{Jenkins} \& {McEllin}}{{Jenkins} \&
  {McEllin}}{1977}]{jenkins77}
{Jenkins} C.~J.,  {McEllin} M.,  1977, \mn@doi [\mnras]
  {10.1093/mnras/180.2.219}, \href
  {http://adsabs.harvard.edu/abs/1977MNRAS.180..219J} {180, 219}

\bibitem[\protect\citeauthoryear{{Kennicutt}}{{Kennicutt}}{1998}]{kennicutt98}
{Kennicutt} Jr. R.~C.,  1998, \mn@doi [\araa] {10.1146/annurev.astro.36.1.189},
  \href {http://adsabs.harvard.edu/abs/1998ARA%26A..36..189K} {36, 189}

\bibitem[\protect\citeauthoryear{{Ker}, {Best}, {Rigby}, {R{\"o}ttgering}  \&
  {Gendre}}{{Ker} et~al.}{2012}]{ker12}
{Ker} L.~M.,  {Best} P.~N.,  {Rigby} E.~E.,  {R{\"o}ttgering} H.~J.~A.,
  {Gendre} M.~A.,  2012, \mn@doi [\mnras] {10.1111/j.1365-2966.2011.20235.x},
  \href {http://adsabs.harvard.edu/abs/2012MNRAS.420.2644K} {420, 2644}

\bibitem[\protect\citeauthoryear{{Kharb}, {Srivastava}, {Singh}, {Gallimore},
  {Ishwara-Chandra}  \& {Ananda}}{{Kharb} et~al.}{2016}]{kharb16}
{Kharb} P.,  {Srivastava} S.,  {Singh} V.,  {Gallimore} J.~F.,
  {Ishwara-Chandra} C.~H.,   {Ananda} H.,  2016, \mn@doi [\mnras]
  {10.1093/mnras/stw699}, \href
  {http://adsabs.harvard.edu/abs/2016MNRAS.459.1310K} {459, 1310}

\bibitem[\protect\citeauthoryear{{Kochanek} et~al.,}{{Kochanek}
  et~al.}{2012}]{kochanek12}
{Kochanek} C.~S.,  et~al., 2012, \mn@doi [\apjs] {10.1088/0067-0049/200/1/8},
  \href {http://adsabs.harvard.edu/abs/2012ApJS..200....8K} {200, 8}

\bibitem[\protect\citeauthoryear{{Komatsu} et~al.,}{{Komatsu}
  et~al.}{2009}]{komatsu09}
{Komatsu} E.,  et~al., 2009, \mn@doi [\apjs] {10.1088/0067-0049/180/2/330},
  \href {http://adsabs.harvard.edu/abs/2009ApJS..180..330K} {180, 330}

\bibitem[\protect\citeauthoryear{{Krause}, {Alexander}, {Riley}  \&
  {Hopton}}{{Krause} et~al.}{2012}]{krause12}
{Krause} M.,  {Alexander} P.,  {Riley} J.,   {Hopton} D.,  2012, \mn@doi
  [\mnras] {10.1111/j.1365-2966.2012.21645.x}, \href
  {http://adsabs.harvard.edu/abs/2012MNRAS.427.3196K} {427, 3196}

\bibitem[\protect\citeauthoryear{{Lacki}}{{Lacki}}{2013}]{lacki13}
{Lacki} B.~C.,  2013, \mn@doi [\mnras] {10.1093/mnras/stt349}, \href
  {http://adsabs.harvard.edu/abs/2013MNRAS.431.3003L} {431, 3003}

\bibitem[\protect\citeauthoryear{{Lacki}, {Thompson}  \& {Quataert}}{{Lacki}
  et~al.}{2010}]{lacki10}
{Lacki} B.~C.,  {Thompson} T.~A.,   {Quataert} E.,  2010, \mn@doi [\apj]
  {10.1088/0004-637X/717/1/1}, \href
  {http://adsabs.harvard.edu/abs/2010ApJ...717....1L} {717, 1}

\bibitem[\protect\citeauthoryear{{Laing} \& {Peacock}}{{Laing} \&
  {Peacock}}{1980}]{laing80}
{Laing} R.~A.,  {Peacock} J.~A.,  1980, \mn@doi [\mnras]
  {10.1093/mnras/190.4.903}, \href
  {http://adsabs.harvard.edu/abs/1980MNRAS.190..903L} {190, 903}

\bibitem[\protect\citeauthoryear{{Lane}, {Cotton}, {van Velzen}, {Clarke},
  {Kassim}, {Helmboldt}, {Lazio}  \& {Cohen}}{{Lane} et~al.}{2014}]{VLSSR}
{Lane} W.~M.,  {Cotton} W.~D.,  {van Velzen} S.,  {Clarke} T.~E.,  {Kassim}
  N.~E.,  {Helmboldt} J.~F.,  {Lazio} T.~J.~W.,   {Cohen} A.~S.,  2014, \mn@doi
  [\mnras] {10.1093/mnras/stu256}, \href
  {http://adsabs.harvard.edu/abs/2014MNRAS.440..327L} {440, 327}

\bibitem[\protect\citeauthoryear{{Magnelli} et~al.,}{{Magnelli}
  et~al.}{2015}]{magnelli15}
{Magnelli} B.,  et~al., 2015, \mn@doi [\aap] {10.1051/0004-6361/201424937},
  \href {http://adsabs.harvard.edu/abs/2015A26A...573A..45M} {573, A45}

\bibitem[\protect\citeauthoryear{{Mahony} et~al.,}{{Mahony}
  et~al.}{2016}]{mahoney16}
{Mahony} E.~K.,  et~al., 2016, \mn@doi [\mnras] {10.1093/mnras/stw2225}, \href
  {http://adsabs.harvard.edu/abs/2016MNRAS.463.2997M} {463, 2997}

\bibitem[\protect\citeauthoryear{{Martin} et~al.,}{{Martin}
  et~al.}{2003}]{martin03}
{Martin} C.,  et~al., 2003, in {Blades} J.~C.,  {Siegmund} O.~H.~W.,  eds,
  \procspie Vol. 4854, Future EUV/UV and Visible Space Astrophysics Missions
  and Instrumentation.. pp 336--350, \mn@doi{10.1117/12.460034}

\bibitem[\protect\citeauthoryear{{Marvil}, {Owen}  \& {Eilek}}{{Marvil}
  et~al.}{2015}]{marvil14}
{Marvil} J.,  {Owen} F.,   {Eilek} J.,  2015, \mn@doi [\aj]
  {10.1088/0004-6256/149/1/32}, \href
  {http://adsabs.harvard.edu/abs/2015AJ....149...32M} {149, 32}

\bibitem[\protect\citeauthoryear{{Mauch}, {Kl{\"o}ckner}, {Rawlings}, {Jarvis},
  {Hardcastle}, {Obreschkow}, {Saikia}  \& {Thompson}}{{Mauch}
  et~al.}{2013}]{mauch13}
{Mauch} T.,  {Kl{\"o}ckner} H.-R.,  {Rawlings} S.,  {Jarvis} M.,  {Hardcastle}
  M.~J.,  {Obreschkow} D.,  {Saikia} D.~J.,   {Thompson} M.~A.,  2013, \mn@doi
  [\mnras] {10.1093/mnras/stt1323}, \href
  {http://adsabs.harvard.edu/abs/2013MNRAS.435..650M} {435, 650}

\bibitem[\protect\citeauthoryear{{Meisenheimer}, {Roser}, {Hiltner}, {Yates},
  {Longair}, {Chini}  \& {Perley}}{{Meisenheimer}
  et~al.}{1989}]{meisenheimer89}
{Meisenheimer} K.,  {Roser} H.-J.,  {Hiltner} P.~R.,  {Yates} M.~G.,  {Longair}
  M.~S.,  {Chini} R.,   {Perley} R.~A.,  1989, \aap, \href
  {http://adsabs.harvard.edu/abs/1989A%26A...219...63M} {219, 63}

\bibitem[\protect\citeauthoryear{{Mohan} \& {Rafferty}}{{Mohan} \&
  {Rafferty}}{2015}]{mohan15}
{Mohan} N.,  {Rafferty} D.,  2015, {PyBDSM: Python Blob Detection and Source
  Measurement}, Astrophysics Source Code Library (\mn@eprint {ascl} {1502.007})

\bibitem[\protect\citeauthoryear{Murphy et~al.,}{Murphy
  et~al.}{2011}]{murphy11}
Murphy E.~J.,  et~al., 2011, The Astrophysical Journal, 737, 67

\bibitem[\protect\citeauthoryear{{Murphy}, {Stierwalt}, {Armus}, {Condon}  \&
  {Evans}}{{Murphy} et~al.}{2013}]{murphy13}
{Murphy} E.~J.,  {Stierwalt} S.,  {Armus} L.,  {Condon} J.~J.,   {Evans} A.~S.,
   2013, \mn@doi [\apj] {10.1088/0004-637X/768/1/2}, \href
  {http://adsabs.harvard.edu/abs/2013ApJ...768....2M} {768, 2}

\bibitem[\protect\citeauthoryear{{Murray} et~al.,}{{Murray}
  et~al.}{2005}]{murray05}
{Murray} S.~S.,  et~al., 2005, \mn@doi [\apjs] {10.1086/444378}, \href
  {http://adsabs.harvard.edu/abs/2005ApJS..161....1M} {161, 1}

\bibitem[\protect\citeauthoryear{{Oke} \& {Gunn}}{{Oke} \&
  {Gunn}}{1983}]{okegunn83}
{Oke} J.~B.,  {Gunn} J.~E.,  1983, \mn@doi [\apj] {10.1086/160817}, \href
  {http://adsabs.harvard.edu/abs/1983ApJ...266..713O} {266, 713}

\bibitem[\protect\citeauthoryear{{Oliver} et~al.,}{{Oliver}
  et~al.}{2012}]{oliver12}
{Oliver} S.~J.,  et~al., 2012, \mn@doi [\mnras]
  {10.1111/j.1365-2966.2012.20912.x}, \href
  {http://adsabs.harvard.edu/abs/2012MNRAS.424.1614O} {424, 1614}

\bibitem[\protect\citeauthoryear{{Pannella} et~al.,}{{Pannella}
  et~al.}{2015}]{pannella15}
{Pannella} M.,  et~al., 2015, \mn@doi [\apj] {10.1088/0004-637X/807/2/141},
  \href {http://adsabs.harvard.edu/abs/2015ApJ...807..141P} {807, 141}

\bibitem[\protect\citeauthoryear{{Planck Collaboration} et~al.,}{{Planck
  Collaboration} et~al.}{2014}]{planck14}
{Planck Collaboration} et~al., 2014, \mn@doi [\aap]
  {10.1051/0004-6361/201321591}, \href
  {http://adsabs.harvard.edu/abs/2014A%26A...571A..16P} {571, A16}

\bibitem[\protect\citeauthoryear{{Polletta} et~al.,}{{Polletta}
  et~al.}{2007}]{polletta07}
{Polletta} M.,  et~al., 2007, \mn@doi [\apj] {10.1086/518113}, \href
  {http://adsabs.harvard.edu/abs/2007ApJ...663...81P} {663, 81}

\bibitem[\protect\citeauthoryear{{Rees} et~al.,}{{Rees} et~al.}{2016}]{rees16}
{Rees} G.~A.,  et~al., 2016, \mn@doi [\mnras] {10.1093/mnras/stv2468}, \href
  {http://adsabs.harvard.edu/abs/2016MNRAS.455.2731R} {455, 2731}

\bibitem[\protect\citeauthoryear{{Rengelink}, {Tang}, {de Bruyn}, {Miley},
  {Bremer}, {Roettgering}  \& {Bremer}}{{Rengelink} et~al.}{1997}]{WENSS}
{Rengelink} R.~B.,  {Tang} Y.,  {de Bruyn} A.~G.,  {Miley} G.~K.,  {Bremer}
  M.~N.,  {Roettgering} H.~J.~A.,   {Bremer} M.~A.~R.,  1997, \mn@doi [\aaps]
  {10.1051/aas:1997358}, \href
  {http://adsabs.harvard.edu/abs/1997A26AS..124..259R} {124}

\bibitem[\protect\citeauthoryear{{Richter}}{{Richter}}{1975}]{richter75}
{Richter} G.~A.,  1975, \mn@doi [Astronomische Nachrichten]
  {10.1002/asna.19752960203}, \href
  {http://adsabs.harvard.edu/abs/1975AN....296...65R} {296, 65}

\bibitem[\protect\citeauthoryear{{Roseboom} et~al.,}{{Roseboom}
  et~al.}{2010}]{roseboom10}
{Roseboom} I.~G.,  et~al., 2010, \mn@doi [\mnras]
  {10.1111/j.1365-2966.2010.17634.x}, \href
  {http://adsabs.harvard.edu/abs/2010MNRAS.409...48R} {409, 48}

\bibitem[\protect\citeauthoryear{{R{\"o}ttgering} et~al.,}{{R{\"o}ttgering}
  et~al.}{2011}]{rottgering11}
{R{\"o}ttgering} H.,  et~al., 2011, \mn@doi [Journal of Astrophysics and
  Astronomy] {10.1007/s12036-011-9129-x}, \href
  {http://adsabs.harvard.edu/abs/2011JApA...32..557R} {32, 557}

\bibitem[\protect\citeauthoryear{{Sargent} et~al.,}{{Sargent}
  et~al.}{2010}]{sargent10b}
{Sargent} M.~T.,  et~al., 2010, \mn@doi [\apjl] {10.1088/2041-8205/714/2/L190},
  \href {http://adsabs.harvard.edu/abs/2010ApJ...714L.190S} {714, L190}

\bibitem[\protect\citeauthoryear{{Scaife} \& {Heald}}{{Scaife} \&
  {Heald}}{2012}]{scaife12}
{Scaife} A.~M.~M.,  {Heald} G.~H.,  2012, \mn@doi [\mnras]
  {10.1111/j.1745-3933.2012.01251.x}, \href
  {http://adsabs.harvard.edu/abs/2012MNRAS.423L..30S} {423, L30}

\bibitem[\protect\citeauthoryear{{Schleicher} \& {Beck}}{{Schleicher} \&
  {Beck}}{2013}]{schleicher13}
{Schleicher} D.~R.~G.,  {Beck} R.,  2013, \mn@doi [\aap]
  {10.1051/0004-6361/201321707}, \href
  {http://adsabs.harvard.edu/abs/2013A26A...556A.142S} {556, A142}

\bibitem[\protect\citeauthoryear{{Schmitt}, {Calzetti}, {Armus}, {Giavalisco},
  {Heckman}, {Kennicutt}, {Leitherer}  \& {Meurer}}{{Schmitt}
  et~al.}{2006}]{schmitt06}
{Schmitt} H.~R.,  {Calzetti} D.,  {Armus} L.,  {Giavalisco} M.,  {Heckman}
  T.~M.,  {Kennicutt} Jr. R.~C.,  {Leitherer} C.,   {Meurer} G.~R.,  2006,
  \mn@doi [\apjs] {10.1086/501529}, \href
  {http://adsabs.harvard.edu/abs/2006ApJS..164...52S} {164, 52}

\bibitem[\protect\citeauthoryear{{Schober}, {Schleicher}  \&
  {Klessen}}{{Schober} et~al.}{2016}]{schober16}
{Schober} J.,  {Schleicher} D.~R.~G.,   {Klessen} R.~S.,  2016, preprint, \href
  {http://adsabs.harvard.edu/abs/2016arXiv160302693S} {} (\mn@eprint {arXiv}
  {1603.02693})

\bibitem[\protect\citeauthoryear{{Seymour} et~al.,}{{Seymour}
  et~al.}{2008}]{seymour08}
{Seymour} N.,  et~al., 2008, \mn@doi [\mnras]
  {10.1111/j.1365-2966.2008.13166.x}, \href
  {http://adsabs.harvard.edu/abs/2008MNRAS.386.1695S} {386, 1695}

\bibitem[\protect\citeauthoryear{{Shapley}}{{Shapley}}{2011}]{shapley11}
{Shapley} A.~E.,  2011, \mn@doi [\araa] {10.1146/annurev-astro-081710-102542},
  \href {http://adsabs.harvard.edu/abs/2011ARA%26A..49..525S} {49, 525}

\bibitem[\protect\citeauthoryear{Simpson et~al.,}{Simpson
  et~al.}{2012}]{simpson12}
Simpson C.,  et~al., 2012, \mn@doi [Monthly Notices of the Royal Astronomical
  Society] {10.1111/j.1365-2966.2012.20529.x}, 421, 3060

\bibitem[\protect\citeauthoryear{{Singh}, {Shastri}, {Ishwara-Chandra}  \&
  {Athreya}}{{Singh} et~al.}{2013}]{singh13}
{Singh} V.,  {Shastri} P.,  {Ishwara-Chandra} C.~H.,   {Athreya} R.,  2013,
  \mn@doi [\aap] {10.1051/0004-6361/201221003}, \href
  {http://adsabs.harvard.edu/abs/2013A26A...554A..85S} {554, A85}

\bibitem[\protect\citeauthoryear{{Smith} et~al.,}{{Smith}
  et~al.}{2014}]{smith14}
{Smith} D.~J.~B.,  et~al., 2014, \mn@doi [\mnras] {10.1093/mnras/stu1830},
  \href {http://adsabs.harvard.edu/abs/2014MNRAS.445.2232S} {445, 2232}

\bibitem[\protect\citeauthoryear{{Smolcic}}{{Smolcic}}{2016}]{smolcic16}
{Smolcic} V.,  2016, preprint, \href
  {http://adsabs.harvard.edu/abs/2016arXiv160305687S} {} (\mn@eprint {arXiv}
  {1603.05687})

\bibitem[\protect\citeauthoryear{{Smol{\v c}i{\'c}} et~al.,}{{Smol{\v c}i{\'c}}
  et~al.}{2008}]{smolcic08}
{Smol{\v c}i{\'c}} V.,  et~al., 2008, \mn@doi [\apjs] {10.1086/588028}, \href
  {http://adsabs.harvard.edu/abs/2008ApJS..177...14S} {177, 14}

\bibitem[\protect\citeauthoryear{{Stevens} et~al.,}{{Stevens}
  et~al.}{2003}]{stevens03}
{Stevens} J.~A.,  et~al., 2003, \nat, \href
  {http://adsabs.harvard.edu/abs/2003Natur.425..264S} {425, 264}

\bibitem[\protect\citeauthoryear{{Tabatabaei} et~al.,}{{Tabatabaei}
  et~al.}{2013}]{tabatabaei13}
{Tabatabaei} F.~S.,  et~al., 2013, \mn@doi [\aap]
  {10.1051/0004-6361/201220249}, \href
  {http://adsabs.harvard.edu/abs/2013A%26A...552A..19T} {552, A19}

\bibitem[\protect\citeauthoryear{{Tasse}, {Le Borgne}, {R{\"o}ttgering},
  {Best}, {Pierre}  \& {Rocca-Volmerange}}{{Tasse} et~al.}{2008}]{tasse08}
{Tasse} C.,  {Le Borgne} D.,  {R{\"o}ttgering} H.,  {Best} P.~N.,  {Pierre} M.,
    {Rocca-Volmerange} B.,  2008, \mn@doi [\aap] {10.1051/0004-6361:20078453},
  \href {http://adsabs.harvard.edu/abs/2008A%26A...490..879T} {490, 879}

\bibitem[\protect\citeauthoryear{{Taylor}}{{Taylor}}{2006}]{taylor06}
{Taylor} M.~B.,  2006, in {Gabriel} C.,  {Arviset} C.,  {Ponz} D.,   {Enrique}
  S.,  eds,  Astronomical Society of the Pacific Conference Series Vol. 351,
  Astronomical Data Analysis Software and Systems XV. p.~666

\bibitem[\protect\citeauthoryear{{Vardoulaki} et~al.,}{{Vardoulaki}
  et~al.}{2015}]{vardoulaki15}
{Vardoulaki} E.,  et~al., 2015, \mn@doi [\aap] {10.1051/0004-6361/201424125},
  \href {http://adsabs.harvard.edu/abs/2015A26A...574A...4V} {574, A4}

\bibitem[\protect\citeauthoryear{{Voelk}}{{Voelk}}{1989}]{voelk89}
{Voelk} H.~J.,  1989, \aap, \href
  {http://adsabs.harvard.edu/abs/1989A%26A...218...67V} {218, 67}

\bibitem[\protect\citeauthoryear{{Whittam}, {Riley}, {Green}, {Davies},
  {Franzen}, {Rumsey}, {Schammel}  \& {Waldram}}{{Whittam}
  et~al.}{2016}]{whittam16}
{Whittam} I.~H.,  {Riley} J.~M.,  {Green} D.~A.,  {Davies} M.~L.,  {Franzen}
  T.~M.~O.,  {Rumsey} C.,  {Schammel} M.~P.,   {Waldram} E.~M.,  2016, \mn@doi
  [\mnras] {10.1093/mnras/stv2960}, \href
  {http://adsabs.harvard.edu/abs/2016MNRAS.457.1496W} {457, 1496}

\bibitem[\protect\citeauthoryear{{Williams} \& {Bower}}{{Williams} \&
  {Bower}}{2010}]{williamsandbower10}
{Williams} P.~K.~G.,  {Bower} G.~C.,  2010, \mn@doi [\apj]
  {10.1088/0004-637X/710/2/1462}, \href
  {http://adsabs.harvard.edu/abs/2010ApJ...710.1462W} {710, 1462}

\bibitem[\protect\citeauthoryear{{Williams}, {Intema}  \&
  {R{\"o}ttgering}}{{Williams} et~al.}{2013}]{williams13}
{Williams} W.~L.,  {Intema} H.~T.,   {R{\"o}ttgering} H.~J.~A.,  2013, \mn@doi
  [\aap] {10.1051/0004-6361/201220235}, \href
  {http://adsabs.harvard.edu/abs/2013A26A...549A..55W} {549, A55}

\bibitem[\protect\citeauthoryear{{Williams} et~al.,}{{Williams}
  et~al.}{2016}]{williams16}
{Williams} W.~L.,  et~al., 2016, \mn@doi [\mnras] {10.1093/mnras/stw1056},
  \href {http://adsabs.harvard.edu/abs/2016MNRAS.460.2385W} {460, 2385}

\bibitem[\protect\citeauthoryear{{Yun} \& {Carilli}}{{Yun} \&
  {Carilli}}{2002}]{yun02}
{Yun} M.~S.,  {Carilli} C.~L.,  2002, \mn@doi [\apj] {10.1086/338924}, \href
  {http://adsabs.harvard.edu/abs/2002ApJ...568...88Y} {568, 88}

\bibitem[\protect\citeauthoryear{{Yun}, {Reddy}  \& {Condon}}{{Yun}
  et~al.}{2001}]{yun01}
{Yun} M.~S.,  {Reddy} N.~A.,   {Condon} J.~J.,  2001, \mn@doi [\apj]
  {10.1086/323145}, \href {http://adsabs.harvard.edu/abs/2001ApJ...554..803Y}
  {554, 803}

\bibitem[\protect\citeauthoryear{{da Cunha} et~al.,}{{da Cunha}
  et~al.}{2015}]{dacunha15}
{da Cunha} E.,  et~al., 2015, \mn@doi [\apj] {10.1088/0004-637X/806/1/110},
  \href {http://adsabs.harvard.edu/abs/2015ApJ...806..110D} {806, 110}

\bibitem[\protect\citeauthoryear{{de Jong}, {Klein}, {Wielebinski}  \&
  {Wunderlich}}{{de Jong} et~al.}{1985}]{dejong85}
{de Jong} T.,  {Klein} U.,  {Wielebinski} R.,   {Wunderlich} E.,  1985, \aap,
  \href {http://adsabs.harvard.edu/abs/1985A%26A...147L...6D} {147, L6}

\bibitem[\protect\citeauthoryear{{de Vries}, {Morganti}, {R{\"o}ttgering},
  {Vermeulen}, {van Breugel}, {Rengelink}  \& {Jarvis}}{{de Vries}
  et~al.}{2002}]{devries02}
{de Vries} W.~H.,  {Morganti} R.,  {R{\"o}ttgering} H.~J.~A.,  {Vermeulen} R.,
  {van Breugel} W.,  {Rengelink} R.,   {Jarvis} M.~J.,  2002, \mn@doi [\aj]
  {10.1086/338906}, \href {http://adsabs.harvard.edu/abs/2002AJ....123.1784D}
  {123, 1784}

\bibitem[\protect\citeauthoryear{{van Haarlem} et~al.,}{{van Haarlem}
  et~al.}{2013}]{LOFAR}
{van Haarlem} M.~P.,  et~al., 2013, \mn@doi [\aap]
  {10.1051/0004-6361/201220873}, \href
  {http://adsabs.harvard.edu/abs/2013A26A...556A...2V} {556, A2}

\bibitem[\protect\citeauthoryear{{van Weeren} et~al.,}{{van Weeren}
  et~al.}{2016}]{vanweeren16}
{van Weeren} R.~J.,  et~al., 2016, \mn@doi [\apjs] {10.3847/0067-0049/223/1/2},
  \href {http://adsabs.harvard.edu/abs/2016ApJS..223....2V} {223, 2}

\makeatother
\end{thebibliography}

\appendix 
\section{Luminosity vs redshift dependence of the IRC }\label{app:IRClum}

\begin{figure}
\centering
    \includegraphics[trim={0cm 1.5cm 0 2cm}, clip, width=1.1\linewidth]{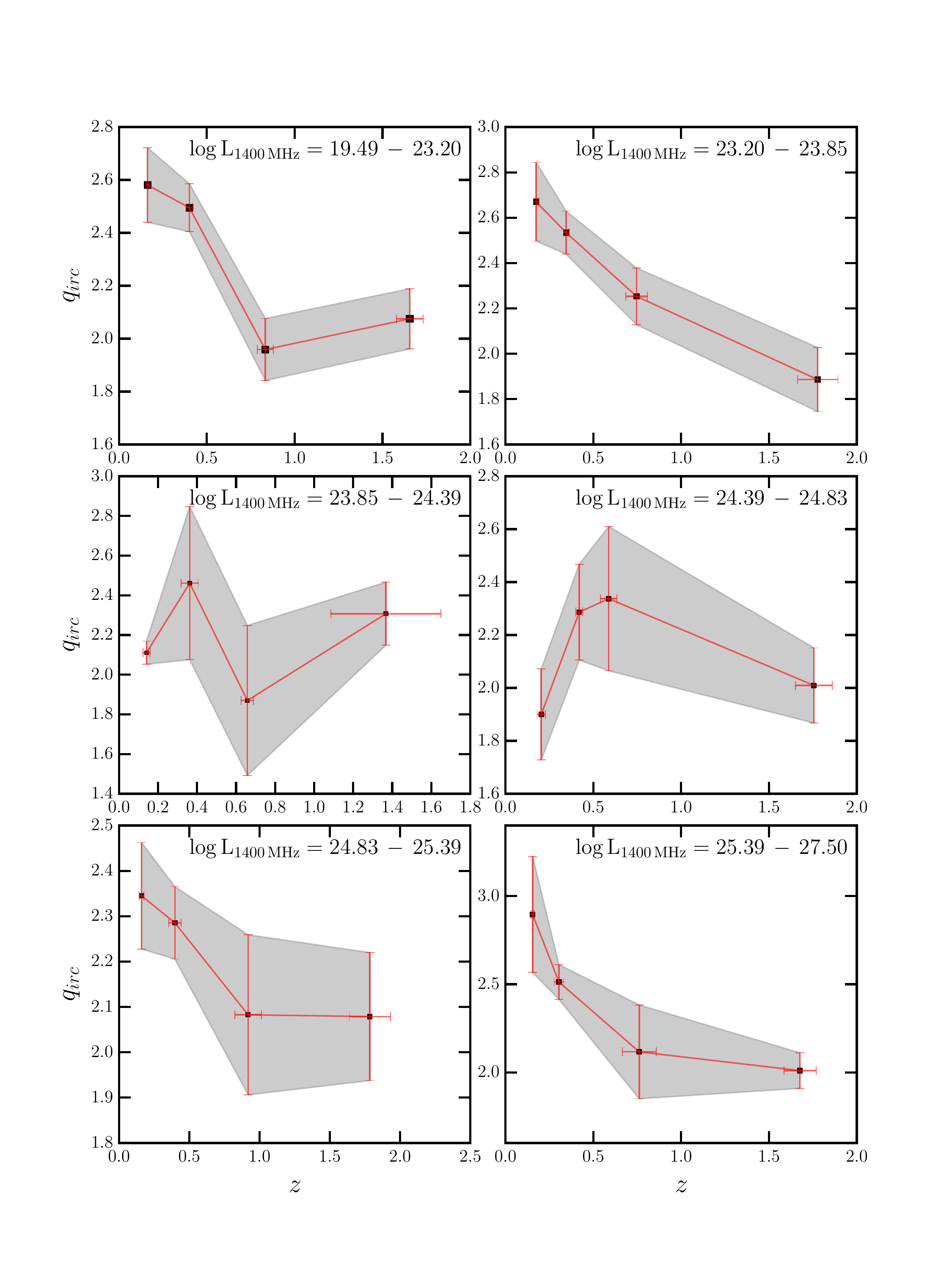}
\caption{q-value for the IRC corresponding to the radio at 1.4 GHz plotted against redshift for different luminosity bins. Error bars correspond to bootstrap errors.}
\label{fig:FRC1.4-lumbins}
\end{figure}

\begin{figure}
\centering
\includegraphics[trim={1cm 1.5cm 0 2cm},clip,width=1.1\linewidth]{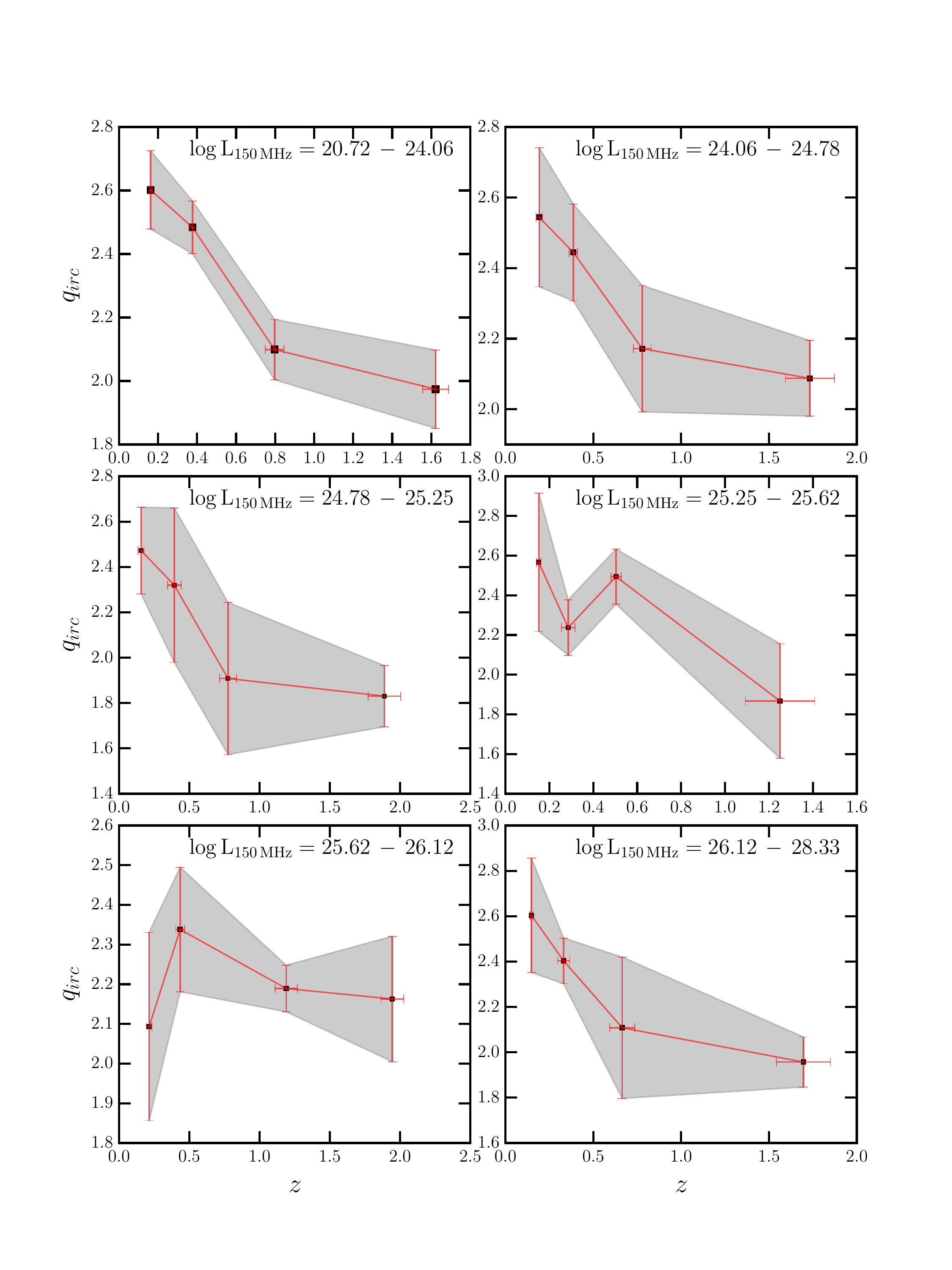}
\caption{q-value for the IRC corresponding to the radio at 150 MHz plotted against redshift for different luminosity bins. Error bars correspond to bootstrap errors.}
\label{fig:FRC150-lumbins}
\end{figure}

Figures \ref{fig:FRC1.4-lumbins} and \ref{fig:FRC150-lumbins}  show that the redshift evolution is found also within subsamples of similar luminosity, suggesting the trend observed in Figs.~\ref{fig:FRC1.4} and \ref{fig:FRClofar} is not the consequence of a dependence on luminosity. 

% % --------------------------------------------------------
% %  FIGURE FRC of 1.4 GHz VS REDSHIFT VS LUM- ALL
% %---------------------------------------------------------
% %----------------------------------------------------------

% --------------------------------------------------------
%  FIGURE FRC of 150GHz VS REDSHIFT VS LUM- ALL
%---------------------------------------------------------
%-------------------------------------------------

\section{SED-fitting examples}\label{app:SED}

Examples of SED fitting of galaxies in our sample using the \textsc{AGNfitter} code.
% --------------------------------------------------------
%  FIGURE \textsc{AGNfitter} output on galaxies
%---------------------------------------------------------

 \begin{figure*}
      \includegraphics[trim={0 0.5cm 0.5cm 0},clip,width=0.48\linewidth]{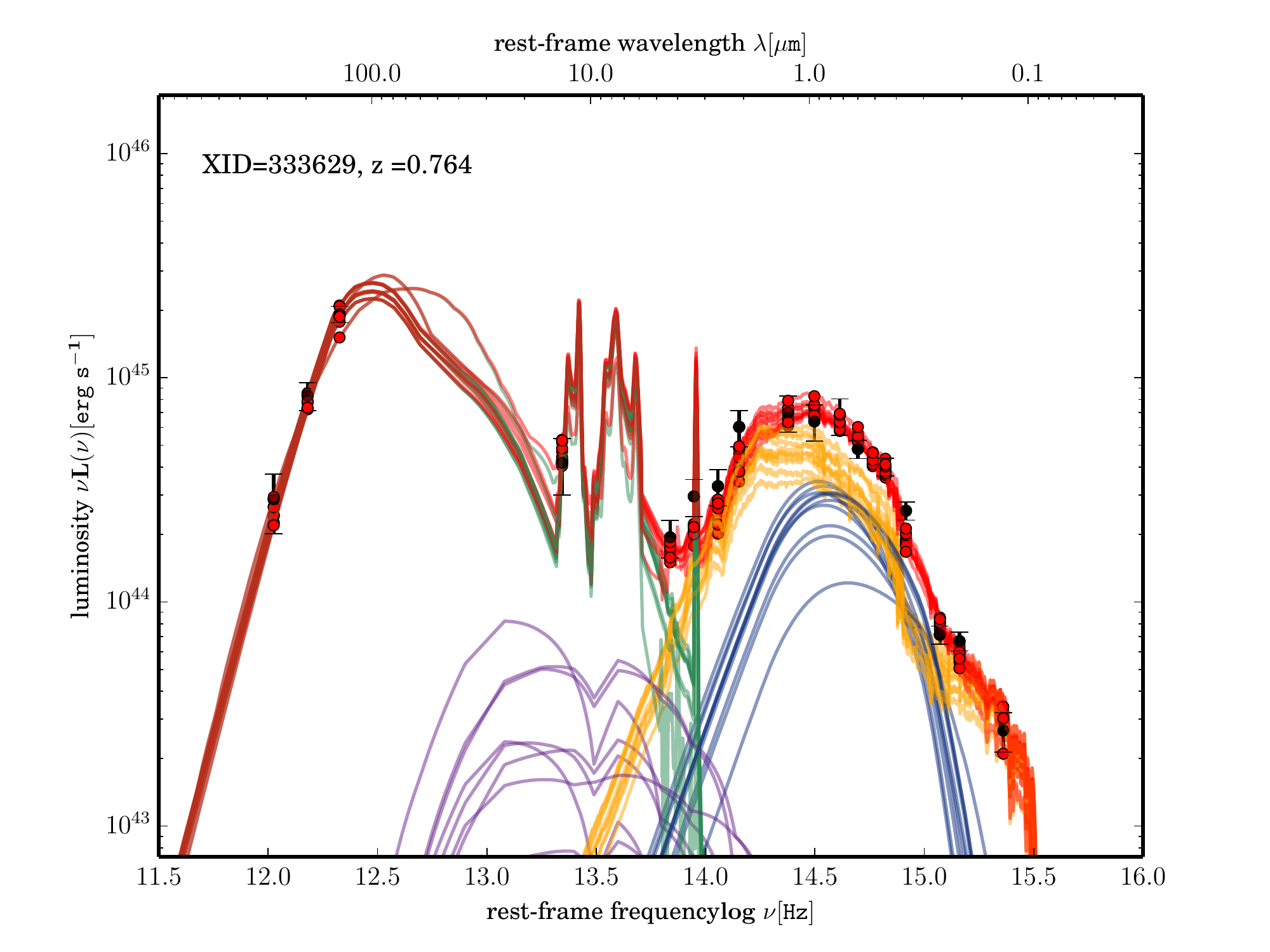}
      \includegraphics[trim={0 0.5cm 0.5cm 0cm},clip,width=0.48\linewidth]{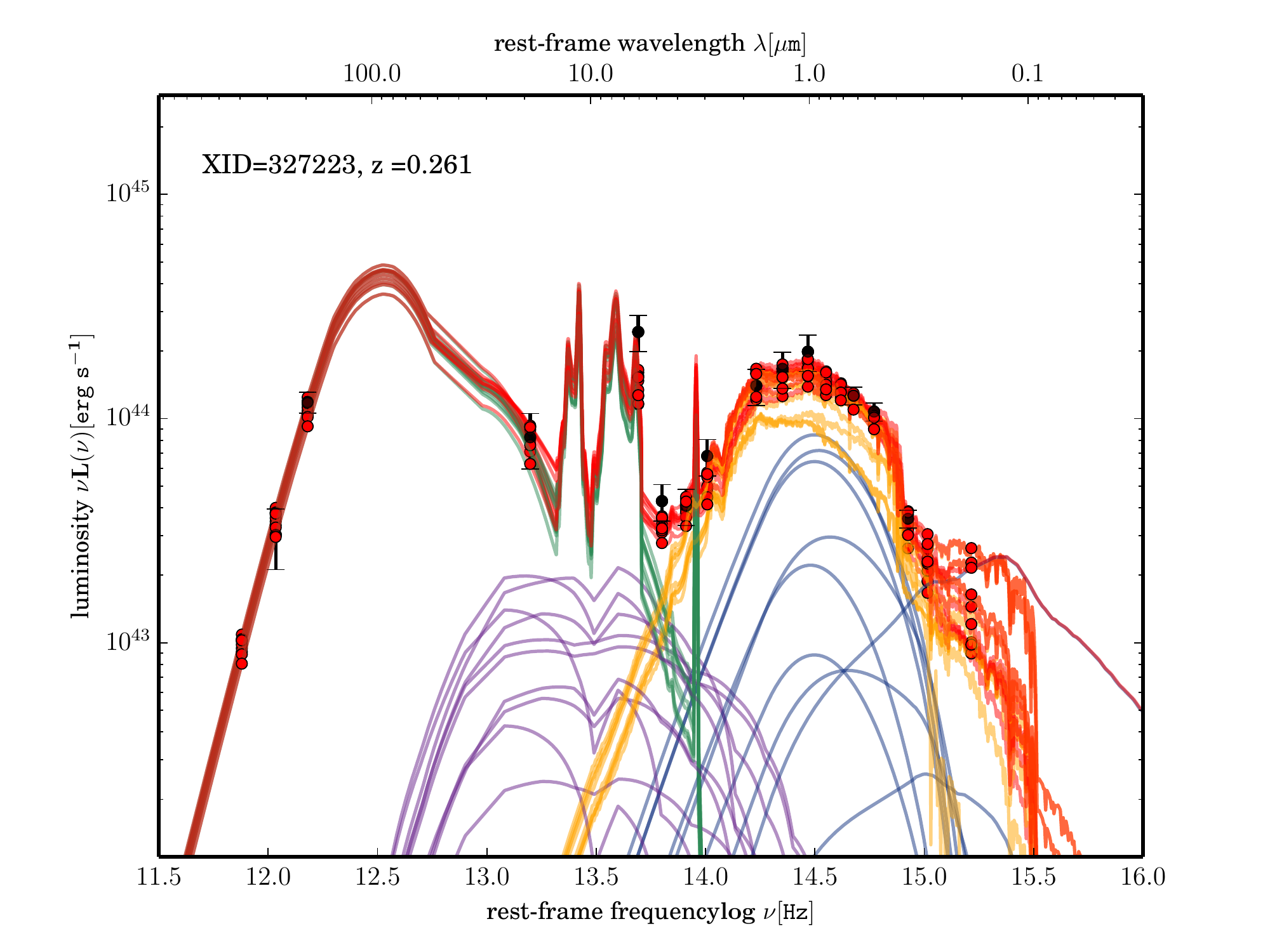}
     \includegraphics[trim={0 0.5cm 0.5cm 0},clip,width=0.48\linewidth]{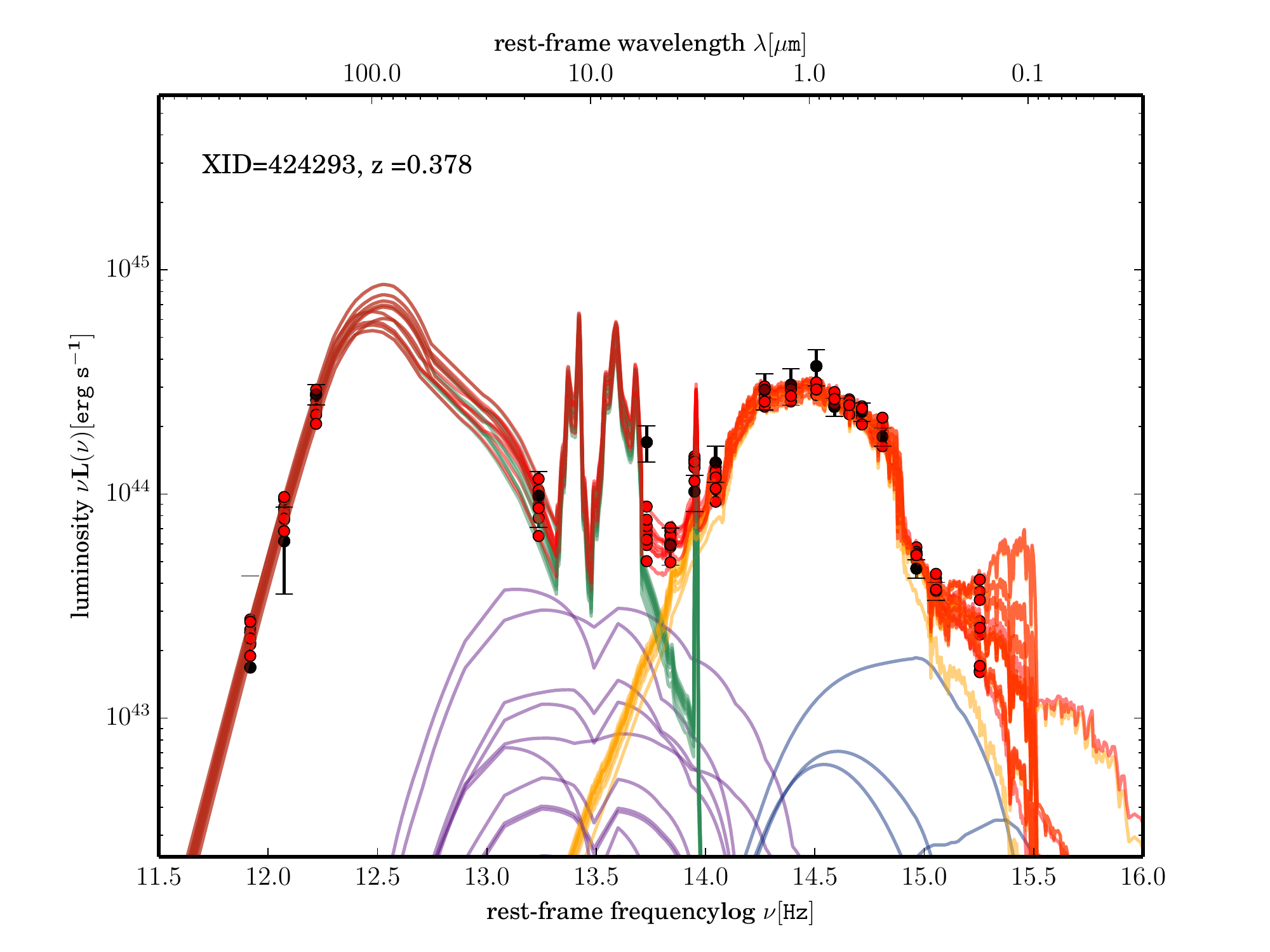}
      \includegraphics[trim={0 0.5cm 0.5cm 0cm},clip,width=0.48\linewidth]{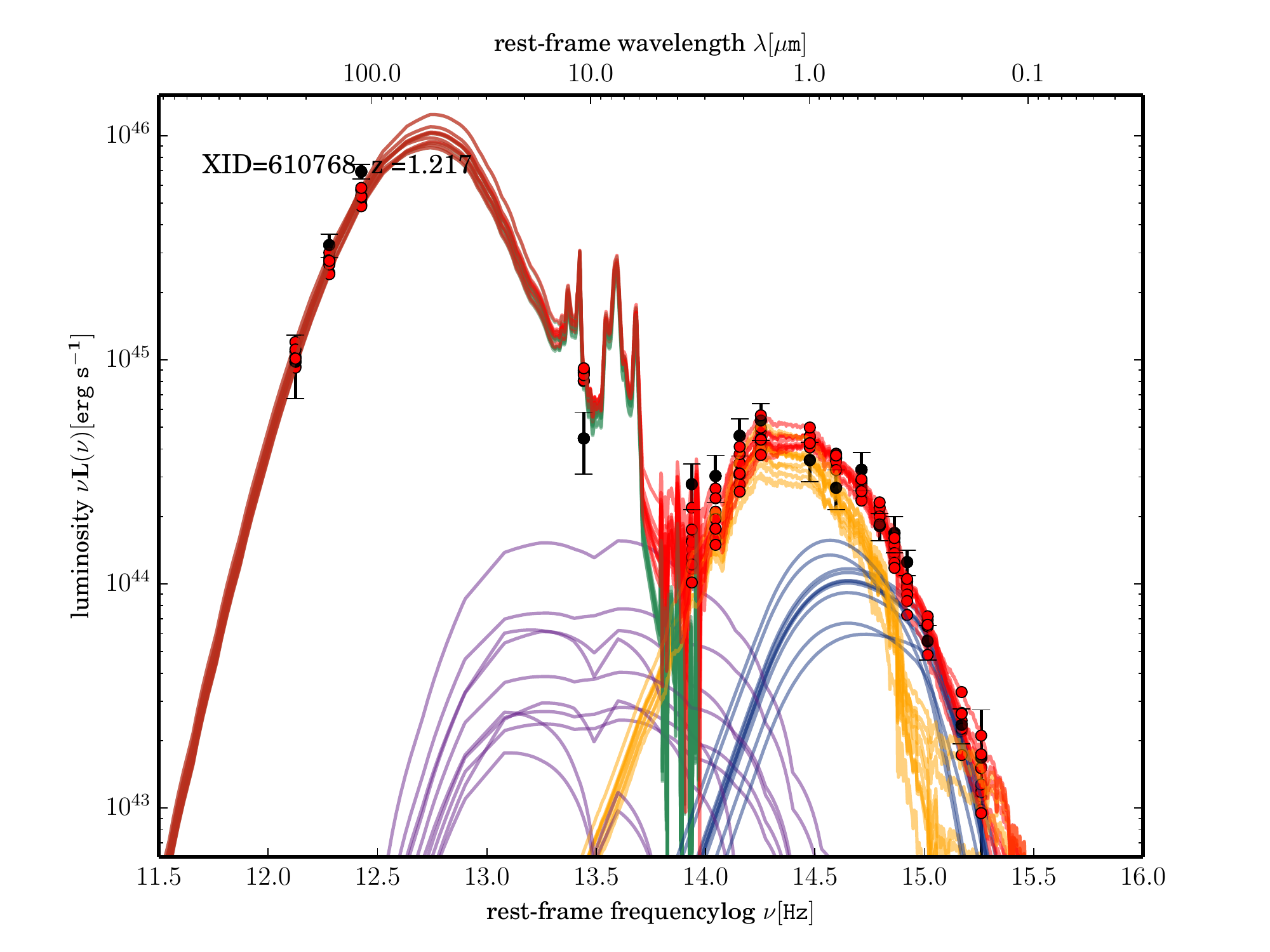}
       \includegraphics[trim={0 0.5cm 0.5cm 0},clip,width=0.48\linewidth]{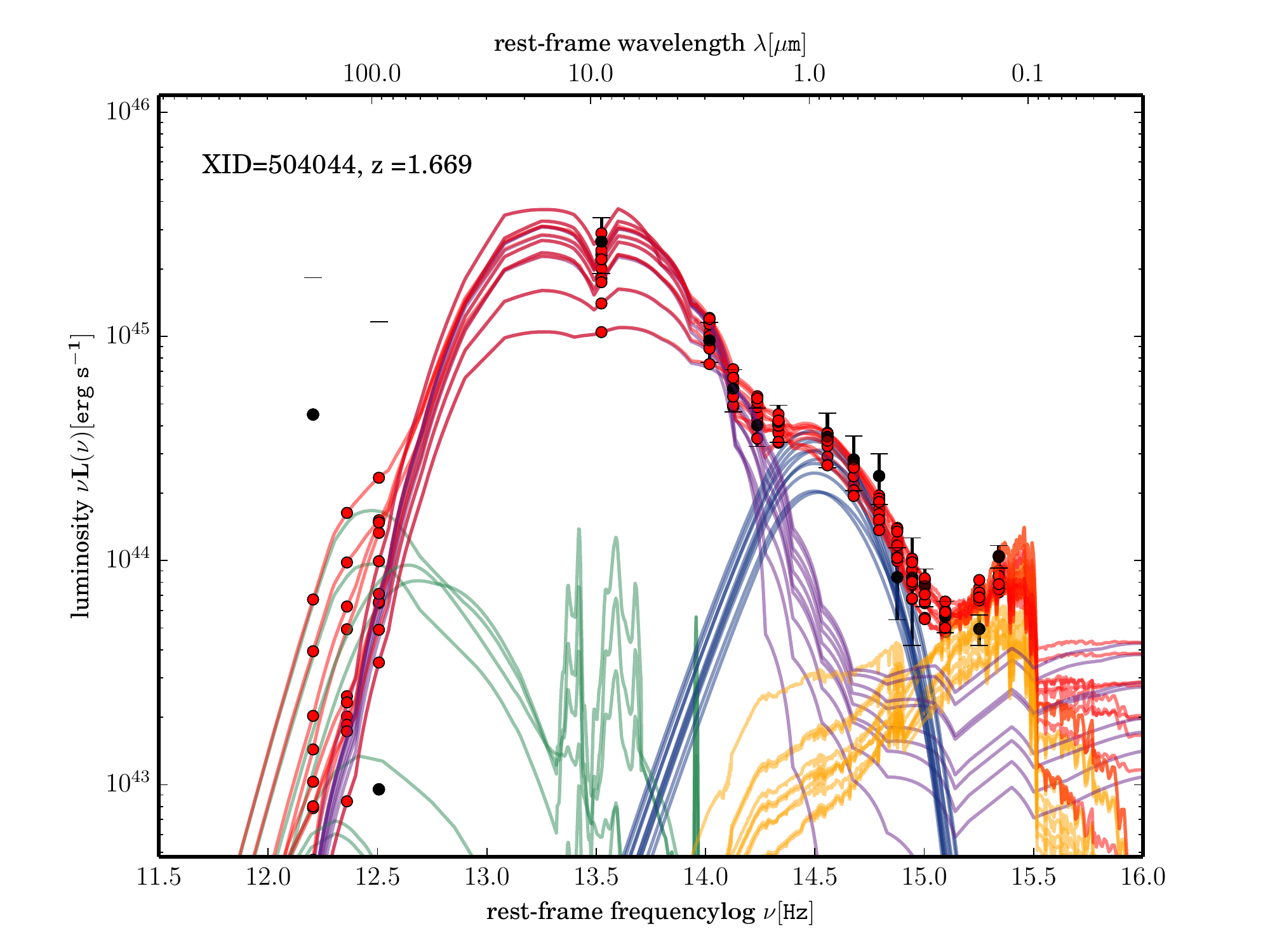}
      \includegraphics[trim={0 0.5cm 0.5cm 0cm},clip,width=0.48\linewidth]{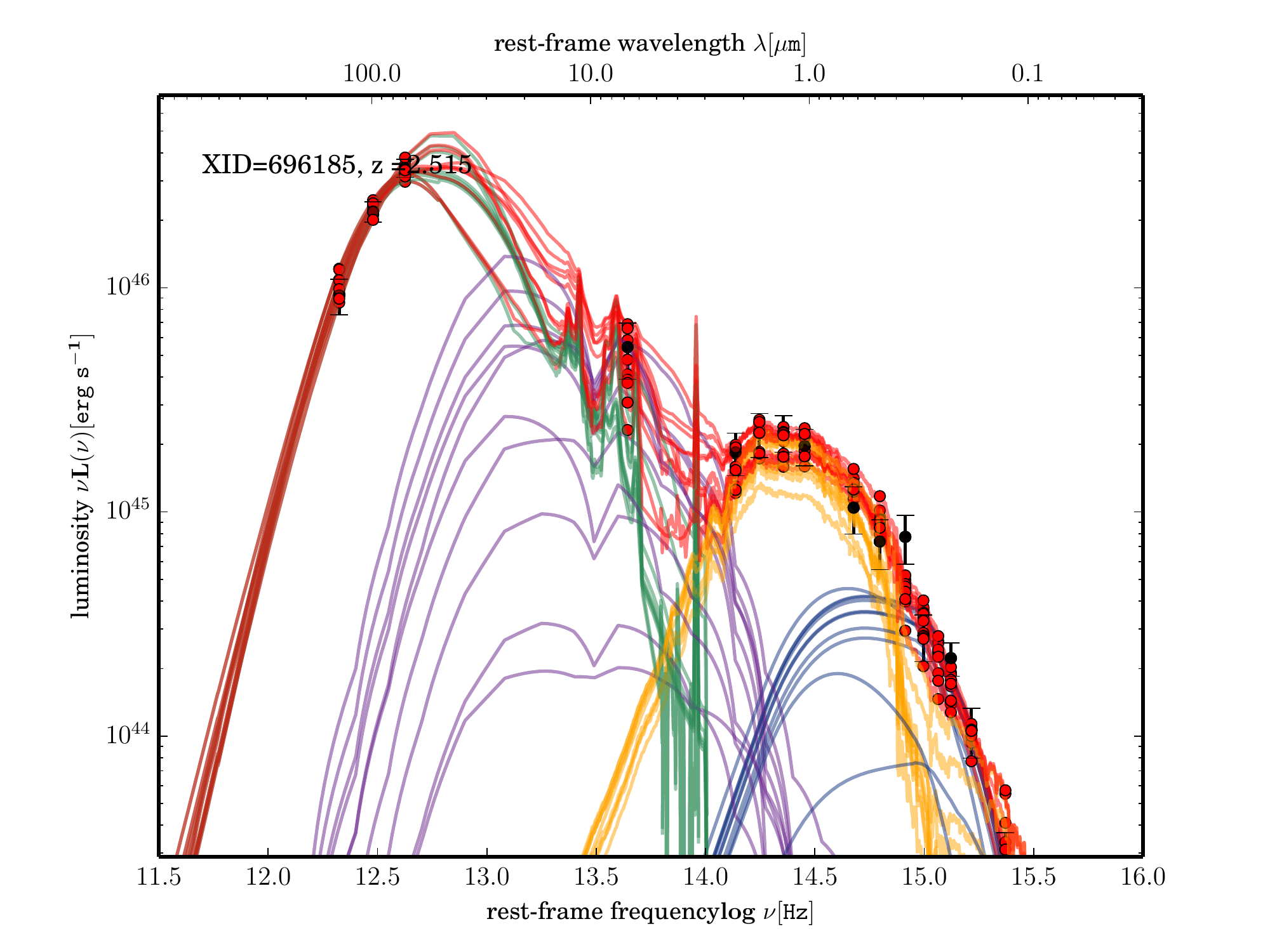}
        
    \caption{SED-fitting examples: Photometric data points for each source are plotted as black circular markers with error bars. We pick 8 different realizations from the parameters posterior probability distributions and over-plot them in order to visualize the effect of the parameters' uncertainties on the SEDS. The SED shapes of the physical components are presented as solid lines: the galactic cold dust emission (green), the hot dust emission from the AGN torus (purple), the stellar emission (orange) and the accretion disk emission (blue).The linear combination of these, the 'total SED', is depicted as a red line.} 
    \label{fig:SEDs}
  \end{figure*}

\end{document}